\documentclass[11pt,a4paper]{article}
\usepackage{jheppub} 
\usepackage{graphicx}
\usepackage{bm}

\usepackage[skip=3pt]{caption}
\usepackage{framed}
\usepackage{xcolor}
\usepackage{rotating}
\usepackage{empheq}
\usepackage{afterpage}

\usepackage[utf8]{inputenc}

\usepackage{tabularray}

\usepackage[skip=3pt]{subcaption}
\usepackage{float}
\usepackage{enumerate}
\usepackage{enumitem}

\newcommand{\lbar}{\lower0.2ex\hbox{$\mathchar'26$}\mkern-10mu \lambda}

\def\Im{\mathrm{Im}}
\def\Re{\mathrm{Re}}

\def\mrmd{{\mathrm{d}}}

\def\th{{\mathrm{th}}}

\def\nd{{\mathrm{nd}}}
\def\exp{\mathrm{exp}}

\def\a{\alpha}
\def\b{\beta}
\def\d{\delta}
\def\D{\Delta}

\def\ve{\varepsilon}

\def\g{\gamma}
\def\G{\Gamma}

\def\k{\kappa}
\def\l{\lambda}

\def\n{\nu}
\def\o{\omega}
\def\O{\Omega}

\def\s{\sigma}
\def\S{\Sigma}
\def\t{\theta}

\def\x{\xi}
\def\z{\zeta}

\def\mfrS{\mathfrak{S}}

\def\nd{{\mathrm{nd}}}

\def\mcalA{\mathcal{A}}
\def\mcalB{{\mathcal{B}}}
\def\mcalC{\mathcal{C}}

\def\mcalF{\mathcal{F}}
\def\mcalH{\mathcal{H}}

\def\mcalV{{\mathcal{V}}}

\def\mcalM{{\mathcal{M}}} 
\def\mcalN{\mathcal{N}}

\def\mcalG{\mathcal{G}}

\def\mcalK{\mathcal{K}}

\def\mcalPT{\mathcal{PT}}

\def\mrmD{{\mathrm{D}}}

\def\mbbZ{\mathbb{Z}}
\def\mbbR{\mathbb{R}}

\def\mbbC{\mathbb{C}}

\def\*{\star}

\def\tA{\tilde A}

\def\tJ{\tilde J}
\def\tK{\tilde K}
\def\tH{\tilde H}
\def\tM{\tilde M}

\def\tV{\tilde V}

\def\tb{\tilde b}

\def\ts{\tilde s}
\def\tu{\tilde u}

\def\mcaltG{\tilde{\mathcal{G}}}

\def\mrmLeft{\mathrm{Left}}
\def\mrmRight{\mathrm{Right}}

\def\mrmTV{\mathrm{TV}}
\def\mrmFV{\mathrm{FV}}
\def\mrmBT{\mathrm{BT}}

\def\mrmA{\mathrm{A}} 

\def\mrmF{\mathrm{F}}

\def\mrmB{\mathrm{B}}

\def\mrmL{\mathrm{L}}
\def\mrmR{\mathrm{R}}

\def\mrmT{\mathrm{T}}

\def\mrmi{\mathrm{i}}
\def\mrmii{\mathrm{ii}}

\def\Res{\mathrm{Res}}
\def\med{\mathrm{med}}
\def\iin{\mathrm{in}}
\def\out{\mathrm{out}}
\def\dual{\mathrm{dual}}

\def\<{\langle}
\def\>{\rangle} 
\def\dee{\partial}

\def\Im{\mathrm{Im}}
\def\Re{\mathrm{Re}}

\def\mrmTDW{\mathrm{TDW}}
\def\mrmATW{\mathrm{ATW}}
\def\mrmSTW{\mathrm{STW}}
\def\mrmDSTW{\mathrm{DSTW}}

\usepackage{slashed}

\definecolor{darkbrown}{rgb}{0.4, 0.26, 0.13}
\definecolor{paleblue}{rgb}{0.69, 0.93, 0.93}
\definecolor{lightskyblue}{rgb}{0.53, 0.81, 0.98}
\definecolor{skyblue}{rgb}{0.53, 0.81, 0.92}
\definecolor{darkred}{rgb}{0.55, 0.0, 0.0}
\definecolor{darkblue}{rgb}{0.0, 0.0, 0.55}
\definecolor{darkpastelgreen}{rgb}{0.01, 0.75, 0.24}
\definecolor{lightgreen}{rgb}{0.56, 0.93, 0.56}

\usepackage{babel}
\usepackage{beramono}
\usepackage[T1]{fontenc}

\definecolor{identifiercolor}{rgb}{.4,.6,.56}
\definecolor{stringcolor}{gray}{0.5}
\definecolor{inactivecolor}{rgb}{0.15,0.15,0.5}

\usepackage{listings}
\title{\boldmath Exact WKB in all sectors II: Potentials with non-degenerate saddles}

\author[1,2]{Tatsuhiro Misumi,}
\emailAdd{misumi@phys.kindai.ac.jp}
\affiliation[1]{Department of Physics, Kindai University, Osaka 577-8502, Japan}
\affiliation[2]{Research and Education Center for Natural Sciences, Keio University, Kanagawa 223-8521, Japan}

\author[3]{Cihan Pazarba\c{s}\i{}}
\emailAdd{cihan.pazarbasi@gmail.com}
\affiliation[3]{Okinawa Institute of Science and Technology, Okinawa 904-0495, Japan}

\abstract{We discuss the exact quantization of general one-dimensional potentials in view of the exact-WKB formalism. Building on our previous work, we perform analytic continuations across different sectors via the complexification to the spectral (energy) parameter $u$ and identify continuous and discontinuous transitions of the exact spectrum for generic potentials. When the transition is discontinuous, it is characterized by the Stokes phenomena, inducing different exact (median) quantization conditions, thereby distinct trans-series structures valid in different sectors. We analyze two illustrative examples, namely asymmetric triple-well (ATW) and tilted double-well (TDW), and verify the general qualitative analysis by deriving exact (median) quantization conditions in each sector. Moreover, by obtaining the trans-series solutions for each system, we identify bion/bounce configurations and show that the trans-series of ATW is organized in accordance with the cluster expansion of the bion gas and there should exist a previously neglected complex saddle in the TDW system. These identifications further strengthen the link between path integral and exact-WKB formalisms, while also demonstrating the predictive power of the latter. In parallel, for the P-NP relations of genus-1 systems, we derive transformation rules between any perturbative and non-perturbative pair of WKB-cycles. Our results show that the entire resurgence data of a genus-1 system transforms only by the change of classical parameters, i.e. frequencies and bion/bounce actions, and the perturbative energy series. This also reveals the underlying reasons of the previously found $S$-duality transformations.

} 

\begin{document}
\maketitle
\flushbottom	

\section{Introduction}\label{Section: Introduction}

Formal WKB (Wentzel-Kramers-Brillouin) methods can be made ``exact'' by incorporating the analytic continuation and the resurgence theory, leading to a framework called Exact WKB (EWKB) \cite{BPV, Voros1983,Silverstone, AKT1, AKT2, Takei1, DDP2, DP1, Takei2, Kawai1, Takei3, AKT3, AKT4, Schafke1, Iwaki1}. In recent years, it has been proven as a powerful framework to understand non-perturbative phenomena in quantum mechanics \cite{Sueishi:2020rug, Sueishi:2021xti, Kamata:2021jrs,Misumi:2024gtf}, yielding exact quantization conditions that include multi-instanton effects, which was also explored in various other methods \cite{Alvarez1,Alvarez2,Alvarez3,Alvarez4,Zinn-Justin:2004vcw,Zinn-Justin:2004qzw, Jentschura:2004jg, Jentschura:2010zza, Jentschura:2011zza, Basar:2013eka, Dunne:2013ada, Dunne:2014bca,Basar:2015xna,Misumi:2015dua,Dunne:2016qix,Dunne:2020gtk, Behtash:2015loa,Behtash:2015zha,Behtash:2017rqj, Gahramanov:2015yxk,Kozcaz:2016wvy, Dunne:2016jsr, Fujimori:2016ljw, Basar:2017hpr, Fujimori:2017oab, Fujimori:2017osz,Behtash:2018voa, Sueishi:2019xcj, Fujimori:2022lng}, and enabling us to explore the resurgent structure, which appears in the analysis of various quantum-mechanical and field-theoretical systems \cite{Zinn-Justin:1981qzi,Zinn-Justin:1983yiy, Unsal:2007vu,Unsal:2007jx,Shifman:2008ja,Poppitz:2009uq,Anber:2011de,Poppitz:2012sw,Misumi:2014raa,Misumi:2014jua,Misumi:2014rsa,Misumi:2014bsa,Misumi:2016fno,Fujimori:2018kqp,Hongo:2018rpy,Fujimori:2018nvz,Fujimori:2019skd,Misumi:2019upg,Fujimori:2020zka,Fujimori:2021oqg,Unsal:2020yeh,Nishimura:2021lno,Pazarbasi:2021ifb,Pazarbasi:2021fey}. Moreover, it has been widely used in analysis of various theoretical and physical setups: 4d $\mcalN=2$ gauge theories \cite{Nekrasov:2009rc,Mironov:2009uv,Kashani-Poor:2015pca,Kashani-Poor:2016edc,Ashok:2016yxz,Yan:2020kkb}, 
wall-crossing phenomena \cite{Gaiotto:2012rg,Allegretti:2020dyt}, 
ODE/IM correspondence \cite{Dorey:2001uw, Dorey:2007zx,Ito:2018eon,Ito:2019jio,Imaizumi:2020fxf,Ito:2025pfo}, 
TBA equations \cite{Emery:2020qqu,Ito:2024nlt,Ito:2025sgq}, 
topological string theory \cite{Grassi:2014cla,Grassi:2014zfa,Codesido:2017dns,Codesido:2017jwp,Hollands:2019wbr,Ashok:2019gee,Coman:2020qgf,Iwaki:2023cek}, 
PT-symmetric quantum mechanics \cite{Kamata:2023opn,Kamata:2024tyb}, time-dependent quantum systems \cite{Taya:2020dco,Enomoto:2020xlf, Enomoto:2022mti,Namba:2025ejw,Fujimori:2025kkc},
black hole quasinormal modes \cite{Miyachi:2025ptm},
the resonance states \cite{Morikawa:2025grx,Morikawa:2025xjq,Morikawa:2025vvs,Morikawa:2025ezl}, and 
other subjects \cite{Duan:2018dvj,Imaizumi:2022dgj,vanSpaendonck:2022kit,Bucciotti:2023trp,vanSpaendonck:2024rin, Ture:2024nbi}.


In the first paper of the series \cite{Misumi:2024gtf}, we developed an approach within EWKB that is based on an analytic continuation of the spectral parameter (i.e. energy eigenvalue) $u$ into the complex plane and continuously connecting exact quantization conditions in all sectors for locally harmonic potentials with degenerate saddles. We analyzed periodic potentials and symmetric double-well potentials, showed that a redefinition of $A$-cycle\footnote{Throughout this paper, we use the canonical definitions of $A$ and $B$ cycles: Around a bottom of one dimensional potentials, an $A$-cycle has a perturbative character, while a $B$-cycle has a non-perturbative one, which is also associated with the quantum mechanical tunneling. We keep these labels even when the perturbative or non-perturbative characters of the cycles changes in different spectral regions of the corresponding potential, e.g. around barrier tops.} keeps quantization conditions real and more importantly intact. We also used the latter to show that the trans-series structure remains intact in all sectors, 
meaning that it encodes the entire spectrum. Moreover, using the exchange of the properties of $A$ and $B$ cycles, which is encoded in the Weber-type EWKB dictionary, we conjectured $S$-duality transformations, mapping P-NP relations between dual potentials. 

In this work, we extend the previous study to general one-dimensional quantum mechanical systems without degeneracy between its saddles. Such systems do not possess symmetries that simplify the quantization procedure as well as the P-NP relations. A completely asymmetric structure of the perturbative saddles prevents any perturbative degeneracy, which is familiar from the symmetric double and triple well as well as periodic potential studies in the literature (See e.g. \cite{Dunne:2014bca,Basar:2015xna,Misumi:2015dua,Dunne:2016qix,Dunne:2020gtk,Sueishi:2020rug,Sueishi:2021xti}). More importantly, such a system possesses multiple linearly independent non-perturbative sectors encoding the trans-series structure, which is a significant difference from the widely studied case of the symmetric potentials.

A generic asymmetric potential can also have minima lying at different classical energy levels, leading to (possibly several) false vacuum sectors and a true vacuum sector\footnote{Of course, a vacuum is a state with no particle in relativistic field theories. Although we only consider non-relativistic systems, we keep the analogy between metastable states and false vacuum for simplicity.}. To our knowledge, an exact quantization procedure handling such potentials has not been discussed in the literature. As an extension of our analytic continuation procedure introduced in \cite{Misumi:2024gtf}, we analyze how false vacuum and true vacuum sectors can be connected to each other and show that such a connection is not a continuous one, leading to different trans-series solutions encoding the spectrum in respective sectors. We note that this property is a significant difference from the case of degenerate perturbative saddles or non-degenerate ones, which are at the same classical level. 

In addition to the exact quantization procedure, the generic potentials also have a more complicated resurgence structure, which is formulated as P-NP relations for genus-1 potentials \cite{Alvarez2,Alvarez3,Alvarez4,Dunne:2013ada,Dunne:2014bca}. Although explicit P-NP relations are not known for higher genus curves, recently the original relations for symmetric potentials have been generalized to deformed genus-1 potentials \cite{Cavusoglu:2023bai,Cavusoglu:2024usn}. In this paper, using their construction, we analyze how the $S$-duality transformations arise from the changes in the saddle frequencies and/or the instanton actions. Moreover, we also show that the same transformation rules link the P-NP relations associated with the different saddles of the original potential.  In this way, we relate the resurgence structures of all WKB cycles of a genus-1 potential.
\\

The paper is constructed as follows: In section~\ref{Section: EWKB_review}, after reviewing the EWKB framework, we present necessary formulae to compute the actions of WKB cycles of generic potentials using the Weber-type EWKB dictionary. Later, in Section~\ref{Section: Transition}, we present the analytic continuation procedure connecting false and true vacuum sectors with emphasis on the discontinuous nature of the transition. In Section~\ref{Section: P_NP_revisit}, starting with the P-NP relation of deformed genus-1 potentials, we derive the transformation rules for both undeformed and deformed genus-1 cases, focusing on the tilted double well potential for the latter one. Then, in the second half of the paper, we focus on two representative examples, namely the asymmetric triple-well in Section~\ref{Section: ATW} and the tilted double-well in Section~\ref{Section: TDW}. We discuss their exact quantization and numerically verify the qualitative arguments of Sections~\ref{Section: EWKB_review} and \ref{Section: P_NP_revisit}. 

The asymmetric triple-well potential is a system that consists of three wells and two barriers. At the classical level, its minima are at the same level but have different frequencies, indicating non-degeneracy. It also has two linearly independent non-perturbative configurations (bions). In Section~\ref{Section: ATW_Transition}, we show that the transitions of the energy level across barrier tops are continuous; thus there is only one trans-series for the entire spectrum. Then, in Section~\ref{Section: ATW_trans-series}, by obtaining the trans-series solutions from the exact quantization conditions, we present how a collaboration between distinct bions is necessary to keep the spectrum real. Our solution also reveals the cluster expansion property of the trans-series, which originates from the weakly interacting instanton gas picture in an analogy to the statistical mechanics and was previously studied in the symmetric cases \cite{Behtash:2018voa,Pazarbasi:2021ifb,Pazarbasi:2021fey}. Later, in Section~\ref{Section: ATW_SymmetricLimit}, focusing on the symmetric limit, which brings back the degeneracy of saddles, we verify the transformation rules for the corresponding P-NP relations for both the triple-well and its dual potential. Finally, we analyze the exact quantization for the PT-symmetric version of the dual symmetric potential, showing the reality of its spectrum.

Finally, in Section~\ref{Section: TDW}, we analyze the exact quantization of the tilted double-well as an example of a quantum system with distinct false vacuum and true vacuum sectors, which are discontinuously connected and have two different trans-series solutions. By analyzing the trans-series solutions, we also identify the necessity of a complex non-perturbative saddle, which has not been noticed in the present setting before, and elaborate it by comparing the similar structure in the SUSY double-well setting \cite{Behtash:2015zha,Behtash:2015loa}.

\section{A Formal Discussion of the EWKB formalism}\label{Section: EWKB_review}

\subsection{Brief Review of EWKB formalism}\label{Section: reviewEWKB}
Let us start with a very quick summary of the Airy-type and Weber-type EWKB approaches. Since they have recently been thoroughly discussed and analyzed in the literature, we only provide the necessary setup for this paper along with the discussion on the general formulas, which do not appear in the literature, for the non-degenerate potentials.

Throughout this paper, we focus on Schroedinger-type equations:
\begin{equation}
	\left(-g^2 \frac{\mrmd^2}{\mrmd x^2} + P(x,u,g) \right) \psi(x) = 0 \, ,\label{SchrodingerEquation_Generic}
\end{equation}
where the algebraic curve $P(x,u,g)$ is given by
\begin{equation}
	P(x,u,g) = P_0(x,u) + g\, P_1(x,u)\, .
\end{equation}
More specifically, for the Airy-type analysis, we consider
\begin{equation}
	P_0(x,u) = 2\left(V(x) - u\right)\, , \quad P_{1}(x,u) = 0\,, \label{CurveTerms_Airy}
\end{equation}
and for the Weber-type analysis, we set
\begin{equation}
	P_0(x,u) = 2\left(V(x) - u_0\right) = \tV(x)\, , \quad P_1(x,u) = 2\tu\, . \label{CurveTerms_Weber}
\end{equation}
We also remind that these two curves are simply related to each other by a rescaling of the spectral parameter, $u\rightarrow u_0 +g \tu$, where $u_0$ is chosen to be the classical energy level for the one of the saddle points of the potential $V(x)$.

Let us first focus on the Airy-type case: We start with the WKB ansatz of the form 
\begin{equation}\label{WKB_ansatz}
	\psi^\pm(x) = \exp\left\{\pm \int^x \mrmd x' s(x',g) \right\}\, ,
\end{equation}
where 
\begin{equation}
	s(x) = \sum_{g=-1}^\infty s_m(x) \, g^m \, .\label{WKB_expansion}
\end{equation}
Then, imposing it to \eqref{SchrodingerEquation_Generic} and solving the resulting Riccati equation separates even and odd terms in \eqref{WKB_expansion}, leading to
\begin{align}
	\psi^\pm(x) & = \frac{1}{\sqrt{\ts(x,g)}}\exp\left\{\pm \int_{x_0}^x \mrmd x' \ts(x,g) \right\} \\&= \frac{e^{\pm \s_0(x)}}{\sqrt{\s_0(x)}}\, \sum_{m=0}^\infty \psi_m^\pm\, g^m\, . \label{FormalSeries_Airy}
\end{align}
where 
\begin{equation}
	\ts(x,g) = \sum_{m=0}^\infty \ts_{2m-1}\, g^{2m-1}\, , 
\end{equation}
is the series of odd terms of \eqref{WKB_expansion} and 
\begin{equation}\label{LeadingOrder_Airy}
	\s_0(x,u) = \frac{1}{g}\int_{x_0}^{x} \mrmd x'\, \sqrt{P_0(x,u)}\, ,
\end{equation}
is the leading order term of the all order WKB expansion of quantum actions, which we express as
\begin{equation}
	\s = \sum_{m=0}^\infty \s_{2m} g^{2m}\, , \label{Expansion_WKBactions}
\end{equation} 
Finally, note that the lower bound $x_0$ in \eqref{LeadingOrder_Airy} is a normalization point and for practical reasons, it is convenient to be chosen as one of the turning points of the potential, i.e. $P_0(x,u)=0$.

It is well-known that the series in \eqref{FormalSeries_Airy} is a factorially divergent one and it is Borel summable unless \cite{AKT1}
\begin{equation}
	\Im\, \s_0(x,u) = 0 \, ,\label{StokesDiagram_condition}
\end{equation} 
which is the equation for Stokes lines. This means that summing the series in \eqref{FormalSeries_Airy} leads to a unique solution except the cases $x$ is on one of the Stokes lines. Such regions are called Stokes regions and two series solutions in different regions can be connected by carefully analyzing the Borel discontinuities on the Stokes lines. 

Another source of discontinuity for $\psi^\pm(x)$ in $x\in\mbbC$ is the branch cut due to the integrand $\sqrt{P_0(x,u)}$. Combining it with the Stokes lines, we obtain a (possibly very complicated) Riemann surface for the exact-WKB solutions for \eqref{SchrodingerEquation_Generic}. Then, choosing a principal sheet fixes the definitions of the physical quantities, which are encoded by the monodromies associated with these discontinuities. We emphasize that choosing a different principal sheet does not change the physics at all since \eqref{SchrodingerEquation_Generic} represents only one physical system. This is ensured by defining the physical observables accordingly. We refer to  Appendix~A of \cite{Misumi:2024gtf} for a detailed argument on the equivalence between different choices. Finally, while changing the Stokes regions, it is possible to change the normalization point and it should be incorporated along with other discontinuous changes when relating two solutions, which are normalized at different turning points, e.g. $x_0$ and $x_1$. 

All these changes that we encounter in the connection problem are encoded by \textit{monodromy} matrices\footnote{Note that what we describe here is nothing but geometrization of the textbook WKB analysis while promoting it to the level of exactness (at least formally) from being only a semi-classical approximation. Hence, the word ``exact'' appears in the formalism. In some sense, this procedure indicates that the semi-classical quantization is an exact one if it is handled properly in view of the resurgence theory.}, which we write as follows:
\begin{equation} \label{Monodromy_Airy}
	M_\mrmA^+ = \begin{pmatrix}
		1  & i \\
		0 & 1
	\end{pmatrix} ,  \quad 
	M_\mrmA^- = \begin{pmatrix}
		1  & 0 \\
		i & 1
	\end{pmatrix} , \quad 	
	M_\mrmA^\mrmB = \begin{pmatrix}
		0  & -i \\
		-i & 0
	\end{pmatrix} \, , \quad 
	N_\mrmA^{x_0,x_1} = \begin{pmatrix}
		e^{\mcalV_{x_0,x_1}}  & 0 \\
		0 & e^{-\mcalV_{x_0,x_1}}
	\end{pmatrix} \, , 
\end{equation}
In \eqref{Monodromy_Airy}, $M_\mrmA^+$ and $M_\mrmA^-$ are for the discontinuities originating in the Stokes lines on which the solution $\psi^+$ and $\psi^-$ are not Borel summable, respectively. $M_\mrmA^\mrmB$ is for the branch cut discontinuity and $N_\mrmA^{x_0,x_1}$ is the matrix encoding the change of the normalization point. Finally, the exponents in $N_\mrmA^{x_0,x_1}$ matrix are given by all order WKB integrals between the turning points $x_0$ and $x_1$ as
\begin{equation}
	\mcalV_{x_0,x_1} = \int_{x_0}^{x_1} \mrmd x\, \ts(x,g)\, . \label{cycleIntegral}
\end{equation}

\subsection{Dictionary for Weber-type expressions}\label{Section: Weber-EWKB}

So far, we have discussed only the analytical properties of the solutions $\psi^\pm$. The physics is provided by boundary conditions that are appropriate for a particular quantum system. As a result, we get an exact quantization condition encoding the physical quantities in terms of WKB cycles, which are the independent cycles, constituting a homology basis associated with the algebraic curve $P_0(x,u)$. 

In the Airy-type procedure, the contribution of the cycles originate from the normalization matrix $\mcalN_\mrmA$ in \eqref{Monodromy_Airy} and they appear in the quantization conditions as exponentiations of the (twice of) cycle integrals in  \eqref{cycleIntegral} as 
\begin{equation}
	\Pi_{x_0, x_1} = e^{2\mcalV_{x_0,x_1}} \, . \label{exponentiatedAction}
\end{equation}
Although the Airy-type approach is elegant and provides a straightforward way to obtain the exact quantization condition, practical computations of the integrals in \eqref{cycleIntegral} are not so simple to carry out. For this purpose, it is better to utilize the Weber-type EWKB, whose expressions are linked to the Airy type EWKB ones via a dictionary \cite{Sueishi:2021xti,Misumi:2024gtf}. 
We review and generalize them to the most general one-dimensional systems.

Let us first recall the canonical versions of the WKB action and its dual:
\begin{equation}\label{WKB_Actions}
	a(u,g) = \frac{1}{2\pi} \oint_A \mrmd x\, p(x,g)\, , \quad a^\mrmD(u,g) = \frac{1}{2\pi} \oint_B \mrmd x\, p(x,g)\,,
\end{equation}
where
\[p(x,g) = \sum_{n=0}^{\infty} p_{2n}(x)\, g^{2n}\, , \] is the momentum with all order quantum corrections. Note that, we observe $p_{2n} = -i \s_{2n}$ in comparison to the expansion in \eqref{Expansion_WKBactions}. 

\begin{figure}
	\centering
	\begin{subfigure}[h]{0.49\textwidth}
		\caption{\underline{2 saddle points at the same level}} \label{Figure: Dictionary1}
		\vspace{6pt}
		\includegraphics[width=\textwidth]{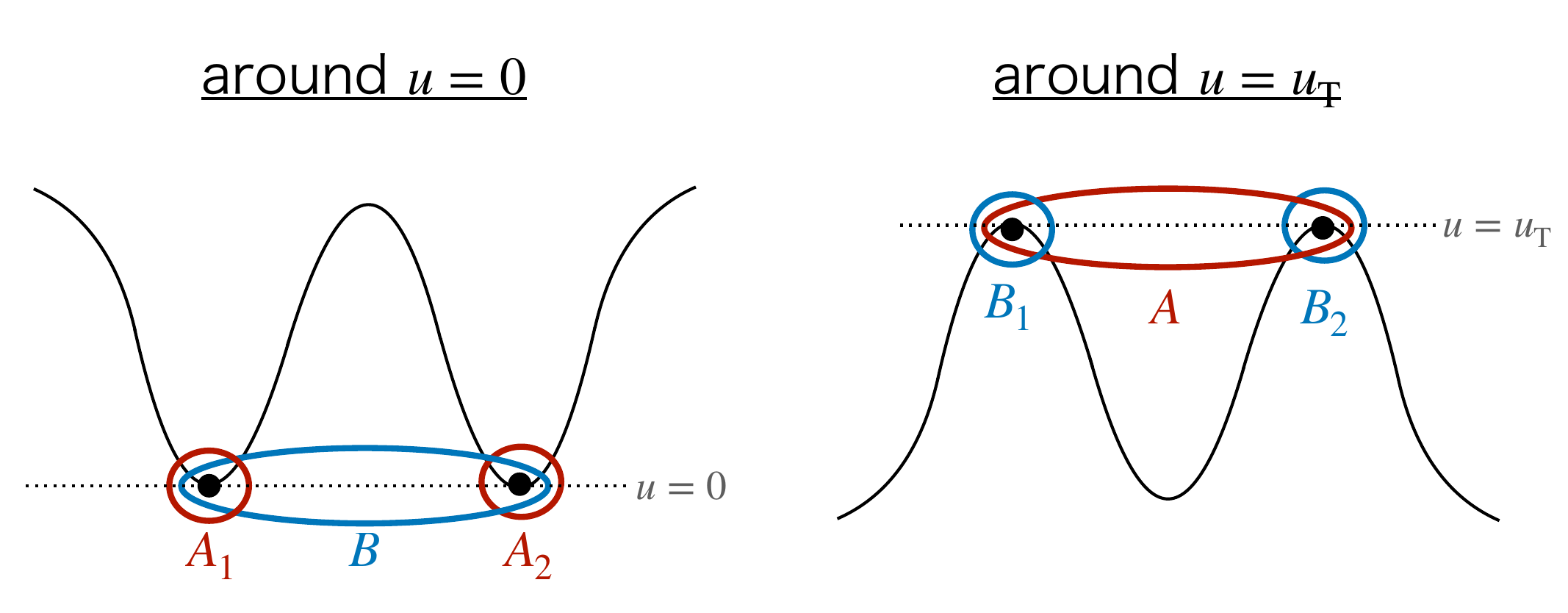}
	\end{subfigure}
	~\hfill 
	\begin{subfigure}[h]{0.49\textwidth}
		\caption{\underline{1 saddle point and 1 simple turning point}}	\label{Figure: Dictionary2}
		\vspace{6pt}
		\includegraphics[width=\textwidth]{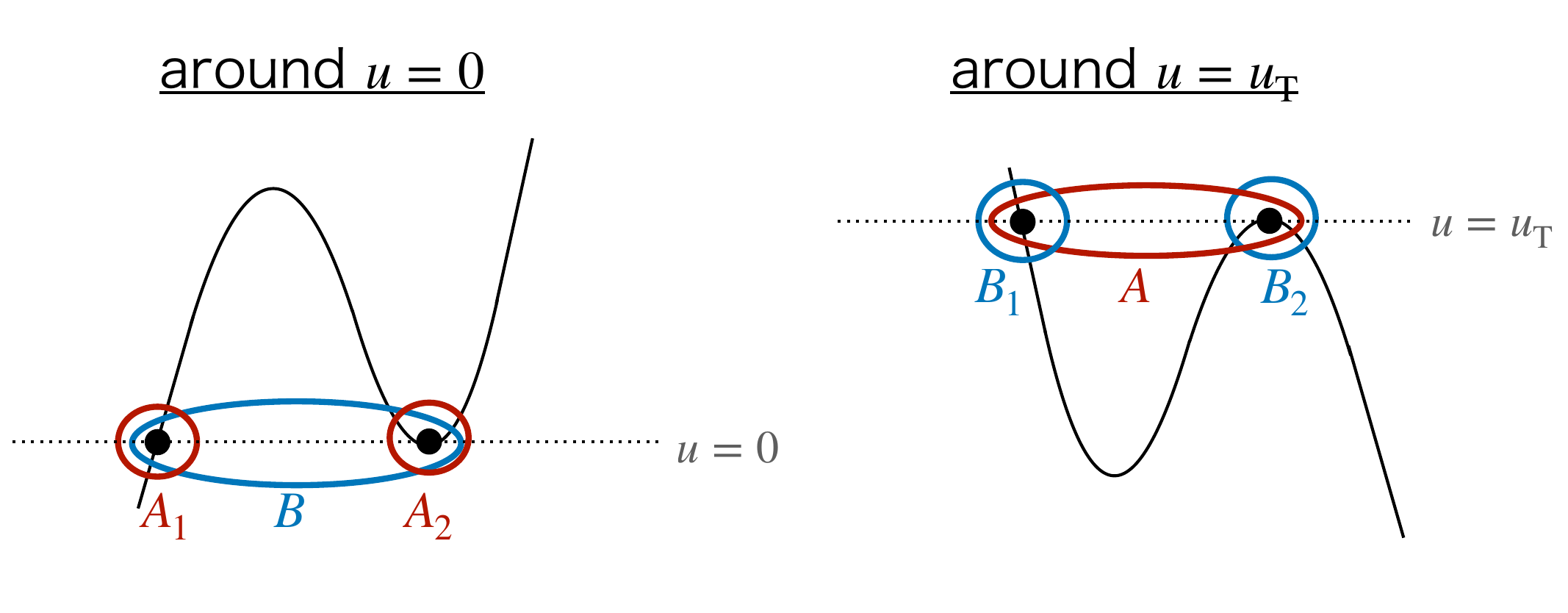}
	\end{subfigure}
	\par\bigskip	
	\caption{Weber type perturbative and non-perturbative cycles for all the possible cases we encounter in one dimensional problems.} \label{Figure: Dictionary}
\end{figure}

In the following, we assume that $u=0$ and $u=u_\mrmT$ are the classical energy levels corresponding to local minima and local maxima, respectively. (See Fig.\ref{Figure: Dictionary}.)  Then, the dictionary between Airy-type and Weber-type expressions is given as follows:
\begin{itemize}
	\item \underline{Perturbative Parts}: This is encoded by $A$-cycles when $u\sim 0$  and $B$-cycles when $u\sim u_\mrmT$.
	\begin{align}
		\Pi_A &= e^{-\frac{2\pi i}{g} a(u, g)} = e^{-2\pi i\mcalF(\frac{u}{g},g)} \, , \qquad \quad \;\; (u \sim 0)
		\, , \label{Dictionary_Perturbative_Well}\\ \nonumber \\
		\Pi_B &= e^{-\frac{2\pi i}{g} a^\mrmD(u,g)} = e^{-2\pi i \mcalF^D(i\frac{u}{g},i g)} \, ,  \qquad \;  (u \sim u_\mrmT) 
		\, . \label{Dictionary_Perturbative_Barrier}
	\end{align}
	where $\mcalF$ and $\mcalF^\mrmD$ are given by the residue of $\ts(x,g)$ around respective saddle points.
	\item \underline{Non-perturbative parts}: This is encoded by $B$-cycles when $u\sim 0$  and $A$-cycles when $u\sim u_\mrmT$ 
	\begin{enumerate}[wide]
		\item \label{Item: WeberCase_Bion}\underline{In the presence of two saddle points}: 
		\begin{align}
			\Pi_B &=  e^{-\frac{2\pi i}{g} a^\mrmD(u,g)} =  \frac{2\pi e^{-\mcalG(\frac{u}{g},g) }\,  \left(\frac{g}{\mcalC_1^{\mrmi}}\right)^{-2\mcalF_1(\frac{u}{g},g)}\left(\frac{g}{\mcalC_2^{\mrmi}}\right)^{-2\mcalF_{2}(\frac{u}{g},g)}}{\G\left(\frac{1}{2} + \mcalF_1\left(\frac{u}{g},g\right) \right)\G\left(\frac{1}{2} + \mcalF_2\left(\frac{u}{g},g\right) \right)}\, ,  \qquad (u \sim 0)
			\, , 	\label{Dictionary_NP_Well_Bion}\\ \nonumber \\
			\Pi_A &=  e^{-\frac{2\pi i}{g} a(u,g)} = \frac{2\pi  e^{-\mcalG^\mrmD(i\frac{u}{g},i g) }\, \left(\frac{g}{\mcalC_1^\mrmi}\right)^{-2\mcalF_1^\mrmD(i\frac{u}{g},ig)}\left(\frac{g}{\mcalC_2^\mrmi}\right)^{-2\mcalF_2^\mrmD(i\frac{u}{g},ig)}}{\G\left(\frac{1}{2} + \mcalF_1^\mrmD\left(i\frac{u}{g},ig\right) \right)\G\left(\frac{1}{2} + \mcalF_2^\mrmD\left(i\frac{u}{g},ig\right) \right)} \, , \quad  (u \sim u_\mrmT) 
			\, ,\label{Dictionary_NP_Barrier_Bion}
		\end{align}
		where the subscripts of $\mcalF_k$, $\mcalF_k^\mrmD$ and $\mcalC^{\mrmi}_k$ refers to the $k^\th$ saddle point. Then, $\mcalF_k$ and $\mcalF^\mrmD_k$ are determined by the residues around $k^\th$ saddle. We provide the explicit expressions for $\mcalC_k^\mrmi$ below.
		\item \label{Item: WeberCase_Bounce} \underline{In the presence of one saddle point and one simple turning point}:
		\begin{align}
			\Pi_B &=  e^{-\frac{2\pi i}{g} a^\mrmD(u,g)} =  e^{-\mcalG(\frac{u}{g},g) }\,\frac{\sqrt{2\pi} \left(\frac{g}{\mcalC^{\mrmi\mrmi}}\right)^{-\mcalF(\frac{u}{g},g)}}{\G\left(\frac{1}{2} + \mcalF\left(\frac{u}{g},g\right) \right)} \, , \qquad \quad \;\; (u \sim 0)
			\, , \label{Dictionary_NP_Well_Bounce} \\ \nonumber \\
			\Pi_A &=  e^{-\frac{2\pi i}{g} a(u,g)} =  e^{-\mcalG^\mrmD(i\frac{u}{g},i g) }\,\frac{\sqrt{2\pi} \left(\frac{g}{\mcalC^{\mrmi\mrmi}}\right)^{-\mcalF^\mrmD(i\frac{u}{g},i g)}}{\G\left(\frac{1}{2} + \mcalF^\mrmD\left(i\frac{u}{g},i g\right) \right)} \, , \qquad  (u \sim u_\mrmT) 
			\, ,\label{Dictionary_NP_Barrier_Bounce}
		\end{align}
		where $\mcalF$, $\mcalF^\mrmD$ and $\mcalC^\mrmii$ are originated from the saddle point in consideration. Again, we provide an explicit formula for $\mcalC^\mrmii$ below.
	\end{enumerate}
\end{itemize}

The equations \eqref{Dictionary_Perturbative_Well} - \eqref{Dictionary_NP_Barrier_Bounce} are valid for any locally harmonic potential and they reveal a universal character for the perturbative and non-perturbative actions. Moreover, the swap between the formulas of $A$ and $B$ cycles indicates a duality between $u\sim 0$ and $u\sim u_\mrmT$ regions for any such potential. We elaborate more on this in Section~\ref{Section: P_NP_revisit}.

In addition to having the general forms summarized in \eqref{Dictionary_Perturbative_Well}-\eqref{Dictionary_NP_Barrier_Bounce}, 
it provides a direct link between WKB and path integral approaches. More specifically, the exponent $\mcalG$ encodes the bounce/bion contributions in a series expansion in $g$, in the following way
\[\mcalG = \frac{S_\mcalB}{g} + \sum_{m=1}^\infty\mcalG_m g^m\, ,\] where $S_\mcalB$ is the classical action of bounce/bion solution and the series encodes the fluctuation around it. 

The $\left(\frac{g}{\mcalC}\right)^{-\mcalF}$ terms, on the other hand, correspond to the leading order of these fluctuations, which is also known as 1-loop determinant term. The Weber-type EWKB approach provides exact formulas of $\mcalC^\mrmi$ and $\mcalC^{\mrmi\mrmi}$ for any non-perturbative configuration which falls into one of the categories above: For both cases-\ref{Item: WeberCase_Bion} and \ref{Item: WeberCase_Bounce}, $\mcalC^\mrmi$ and $\mcalC^{\mrmi\mrmi}$ can be expressed in the same form as
\begin{align}
	\mcalC^\mrmi_l &= 2\left(x_2 - x_1\right)^2 \o_l \, e^{\S_l^\mrmi}\, , \qquad \qquad l=1,2\, , \label{1LoopConstant_Bion}\\
	\mcalC^{\mrmii} & = 2\left(x_2 - x_1\right)^2 \o_1 \, e^{\S^\mrmii_1}\, , \label{1LoopConstant_Bounce}
\end{align}
where the label $l$ in \eqref{1LoopConstant_Bion} refers to one of the saddles at $x=x_1$ or $x=x_2$, while in \eqref{1LoopConstant_Bounce} we only use $l=1$ as it is the only saddle point in this case. These two cases only differ in the details of the formula for $\S_l^\mrmi$ and $\S_1^\mrmii$, which can be obtained by following the discussion in \cite{Zinn-Justin:2004qzw}. Then, for the case-\ref{Item: WeberCase_Bion}, we obtain $\S_{1,2}^{\mrmii}$ as
\begin{align}
	\S_1^\mrmi & = \int_{x_1}^{x_2} \mrmd x\, \left[\frac{\o_1}{\sqrt{2\tV(x)}} - \frac{1}{x - x_1} - \frac{\o_1}{\o_2(x_2-x)}\right] \, , \label{OneLoopDeterminant_Integral_Bion1} \\ \nonumber \\
	\S_2^\mrmi & = \int_{x_1}^{x_2} \mrmd x\, \left[\frac{\o_2}{\sqrt{2\tV(x)}} - \frac{\o_2}{\o_1(x-x_1)} - \frac{1}{x_1 - x}\right]\, , \label{OneLoopDeterminant_Integral_Bion2}
\end{align}
and for the case-\ref{Item: WeberCase_Bounce}, we get $\S^\mrmi_1$ as
\begin{equation}
	\S_1^\mrmii = \exp\left\{2 \int_{x_1}^{x_2} \mrmd x\, \left[\frac{\o_1}{\sqrt{2\tV(x)}} - \frac{1}{(x-x_1)} \right] \right\} \, . \label{OneLoopDeterminant_Integral_Bounce}
\end{equation}

\subsection{Transition Below A Local Minimum}\label{Section: Transition}
Having all the tools that the EWKB formalism provides us, we now turn our focus on the transition of the Stokes diagrams across different sectors determined by the classical energy levels corresponding to a maximum or a minimum of the potential. In our previous work \cite{Misumi:2024gtf}, we discussed the transition across barrier tops of a locally harmonic oscillator by performing an analytic continuation of the spectral parameter as $u = |u| e^{i\t_u}$. In that setting, we showed changing the phase $\t_u$ also rotates the Stokes diagrams, leading to the analytic continuation of the entire Stokes geometry. 

\begin{figure}
	\centering
	\includegraphics[width=0.9\textwidth]{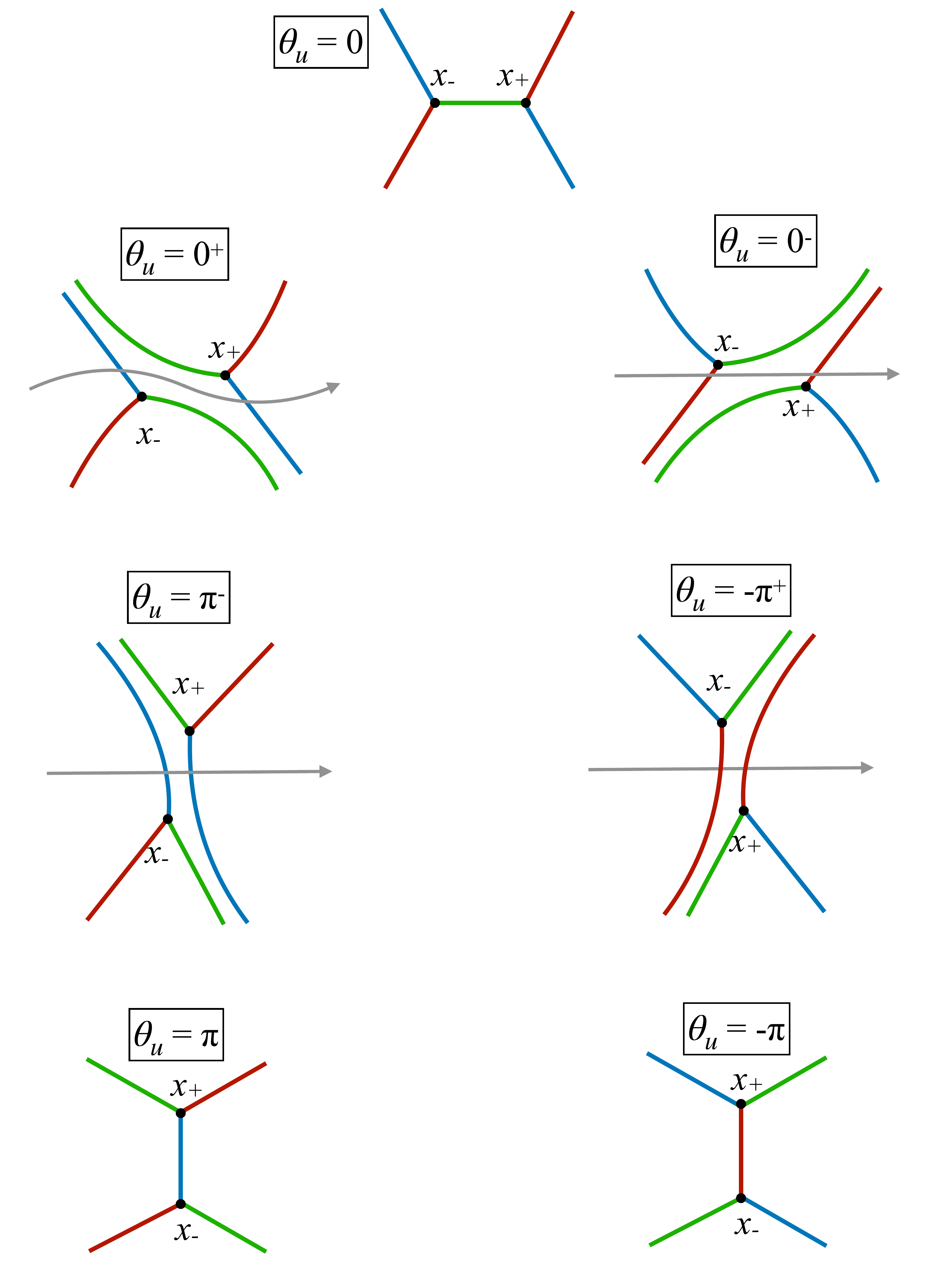}
	\vspace{5pt}
	\caption{The change of the Stokes diagrams during the transition across a barrier top which is induced by $\t_u: 0 \rightarrow \mp\pi$. It demonstrates the continuous character of the transition as no Stokes phenomenon is encountered. As a result, the transition matrices and the quantization conditions stay intact. (This figure first appeared in \cite{Misumi:2024gtf}.)}\label{Figure: Transition_aboveBarrierTop}
\end{figure}

For the transition across a barrier top, the evolution of the Stokes diagrams during the analytic continuation is illustrated in Fig.~\ref{Figure: Transition_aboveBarrierTop}, which first appeared in \cite{Misumi:2024gtf}. An important aspect of this transition is that except $\t_u=0$, there is no Stokes phenomenon occurring; and as a result the transition matrices connecting any two points, for example along the gray lines in Fig.~\ref{Figure: Transition_aboveBarrierTop}, stay the same. This leads to having the same quantization conditions in different sectors connected by this transition. 


Now, we adapt the same techniques to analyze the transition across a level corresponds to a locally harmonic minimum. Let us start with the simple harmonic oscillator. The Airy-type algebraic curve written as
\begin{equation}
	P_\mrmA = 2\left(x^2 - |u| e^{i\t_u}\right) \, , \label{Curve_AiryType}
\end{equation}
where we introduce a phase to the parameter $u$, and $\t_u=0$ and $\t_u = \pm \pi$ phases correspond to classical energy levels above and below the harmonic minima at $x=0$. The corresponding Stokes lines arising from the turning points $x_\pm = \pm \sqrt{ |u|e^{i\t_u}}$ are given by 
\begin{align}
	\s_0^\pm &= \frac{1}{g}\int_{x_\pm}^x \mrmd x\, \sqrt{P_\mrmA} \nonumber \\ 
	& \simeq \pm\frac{4}{3} \frac{|u|e^{\frac{i\t_u}{4}}}{g} \left[\pm \left(x - x_\pm\right)\right]^{3/2} + O\Big(\left(x - x_\pm\right)^{5/2}\Big). \label{StokesLineIntegral_HarmonicOscillator}
\end{align} 
To get the Stokes lines, we also set $x - x_\pm = |x - x_\pm|e^{i\t_\pm}$ and identify the direction of the Stokes lines with $\t_\pm$ which is the angle between the positive real axis and individual Stokes line. Then, the direction of the Stokes lines is expressed as
\begin{equation}
	\t_+ = \frac{2n\pi}{3} - \frac{\t_u}{6}\, , \quad \t_- = \frac{2\left(n + \frac{1}{2}\right)\pi}{3} - \frac{\t_u}{6}\, , \label{StokesLine_Rotation}
\end{equation}
for $n \in \mbbZ $. Note that specific choices of $n$ determine the continuous and discontinuous WKB solutions along a Stokes line. Then, it determines the properties of the WKB action in the principal Riemann sheet, but as we stated above, the physical quantities are not altered by such choices.

\begin{figure}
	\centering
	\begin{subfigure}[h]{0.46\textwidth}
		\caption{\underline{Rotation for $x=x_-$}} \label{Figure: Rotationx_-}
		\includegraphics[width=\textwidth]{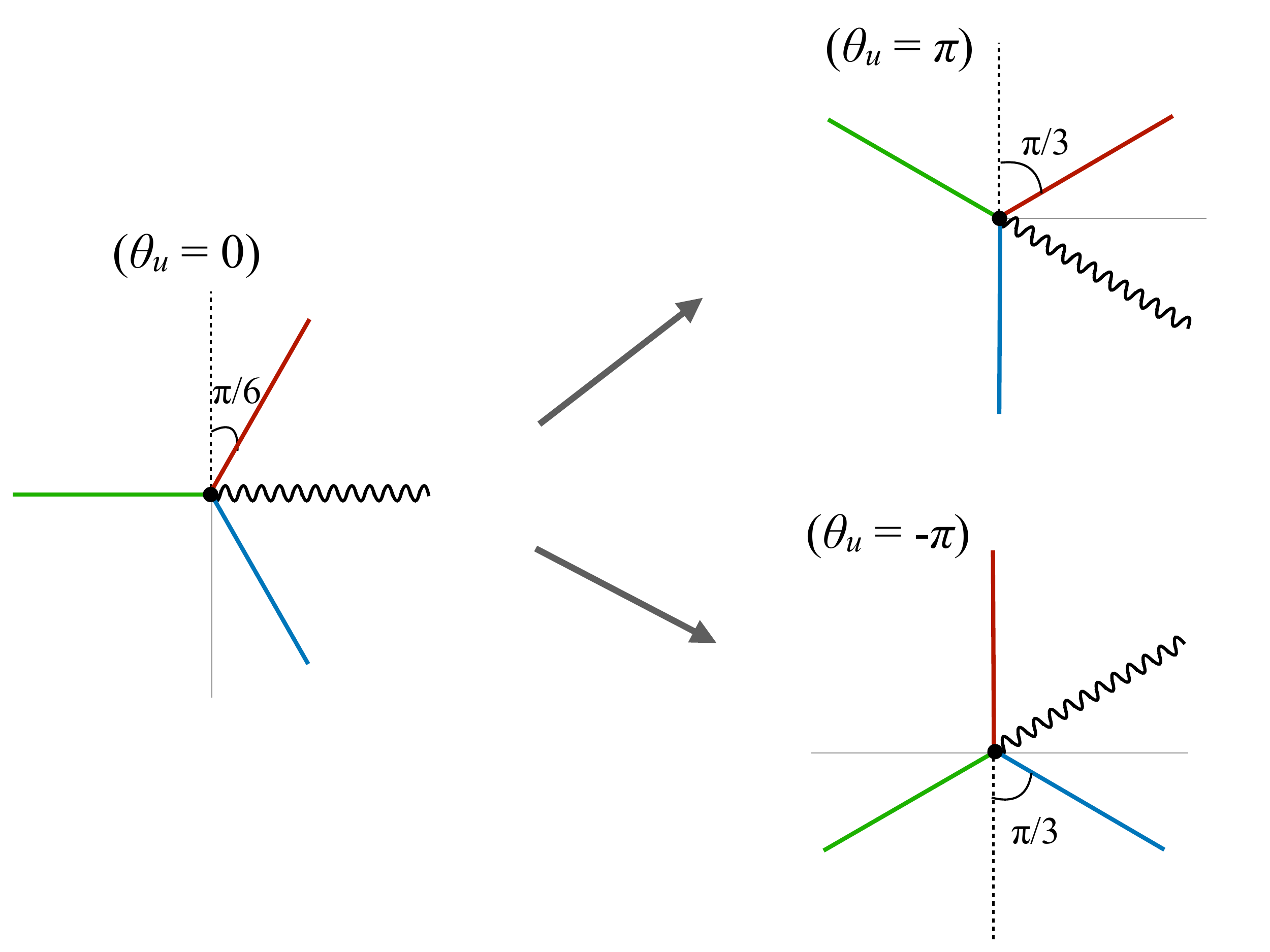}
	\end{subfigure}
	~\hfill 
	\begin{subfigure}[h]{0.46\textwidth}
		\caption{\underline{Rotation for $x=x_+$}}	\label{Figure: Rotationx_+}
		\includegraphics[width=\textwidth]{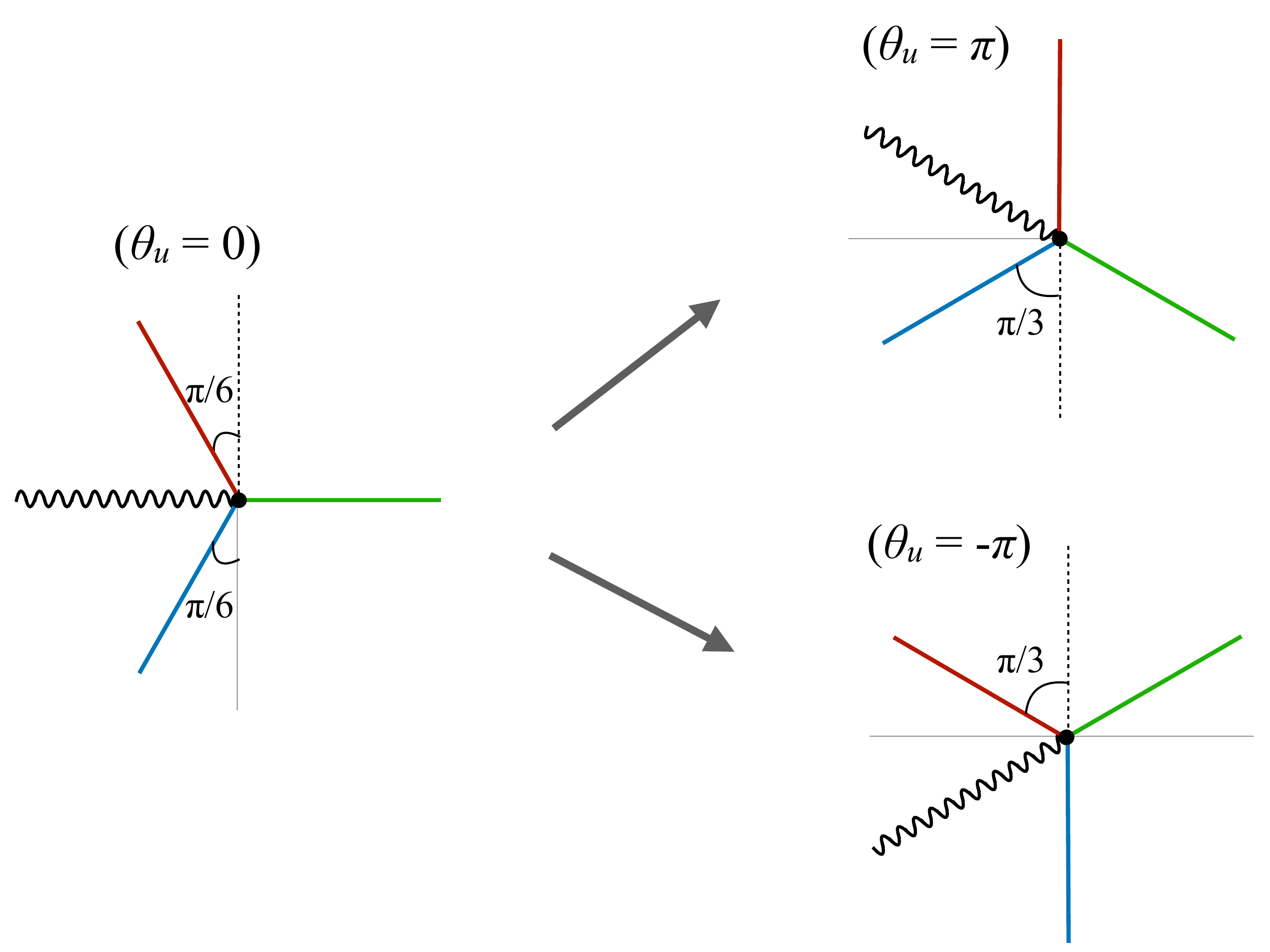}
	\end{subfigure}
	\par\bigskip	
	\caption{The rotation of Stokes line emerging from the turning points at $x=x_\pm$. Note that these figures first appeared in \cite{Misumi:2024gtf}. Here, only the labels for $x=x_\pm$ is altered.} \label{Figure: RotationStokesDiagrams}
\end{figure}

We illustrate the analytic continuation $\t_u: 0 \rightarrow \pm \pi$ for both the turning points at $x=x_\pm$ and the rotation of the Stokes diagrams in Fig.~\ref{Figure: RotationStokesDiagrams}. Note that the rotations given in \eqref{StokesLine_Rotation} are equivalent to the ones describing the transition through the barrier tops of locally harmonic oscillators and they form the basis of the transition illustrated in Fig.\ref{Figure: Transition_aboveBarrierTop}. Then, the transition below a locally harmonic minimum is expected to take place in the same way as well. 

\begin{figure}[h]
	\centering
	\includegraphics[width=0.75\textwidth]{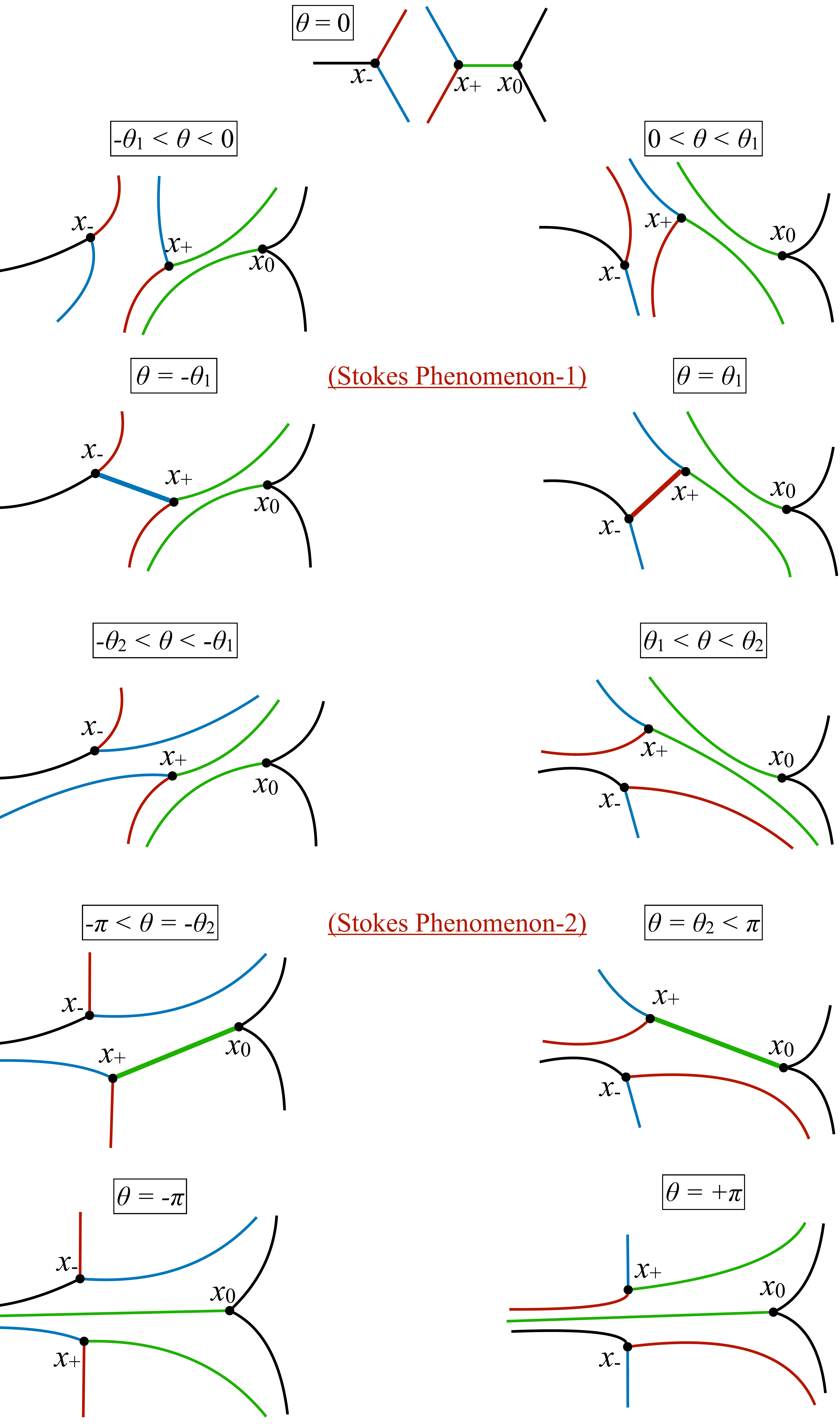}
	\caption{Change of the Stokes geometry during the transition below a locally harmonic well with a neighbouring barrier. Contrary to the transition across a barrier top, we encounter two Stokes automorphisms at $\t_2=\t_1$ and $\t_u =\t_2$, which leads to distinct exact quantization conditions, namely trans-series solutions at $\t_u=0$ and $\t=\mp\pi$.
    } \label{Figure: Transition_BelowWell}
\end{figure}

Naively, the transition across a locally harmonic minima might be expected to take place in the same way. However, this would not be correct. The main reason is the differences at the starting points $\t_u=0$. As we observe in Fig.~\ref{Figure: Transition_aboveBarrierTop}, around a barrier top, $\t_u=0$ corresponds to a case of degenerate Stokes diagrams and turning on $\t_u$ breaks the degeneracy. Then, during the analytic continuation $\t_u: 0 \rightarrow \mp \pi^\pm$, we encounter no degeneracy except the limits $\t_u=\mp \pi$. 

Around the minimum of the harmonic oscillator, on the other hand, the diagram is not the same at $\t_u=0$. We illustrate such a diagram at the top of Fig.~\ref{Figure: Transition_BelowWell}, where the harmonic oscillator well corresponds to the region between $x_-$ and $x_+$. Then, we observe that rotating the Stokes diagrams in view of \eqref{StokesLine_Rotation}, we encounter a Stokes phenomenon at $\t_u = \mp \t_1 $, where the sign is determined by the direction of the analytic continuation. Then, the diagrams on both sides of $\t_u = \mp \t_1$ are connected via Stokes automorphisms, which we discuss shortly. For the simple harmonic oscillator case, except this single discontinuity, the analytic continuation $\t_u: 0 \rightarrow \mp\pi$ is  continuous. In Fig.~\ref{Figure: Transition_BelowWell}, this can be observed by ignoring the Stokes lines arising from $x_0$.


Let us now consider the case with a barrier right to the locally harmonic well and assume it extends below the minimum of the well. At $\t_u = 0$, the barrier corresponds to the degenerate Stokes line between $x_+$ and $x_0$ in Fig.~\ref{Figure: Transition_BelowWell}.  The degeneracy is lifted once we start the analytic continuation as the degenerate lines turn in different directions. However, the line emerging from $x_0$ turns back to its original position at $\t_u=\pm \pi$. This means that we encounter another Stokes phenomenon\footnote{Note that, we don't have analytic solutions for $t_1$ and $t_2$ in general. However, plotting the Stokes diagrams and deforming the phase between $\t_u=0$ and $\t_u=\pm \pi$, we observe that $\frac{\pi}{2}=|\t_1| < |\t_2|<\pi$, meaning that we always encounter the degenerate diagram at $\t_1$ before the one at $\t_2$. Qualitatively, $|\t_1| = \frac{\pi}{2}$ means it is related to passing across the level of minimum; hence we observe it in simple harmonic oscillator case. The degenerate diagram at $\t_2$, on the other hand, is closely related to the complex instantons governing the Borel summable perturbation series associated with the neighbouring well, if there is one.} at $\t_u=\pm \t_2$, which induces a Stokes automorphism in addition to the one at $\t_u=\pm \t_1$.

Then, we observe that for a generic locally harmonic oscillator, there are multiple Stokes phenomena that occur during the analytic continuation across the bottom of a well. This means that the WKB actions along the degenerated lines are discontinuous at $\t_u = \pm\t_1$ and $\t_u = \pm \t_2$ and the discontinuities are represented by Stokes automorphisms via DDP formula \cite{DDP2,DP1,Iwaki1}. 

Specifically, at $\t_u=\pm \t_1$, the WKB cycle connecting $x_+$ and $x_0$ is discontinuous and the jump of the corresponding actions becomes
\begin{equation}
	\mfrS_{\pm \t_1}^\n: e^{2\mcalV_{x_+,x_0}} \mapsto e^{2\mcalV_{x_+,x_0}}\left(1 + e^{2\mcalV_{x_-,x_+}}\right)^{\n} \, .\label{StokesJump1}
\end{equation}
At $\t_u=\pm \t_2$, on the other hand, the cycle connecting $x_-$ and $x_+$ becomes discontinuous and the corresponding jump is expressed as
\begin{equation}
	\mfrS_{\pm \t_2}^\n: e^{2\mcalV_{x_-,x_+}} \mapsto e^{2\mcalV_{x_-,x_+}}\left(1 + e^{2\mcalV_{x_+,x_0}}\right)^{\n} \, . \label{StokesJump2}
\end{equation}
In both cases of $\n=\pm 1$, whose signs are determined by the orientation of the cycles and the direction of the transitions.

Finally, we stress that, for more general potentials, the number of cycles contributing to the Stokes phenomena might be two. This would not change the general transition procedure we discuss above. In such cases, \eqref{StokesJump1} and \eqref{StokesJump2} would be modified accordingly.

\begin{figure}
	\centering
	\includegraphics[width=\textwidth]{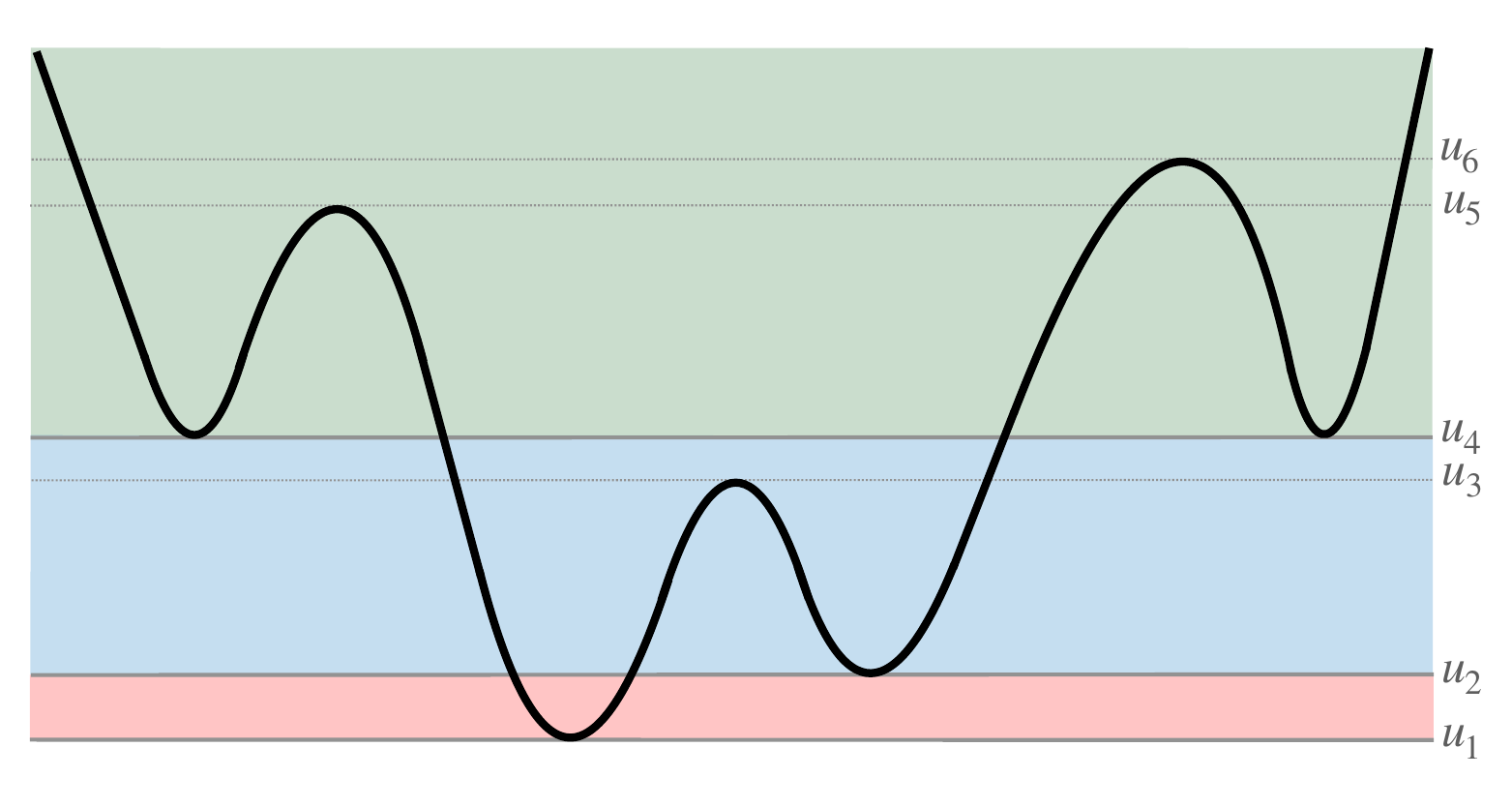}
	\caption{An illustration of a locally harmonic potential with a spectrum contains disconnected sectors associated with different trans-series representations. Each distinct sector is associated with a different trans-series and is represented by a different color, which are separated by gray lines. Critical values of the discontinuous transitions are labeled with $u_1$, $u_2$ and $u_4$. At other critical values $u_3$, $u_5$ and $u_6$, the transition is continuous as they correspond to barrier tops.} \label{Figure: GenericPotential_DifferentSectors}
\end{figure}

\noindent {\bf \underline{Remark}:} The discontinuous changes play an important role in determining properties of the spectrum of the quantum system. In \cite{Misumi:2024gtf}, we argued that the continuous transition across a barrier top yields a single trans-series for the sectors corresponding to below and above it. This fact is based on having a single transition matrix connecting the $x\rightarrow \pm\infty$ regions, which leads to the same quantization condition. In the case of the transition below the bottom of a well, on the other hand, the Stokes automorphisms change the structure of the transition matrices, i.e. monodromy matrices acting along grey lines in Fig.~\ref{Figure: Transition_BelowWell}, resulting in different quantization conditions, so that different trans-series representing the spectrum. Then, for a generic potential, we observe that the number of distinct trans-series is equal to the number of energy levels corresponding to local minima. We illustrate this conclusion in Fig.~\ref{Figure: GenericPotential_DifferentSectors}, where each colored region is associated with a different trans-series.

\section{Revisiting P-NP relation and duality}\label{Section: P_NP_revisit}

\subsection{Basics of P-NP relation}
In this section, we discuss the P-NP relation for generic genus-1 potentials and we illuminate the source of the $S$-duality transformations we recently found in \cite{Misumi:2024gtf}. We also extend it to relate P-NP relations associated with the different saddles of the original quantum system. This is particularly important when the perturbative saddles are not degenerate but linearly dependent on each other, with keeping the genus-1 geometry properties. Finally, we also discuss the generalization of these transformations to the deformed genus-1 cases by focusing on the tilted double-well (TDW) potential. 

Before discussing the transformation rules, we first set the stage by reviewing the deformed P-NP relation, which was recently introduced in \cite{Cavusoglu:2023bai,Cavusoglu:2024usn}: Let us consider a generic genus-1 potential $V(x,\g)$ where $\g$ is a deformation parameter. In the following, we assume that $V(x,\g=0)$ corresponds to the undeformed potential, which has only two linearly independent cycles, i.e. one perturbative and one non-perturbative cycles. Then, the associated P-NP relation for generic $\g$ is written as
\begin{equation}\label{P_NP_deformed_general}
	-\frac{\dee \tu^{(i)}}{\dee \mcalF} = g\left[f_1^{(i)}(\g) \mcalF + f_2^{(i)}(\g) g\frac{\dee \mcalG^{(i)}}{\dee g} + f_3^{(i)}(\g) \frac{\dee \mcalG^{(i)}}{\dee \g} \right]\, ,
\end{equation}
where the superscript labels different perturbative saddles with $\tu^{(i)}$ being the associated perturbative energy. Finally, in the undeformed limit, i.e. $\g\rightarrow 0$, \eqref{P_NP_deformed_general} must reduce to the original P-NP relation \cite{Dunne:2013ada,Alvarez1,Alvarez3}, which means $f_3^{(i)}(0) = 0$. 

\begin{figure}[H]
	\centering
	\begin{subfigure}[h]{0.49\textwidth}
		\caption{\underline{Cycles with cubic deformation parameter}} \label{Figure: Deformation_InstantonCycle}
		\vspace{5pt}
		\includegraphics[width=\textwidth]{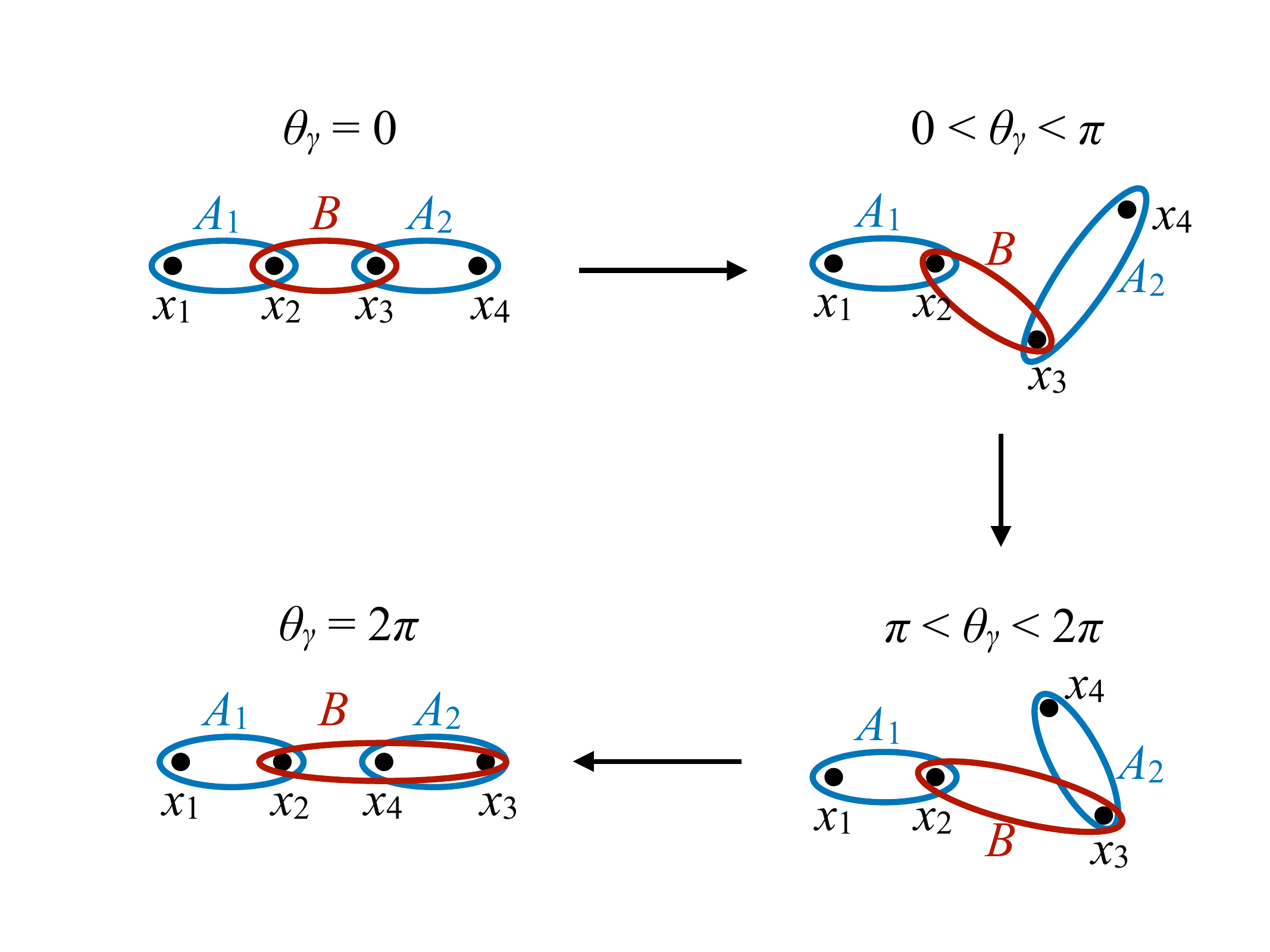}
	\end{subfigure}
	~\hfill 
	\begin{subfigure}[h]{0.49\textwidth}
		\caption{\underline{Cycles with linear deformation parameter}}	\label{Figure: Deformation_Linear}
		\vspace{5pt}
		\includegraphics[width=\textwidth]{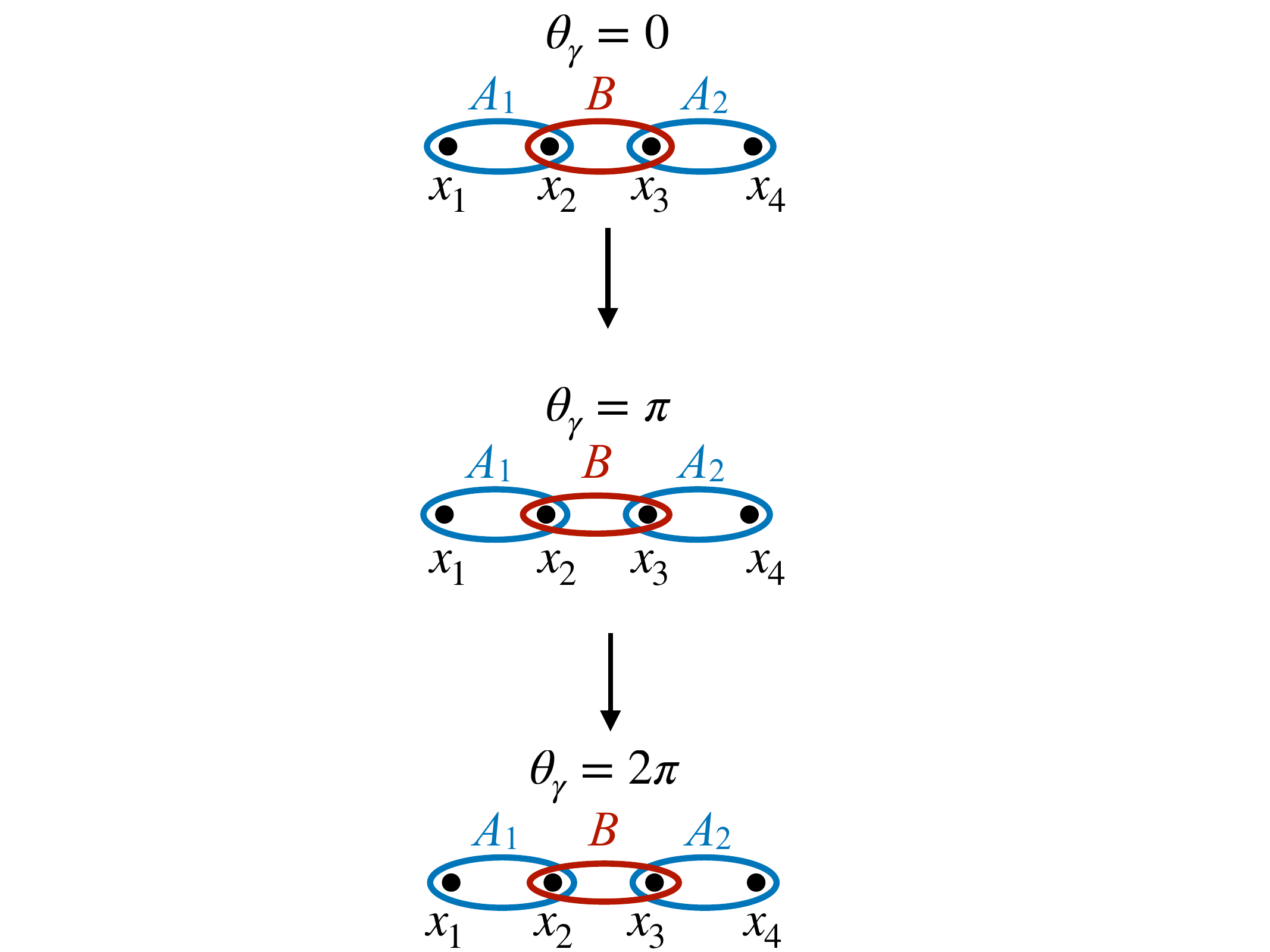}
	\end{subfigure}
	\vspace{5pt}
	\caption{Cycle geometry under analytic continuation of the deformation parameter $\gamma$ for deformations of the symmetric double-well potential. {\bf (a)} In the cubic (tilted) deformation $V_{\mathrm{TDW}}=V_{\mathrm{STW}}-\gamma x^3$ case, the non-perturbative cycle at $\gamma e^{2\pi i}$ encloses the neighbouring perturbative cycle, shifting the instanton action by $2\pi R^{(i)}(\gamma)$. {\bf (b)} For the linear deformation $V_{\mathrm{lin}}=V_{\mathrm{STW}}+\gamma x$ , the cycle geometry is unchanged under $\gamma\to\gamma e^{2\pi i}$, so the undeformed P-NP relation remains valid.} \label{Figure: TurningPoints_LinearVsCubic}
\end{figure}

We note that $V(x,\g=0)$ has certain symmetries, such as parity or periodicity, and this is the main reason of having only two independent cycles. However, not all deformations break such symmetries of the classical potential and lead to more independent cycles. An illuminating comparison in this line can be done between linear and cubic deformations of the symmetric double well (SDW) potential: 

The SDW system is represented by a quartic curve which does not have linear or cubic terms, e.g. $V_\mathrm{SDW}(x) = x^2(1-x)^2$. It is well-known that its perturbative and non-perturbative sectors are linked via the undeformed P-NP relations \cite{Dunne:2013ada,Alvarez1,Alvarez3}. A cubic deformation on this curve, e.g. $V_\mrmTDW = V_\mrmSTW -\g x^3$, breaks the parity symmetry of STW and the resurgence structure of this deformed system is now governed by \eqref{P_NP_deformed_general} \cite{Cavusoglu:2023bai,Cavusoglu:2024usn}. The underlying reason for the additional term in \eqref{P_NP_deformed_general} is the deformation of the cycle geometry of $V_\mrmTDW$ under the analytic continuation of the deformation parameter as $\g \rightarrow \g e^{2\pi i}$. Clearly, this analytic continuation recovers the same potential at the end. However,  at $\g e^{2\pi i}$, the original non-perturbative cycle encapsulates the neighbouring perturbative cycle as illustrated in Fig.\ref{Figure: Deformation_InstantonCycle}, leading to the associated instanton action changing as
\begin{equation}
	S^{(i)}_\mcalB \mapsto S^{(i)}_\mcalB + 2\pi R^{(i)}(\g) \,  , \label{Deformation_InstantonAction}
\end{equation}
where $R^{(i)}(\g)$ is called the residue of the WKB cycle.

On the other hand, if we consider a linear deformation, e.g. $V_\mathrm{lin} = V_\mrmSTW + \g x$, the geometry of the cycles remains the same after the analytic continuation of $\g$ parameter as $\g \rightarrow \g e^{\pi i} \rightarrow \g e^{2\pi i}$. See Fig.~\ref{Figure: Deformation_Linear}. Then, the associated resurgence structure is governed by the undeformed P-NP relation. A famous example of the linearly deformed double-well is the SUSY double-well system, which was shown to obey the undeformed P-NP relation \cite{Kamata:2021jrs}.

In the SUSY system\footnote{In fact, the generic deformed potential does not have to preserve SUSY even at the classical level. The resurgence structure stays the same regardless of the details of $\g$ as discussed in \cite{Kamata:2021jrs}. We keep the term SUSY, however, for the simplicity in the language.}, the analytic continuation of $\g$ is linked to a complex saddle (bion), leading to the non-perturbatively broken SUSY. In Section~\ref{Section: TDW}, performing an exact quantization of the cubic deformation case, we present the existence of a complex saddle, which is again linked to the analytic continuation $\g\rightarrow \g e^{2\pi i}$ and the geometry illustrated in Fig.\ref{Figure: Deformation_InstantonCycle}. We also argue that this complex saddle should be different than the complex bounces associated with the global minima of the corresponding system, which is known to be related to the Borel summable series around the minima \cite{Cavusoglu:2023bai}.

In the remainder of this section, we discuss the deformed P-NP relation \eqref{P_NP_deformed_general} and its generic series solution. In particular, we show how the $f_2(\g)$ and $f_3(\g)$ depend only on the classical parameters, i.e. frequency and bion/bounce actions for any genus-1 system. In the $\g\rightarrow 0$ limit, this leads to the transformation rules that connect the P-NP relations of different perturbative/non-perturbative saddles. Then, we focusing on the TDW potential, we generalize the transformation rules of $f_2(\g)$ and $f_3(\g)$ to this deformed case. 

Moreover, we also discuss the parameter $f_1(\g)$, which has a different property: In the undeformed limit $\g=0$, it simply scales with the classical frequency $\o_i$, which is no longer the case for $\g\neq 0$. However, we argue that this deviation does not affect the instanton function $\mcalG^{(i)}$ and show that the transformation of the fluctuations depends only on the classical parameters.

\subsection{Solving P-NP relation}
To explore the explicit $\o_i$ and $S_\mcalB^{(i)}$ dependence of $f_{1,2,3}^{(i)}(\g)$ terms, we first solve \eqref{P_NP_deformed_general} using the series ansatz for the functions $\tu^{(i)}$, $\mcalF^{(i)}$ and $\mcalG^{(i)}$:
\begin{align}
	\tu^{(i)}\left(\mcalF,g\right) &= \sum_{n=0}^{\infty}\tu_n(\mcalF)g^n\quad , \quad \mcalF^{(i)}(\tu,g) = \sum_{n=0}^\infty \mcalF^{(i)}_n(\tu,g)g^n \, , \label{P_NP_SeriesAnsatz_Perturbative}\\
	\mcalG^{(i)}(\mcalF,g)& = \frac{S^{(i)}_\mcalB(\g)}{g} + \sum_{n=1}^\infty \mcalG^{(i)}_n(\mcalF) g^n \, ,\label{P_NP_SeriesAnsatz_Nonperturbative}
\end{align}
where the series for $\tu^{(i)}$ and $\mcalF^{(i)}$ are inverse of each other and we drop the superscript when they are used as variables. Note also that $\mcalG^{(i)}(\mcalF,g)$ can be transformed to $\mcalG^{(i)}(\tu,g)$ using this inversion. Then,
inserting the series ansatzes in \eqref{P_NP_deformed_general}, we get
\\

\underline{at $O\left(g^{-1}\right)$}:
\begin{equation}
	\qquad -\frac{\dee \tu_0^{(i)}}{\dee \mcalF} = f_2^{(i)}(\g) S_\mcalB^{(i)}(\g) + f_3^{(i)}(\g)\frac{\dee S^{(i)}_\mcalB}{\dee \g}\, , \label{PNP_LeadingOrder}
\end{equation}

\underline{at $O\left(g^0\right)$}:
\begin{equation}
	f_1^{(i)}(\g) = -\frac{1}{\mcalF}\frac{\dee \tu^{(i)}_1}{\dee\mcalF}\, ,\label{PNP_VanishingOrder}
\end{equation}

and \underline{at $O\left(g^n\right)$} for $n\geq 1$:
\begin{equation}
	\frac{\dee u^{(i)}_{n+1}}{\dee \mcalF}=  n\, f^{(i)}_2(\g)\mcalG^{(i)}_n + f^{(i)}_3(\g)\frac{\dee \mcalG^{(i)}_n}{\dee\g}\,,\label{PNP_HigherOrders}
\end{equation}
The meaning of \eqref{PNP_VanishingOrder} is simple: The only role of the term $f_1^{(i)}\mcalF$ in \eqref{P_NP_deformed_general} is to cancel the $O(g^0)$ term in the perturbative expansion of $\tu^{(i)}$. In our discussion, this is forced by the ansatz \eqref{P_NP_SeriesAnsatz_Nonperturbative}, but it is a generic property\footnote{In the Weber-type expressions, the $O(g^0)$ term appear as the constant $\mcalC$ in the prefactors, i.e. 1-loop determinant terms, in \eqref{Dictionary_NP_Well_Bion} - \eqref{Dictionary_NP_Barrier_Bounce}. In the Airy-type expressions, they appear as logarithm terms in $a^\mrmD$, explaining the drop of from the exponent.} of the instanton function $\mcalG(\mcalF,g)$ \cite{Zinn-Justin:2004vcw,Zinn-Justin:2004qzw}.

Other parameters $f_2$ and $f_3$ can be obtained from \eqref{PNP_LeadingOrder} and the deformation of the  actions $S^{(i)}_\mcalB$ under the analytic continuation $\g\rightarrow \g e^{2\pi i}$, which we introduce in \eqref{Deformation_InstantonAction}. Note that since this analytic continuation leads to the same quantum system, \eqref{PNP_LeadingOrder} should still hold, leading to an equation for the residue $R^{(i)}(\g)$ as
\begin{equation}
	0 = f_2^{(i)}(\g) R^{(i)}(\g) + f_3^{(i)}(\g) \frac{\dee R^{(i)}}{\dee \g}\, . \label{PNP_DeformationEquation}
\end{equation}
Together with \eqref{PNP_LeadingOrder}, \eqref{PNP_DeformationEquation} forms a system of linear equations. If $f_2^{(i)}$ and $f_3^{(i)}$ were known, it would be possible to obtain $S_\mcalB^{(i)}$ and $R^{(i)}$. Or if $S_\mcalB^{(i)}$ is known, $R^{(i)}$ is simply derived from it, and \eqref{PNP_LeadingOrder} and \eqref{PNP_DeformationEquation} yield the parameters $f_2$ and $f_3$. Finally, \eqref{PNP_HigherOrders} is used to obtain quantum corrections to $\mcalG^{(i)}(\mcalF,g)$. 

In our discussion, we assume that the classical quantities, $S^{(i)}_\mcalB(\g)$, $R^{(i)}(\g)$ and $\o_i(\g)$ are known. The last one is the frequency of the corresponding perturbative saddle and it fixes the left-hand-side of \eqref{PNP_LeadingOrder} via \[\frac{\dee \tu^{(i)}_0}{\dee \mcalF} = \o_i\, . \] Note that this is a generic relationship for any locally harmonic case. Then, using \eqref{PNP_LeadingOrder} and \eqref{PNP_DeformationEquation}, we obtain 
\begin{align}
	f_3^{(i)}(\g) = \frac{\o_i(\g) R^{(i)}(\g)}{\det M^{(i)}(\g)}\quad , \quad f_2^{(i)}(\g) = \frac{\o_i(\g)\, \dee_\g R^{(i)}}{\det M^{(i)}(\g)}\, , \label{PNP_parameters}
\end{align}
where $M(\g)$ is a $2\times 2$ matrix given as
\begin{equation}
	M^{(i)}(\g) = \begin{pmatrix}
		S_\mcalB^{(i)} & -\dee_\g S_\mcalB^{(i)} \\
		R^{(i)} & - \dee_\g R^{(i)}\label{matrix_2x2}
	\end{pmatrix}\, .
\end{equation}
The expressions in \eqref{PNP_parameters} shows that for any potential, for which \eqref{P_NP_deformed_general} represents the P-NP relation, $f_2$ and $f_3$ always differ by only one term. In the limit of undeformed potential, i.e. $\g\rightarrow 0$, we have $R^{(i)}(0)$, so that $f_3^{(i)}$ vanishes as expected. However, since $f_2^{(i)}(0)$ exists in the undeformed P-NP relation (see \eqref{P_NP_undeformed}), the derivative $\dee_\g R^{(i)}$ must be non-zero as $\g\rightarrow 0$. Finally, \eqref{PNP_parameters} also reveals the dependence on the classical frequency $\o_i$ of $f_2$ and $f_3$. We note that the determinant part hides another $\o_i$ dependence, which plays an important role in uncovering the transformation rules of both deformed and undeformed cases. 

Before elaborating on the $\o_i$ dependence of \eqref{PNP_parameters}, let us first finish solving the P-NP equation \eqref{P_NP_deformed_general}: Given the expressions of $f_2$ and $f_3$ in \eqref{PNP_parameters}, a generic expression for $\mcalG_n^{(i)}$ can be obtained easily. The homogeneous solution to \eqref{PNP_parameters} is
\begin{equation}
	\mcalH_n^{(i)}(\g) = \left(R^{(i)}(\g)\right)^{-n}\, ,  \label{HigherOrders_Homog}
\end{equation}
where the simple dependence on the residue function arises from $\frac{f^{(i)}_2(\g)}{f^{(i)}_3(\g)} = \dee_\g \log R^{(i)}(\g)$. Similarly, for the inhomogeneous solution $\mcaltG_n^{(i)}$, using \[n \frac{\dee R^{(i)}}{\dee \g} \mcaltG^{(i)}_n + R^{(i)} \frac{\dee \mcaltG^{(i)}_n}{\dee \g} = \left(R^{(i)}\right)^{-n+1} \frac{\dee}{\dee \g}\left[\left(R^{(i)}\right)^n \mcaltG_n\right]\, ,\] we get
\begin{equation}
	\mcaltG^{(i)}_n = \left(R^{(i)}(\g)\right)^{-n} \int \mrmd g\, \left(R^{(i)}(\g)\right)^n \frac{\det M^{(i)}(\g)}{\o_i(\g)} \frac{\dee \tu^{(i)}_{n+1}}{\dee \mcalF} \,,\label{HigherOrders_Inhomog}
\end{equation}
where the integral means an anti-derivative. 

We note that this procedure was in part argued in \cite{Cavusoglu:2023bai,Cavusoglu:2024usn} recently. Our argument, however, reveals the explicit dependence on the classical parameters $\o_i$, $S_\mcalB^{(i)}(\g)$ and $R^{(i)}(\g)$. For the instanton fluctutaions $\mcaltG^{(i)}_n$, the only remaining term is $\frac{\dee \tu^{(i)}}{\dee \mcalF}$. Interestingly enough, for TDW case, the combination of the classical parameters in \eqref{PNP_parameters} and \eqref{HigherOrders_Inhomog} becomes invariant quantities, leaving $\frac{\dee \tu^{(i)}_{n+1}}{\dee \mcalF}$ for $n\geq 1$ as the only varying quantity at each order. 

Before discussing the TDW case, we now take a $\g\rightarrow 0$ limit and show how the explicit expressions in \eqref{PNP_parameters} and the symmetric property of the undeformed genus-1 potentials lead to the duality transformations introduced in \cite{Misumi:2024gtf}.


\subsection{Symmetric Limit}
In the $\g\rightarrow 0$ limit, we have $f_3^{(i)}(0) = 0$ and \eqref{P_NP_deformed_general} is reduced to the original form for undeformed genus-1 potentials\footnote{In this limit, P-NP relation is closely related to the Matone relation \cite{Matone:1995rx,Flume:2004rp} via the dictionary we present in Section~\ref{Section: Weber-EWKB} \cite{Basar:2015xna,Gorsky:2014lia}.}:
\begin{equation}
	-\frac{\dee \tu}{\dee \mcalF} = g\left[f_1^{(i)}(0) \mcalF + f_2^{(i)}(0)g\frac{\dee \mcalG^{(i)}}{\dee g}\right]\, , \label{P_NP_undeformed}
\end{equation}
where 
\begin{equation}
	f_2^{(i)}(0) = \frac{\o_i(0)}{S_\mcalB^{(i)}(0)}\, .  \label{f2_NonDeformed}
\end{equation}
Note that the simplification in $f_2^{(i)}$ is due to the cancellation between $\dee_\g R^{(i)}$ in the numerator and the denominator in \eqref{PNP_parameters}.

In the undeformed limit, the genus-1 potentials enjoy symmetries leading to simple proportionality between the classical actions and their duals associated with different saddles\footnote{In fact, this property extends beyond the genus-1 cases when the potential has enough symmetry. For example, Chebyshev potentials $T_n^2(x)$ for $n\in \mbbZ^+$ fall into this class and can be seen as an example for generic one dimensional parity symmetric potentials.}. For example, we can express such classical actions as
\begin{equation}
	a_0 = \frac{a_\mathrm{base}}{\o_i} \, , \quad a_0^\mrmD = \o_i a_\mathrm{base}^\mrmD\, ,\label{proportional_ClassicalActions}
\end{equation}
where $a_\mathrm{base}$ and $a_\mathrm{base}^\mrmD$ are the actions for a well and barrier with $\o=1$, indicating overall constants (bases) for any classical action in the system. In fact, this is the basis between the classical $S$-duality between the perturbative and non-perturbative actions, which we express as
\begin{equation}
	a_0^\mrmD(u) = -i \sqrt{\k}\, a_0(u_0-u) \label{S-duality_ClassicalActions}\, .
\end{equation}
In terms of the Weber-type (scaled) actions, this duality relation turns into
\begin{equation}
	\mcalF_0^\mrmD(\tu) = \sqrt{\k} \mcalF_0(\tu) \qquad , \qquad \mcalG_0^\mrmD(\tu)= \frac{1}{\sqrt{\k}} \mcalG_0^(\tu) \, . \label{DualityRelations_original}
\end{equation}
These are the $S$-duality transformations we discussed in \cite{Misumi:2024gtf}, where we showed how the P-NP relations transform in view of \eqref{DualityRelations_original} connecting the resurgence structures of the original and dual quantum systems. Note that for undeformed genus-1 potentials, the source of this resurgent duality can be traced back to the simple EWKB integrals in the Weber-type approach, which we discuss below:

Recall, around a saddle point $x=x_i$ with curvature $\o_i$, the leading order of $\mcalF^{(i)}$ is simply given as
\begin{equation}
	\mcalF^{(i)}_0 = \tu\,  \Res\left(\sqrt{2 \tV(x)}\right)^{-1}\bigg|_{x=x_i}\,,
\end{equation}
where $\tV(x) = V(x) - u_i$ with $u_i$ being the classical energy level of the saddle points. Then, with the local harmonicity assumption of the potential, we simply get\footnote{Note that this is also a verification of $\frac{\dee \tu_0^{(i)}}{\dee \mcalF} = \o_i$, which we use to get \eqref{PNP_parameters}.} \[\mcalF^{(i)}_0= \frac{\tu}{\o_i}\, . \] Note that this $\o_i$ is valid for any locally harmonic potential and it relates any two perturbative actions at the classical level as
\begin{equation}
	\mcalF_0^{(j)} = \frac{\o_j}{\o_i}\mcalF^{(i)}_0 \, .\label{FreqDependence_Perturbative}
\end{equation}
Comparing with the duality relations in \eqref{DualityRelations_original}, we deduce $\sqrt{\k} = \frac{\o(0)}{\o^\mrmD(0)}$ where we set $\o(0)$ and $\o^\mrmD(0)$ as the frequencies associated with the actions $\mcalF_0$ and $\mcalF^\mrmD_0$. Then, using $\mcalG_0=S_\mcalB$ and $\mcalG_0^\mrmD = S_\mcalB^\mrmD$, we get
\begin{equation}
	S_\mcalB^\mrmD = \frac{\o^\mrmD(0)}{\o(0)}S_\mcalB \, , \label{FreqDependence_Instanton}
\end{equation}
which is the duality of the bion/bounce actions corresponding to the first term of the non-perturbative (Weber-type) action $\mcalG_0^\mrmD$. Finally, combining these relations with \eqref{f2_NonDeformed}, we find that \eqref{FreqDependence_Instanton} leads to the invariance of $f_2(0)$ under the duality transformation: 
\begin{equation}
	f_2^\mrmD(0)=\frac{\o^\mrmD}{S_\mcalB^\mrmD} = \frac{\o}{S_\mcalB} = f_2(0)\, . \label{f2_NonDeformed_Dual}
\end{equation}

\begin{figure}[t]
	\centering
	\begin{subfigure}[h]{0.48\textwidth}
		\includegraphics[width=\textwidth]{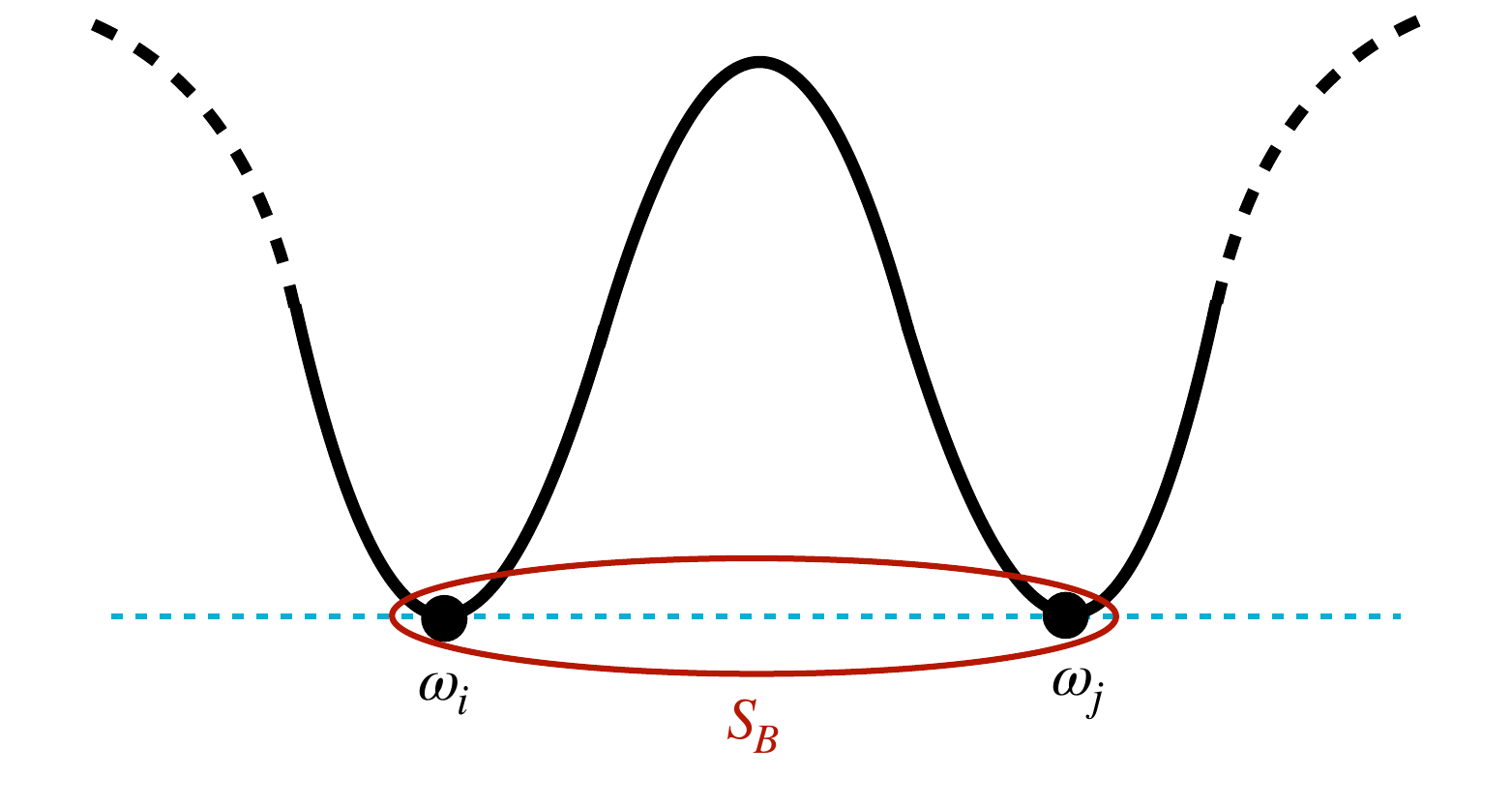}
		\caption{
		}	\label{Figure: PNP_transformation1}
	\end{subfigure}
	~\hfill 
	\begin{subfigure}[h]{0.48\textwidth}
		\includegraphics[width=\textwidth]{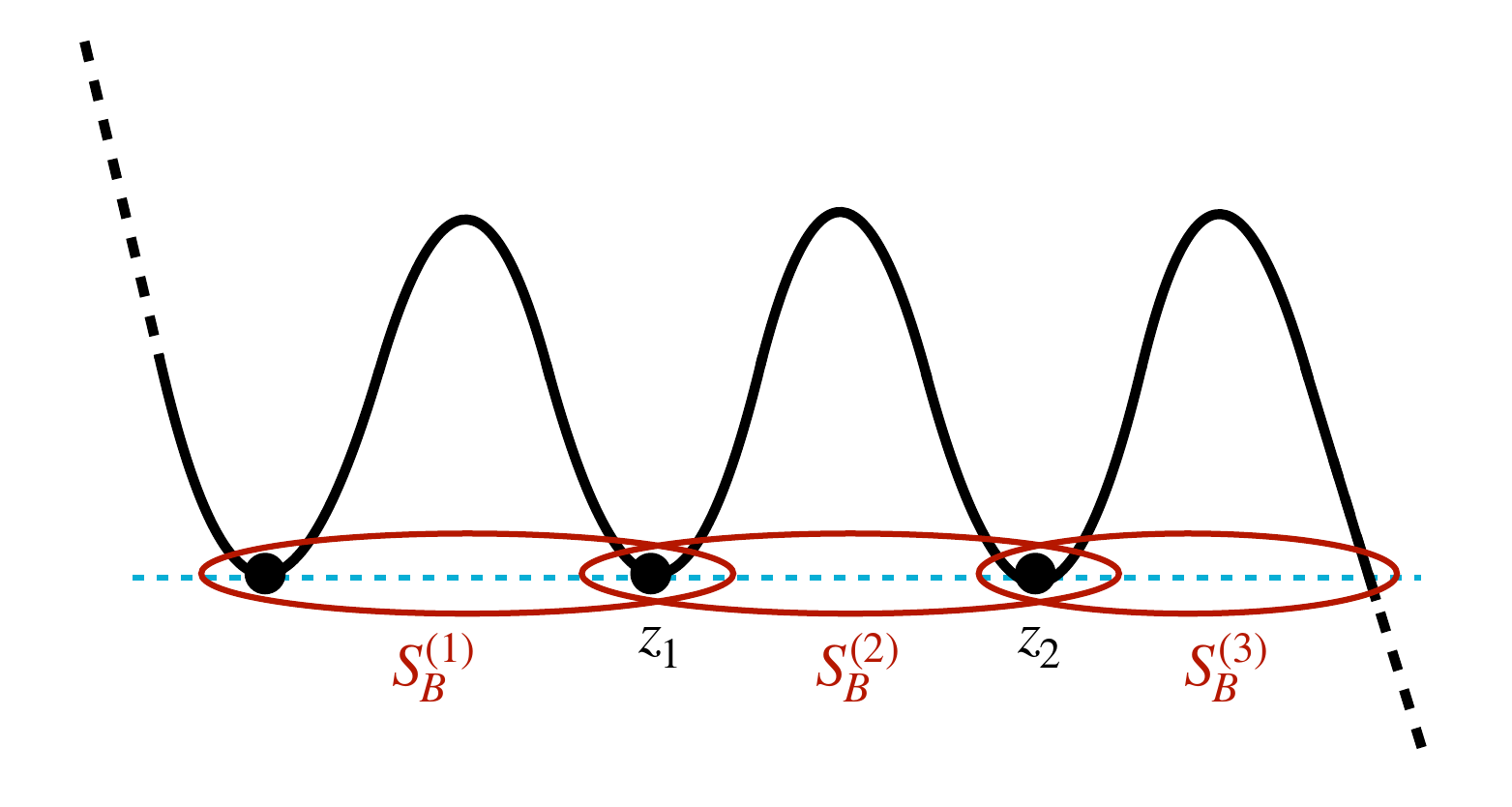}
		\caption{
		}	\label{Figure: PNP_transformation2}
	\end{subfigure}
	\par\bigskip	
	\caption{Illustrations for possible cases leading to the transformation formula \eqref{f2_NonDeformed_DifferentPerturbative} and \eqref{f2_NonDeformed_DifferentNP_1}. {\bf (a)} A bion with action $S_\mcalB$ connects two perturbative wells with frequencies $\o_i$ and $\o_j$. {\bf (b)} Two perturbative wells that are neighbouring to indenpendent bion/bounce configurations. The well around $x=z_1$ is linked to two bions with actions $S_\mcalB^{(1)}$ and $S_\mcalB^{(2)}$. The one at $x=z_2$ is linked to a bion and a bounce with actions $S_\mcalB^{(2)}$ and $S_\mcalB^{(3)}$.} \label{Figure: PNP_transformations}
\end{figure}


This explains $f_2$ part of the $S$-duality transformations of \cite{Misumi:2024gtf}. This procedure can also be extended to relate $f_2^{(i)}$ terms associated with the different perturbative/non-perturbative saddles of the original potential, which are not connected via an $S$-duality. Let us consider two possible cases:

\begin{enumerate}[wide]
	\item \underline{Two perturbative saddles connected via non-perturbative one}:\\ 
	This is the case when a bion connects two classically degenerate perturbative minima with frequencies $\o_i$ and $\o_j$ (See e.g. Fig.~\ref{Figure: PNP_transformation1}.) Then, \eqref{f2_NonDeformed} directly implies that, the $f_2(0)$ terms of these perturbative saddles are related by 
	\begin{equation}
		f_2^{(j)}  = \frac{\o_j}{\o_i}f_2^{(i)} \, . \label{f2_NonDeformed_DifferentPerturbative}
	\end{equation}
	\item \underline{Two non-perturbative saddles connected to a single perturbative one}:\\
	In this case, there are more than one non-perturbative saddles which are linked to the one perturbative saddle. The non-perturbative ones can be either bion or bounce configurations. For example in Fig.~\ref{Figure: PNP_transformation2}, $S_\mcalB^{(1)}$ and $S_\mcalB^{(2)}$ are bions and either of them can be used to write P-NP relation of the perturbative saddle at $z_1$. Similarly, the P-NP relation for the perturbative saddle at $z_2$ can be written in terms of the bion $S_\mcalB^{(2)}$ or the bounce $S_\mcalB^{(3)}$. In both cases, the $f_2(0)$ term changes depending on which non-perturbative configuration is used, and the different choices are related with each other as
	\begin{equation}
		f_2^{(j)} = \frac{S_\mcalB^{(i)}}{S_\mcalB^{(j)}}f_2^{(i)}\, .\label{f2_NonDeformed_DifferentNP_1}
	\end{equation}
	
\end{enumerate}

\vspace{5pt}
The expressions in \eqref{f2_NonDeformed_Dual}, \eqref{f2_NonDeformed_DifferentPerturbative} and \eqref{f2_NonDeformed_DifferentNP_1} covers the relationships between all saddles for $f_2^{(i)}$ term. Let us now discuss the transformation of $f_1(0)$ term. As we noted above, the only role of $f_1(0)\mcalF$ in the P-NP relations is to cancel the $O(g^0)$ term appearing in $\frac{\dee \tu^{(i)}}{\dee \mcalF}$. For all genus-1 cases, the perturbative expansion for $\tu^{(i)}(\mcalF,g)$ has the form of
\begin{equation}
	\tu_i(\mcalF,g) = \o_i \mcalF^{(i)} + \left(\a(\o_i) + \b(\o_i) \mcalF^2 \right)g +O(g^2)\, , \label{PerturbativeEnergy_Generic} 
\end{equation}
which sets $f_1^{(i)}(0) = 2\b(\o_i)$; so we need to figure out the $\o_i$ dependence of the coefficient of $\mcalF^2 g$ term. 

Recall that for a symmetric potential, the classical action $a_0(u)$ of a perturbative saddle scales with the associated frequency as in \eqref{proportional_ClassicalActions}.
Then, we get the scaled perturbative actions derived via $a(u,g) = \mcalF\left(\frac{\tu}{g},g\right)$ as
\begin{equation}
	\mcalF^{(i)} = \frac{\tu}{\o_i} + \left(\tilde \a(\o_i) + \frac{\tilde\b}{\o_i} \tu^2 \right)g +O(g^2) \, , \label{ScaledPerturbativeAction_Generic}
\end{equation}
where $\tu^2$ term on the right hand side comes from the $\frac{a_0(u)}{g}$ term of the quantum action $a(u,g)$ after the rescaling $u\rightarrow \frac{\tu}{g}$. Then, we observe that although the $\o_i$ dependence of $\tilde \a$ is not clear from the classical action and simple scaling relations disappear at the quantum level, \eqref{ScaledPerturbativeAction_Generic} is enough to determine the $\o_i$ dependence of $f_1(0)$ term. 

For this purpose, we invert the series \eqref{ScaledPerturbativeAction_Generic} and rewrite \eqref{PerturbativeEnergy_Generic} as
\begin{equation}
	\tu_i(\mcalF,g) = \o_i \mcalF - \left(\o_i \tilde \a(\o_i) + \o_i^2 \tilde \b \mcalF^2  \right)g + O(g^2)\, .\label{PerturbativeEnergy_FreqDependence}
\end{equation}
Then, for any perturbative saddle, we get $f_1^{(i)}(0) = 2 \o^2_i \tilde \b$ and find that $f_1(0)$ associated with different saddles are related by
\begin{equation}
	f_1^{(j)} = \frac{\o^2_j}{\o^2_i} f_1^{(i)}\, . \label{f1_FreqDependence}
\end{equation}
Note that in this transformation the bion/bounce actions play no role and the generality of \eqref{PerturbativeEnergy_FreqDependence} indicates that \eqref{f1_FreqDependence} also relates the P-NP relation of the original theory to the one for the dual theory in the same way.

With sufficient symmetry in the classical potential, using the relations \eqref{f2_NonDeformed_Dual}, \eqref{f2_NonDeformed_DifferentPerturbative}, \eqref{f2_NonDeformed_DifferentNP_1} and \eqref{f1_FreqDependence} it is possible to transform P-NP relations between two saddles to another couple of perturbative and non-perturbative saddles. We verify our claim in the example of symmetric triple-well potential in Section~\ref{Section: ATW_SymmetricLimit}. 

We finally note that the generalization of \eqref{P_NP_undeformed}(or \eqref{P_NP_deformed_general}) to higher genus cases is not clear and is still an open problem. Providing the generality of the classical duality in EWKB formalism, we anticipate similar transformations depending only on the classical terms to exist, for the higher genus cases as well. 

\subsection{Extension to Tilted Double-Well}
Let us now extend the argument on the transformation of the P-NP relation to the deformed double-well potential. At a first glance, the classical actions for different perturbative saddles $a^{(i)}_0(u)$ do not scale with the frequency any more\footnote{The simple scaling relation still holds at the leading order, i.e. $\tu_0 \sim \o_i$, as expected from the classical mechanics.}. This means the $f_1(\g)$ term does not transform according to \eqref{f1_FreqDependence} anymore. Similarly, the bion/bounce actions do not scale as in \eqref{FreqDependence_Instanton}, which suggests that the invariance of $f_2$ terms is lost in the deformed case as well. 

However, \eqref{PNP_parameters} indicates that the residue $R^{(i)}$ and the determinant $\det M^{(i)}$ are central objects for the parameters $f_2^{(i)}(\g)$ and $f_3^{(i)}(\g)$. In the following, we show that for the tilted double-well potential given by the potential,
\begin{equation}
	V_\mrmTDW(x,\g) = \frac{1}{2}x^2(1-x)^2 - \g x^3\, ,  \label{TDW_generic}
\end{equation}  
$R^{(i)}$ is an invariant object and $\det M^{(i)}$ scales with $\o_i$. As a result, both $f_2^{(i)}(\g)$ and $f^{(i)}_3(\g)$ terms in the P-NP relations are invariant parameters. 

\begin{figure}
	\centering
	\includegraphics[width=0.8\textwidth]{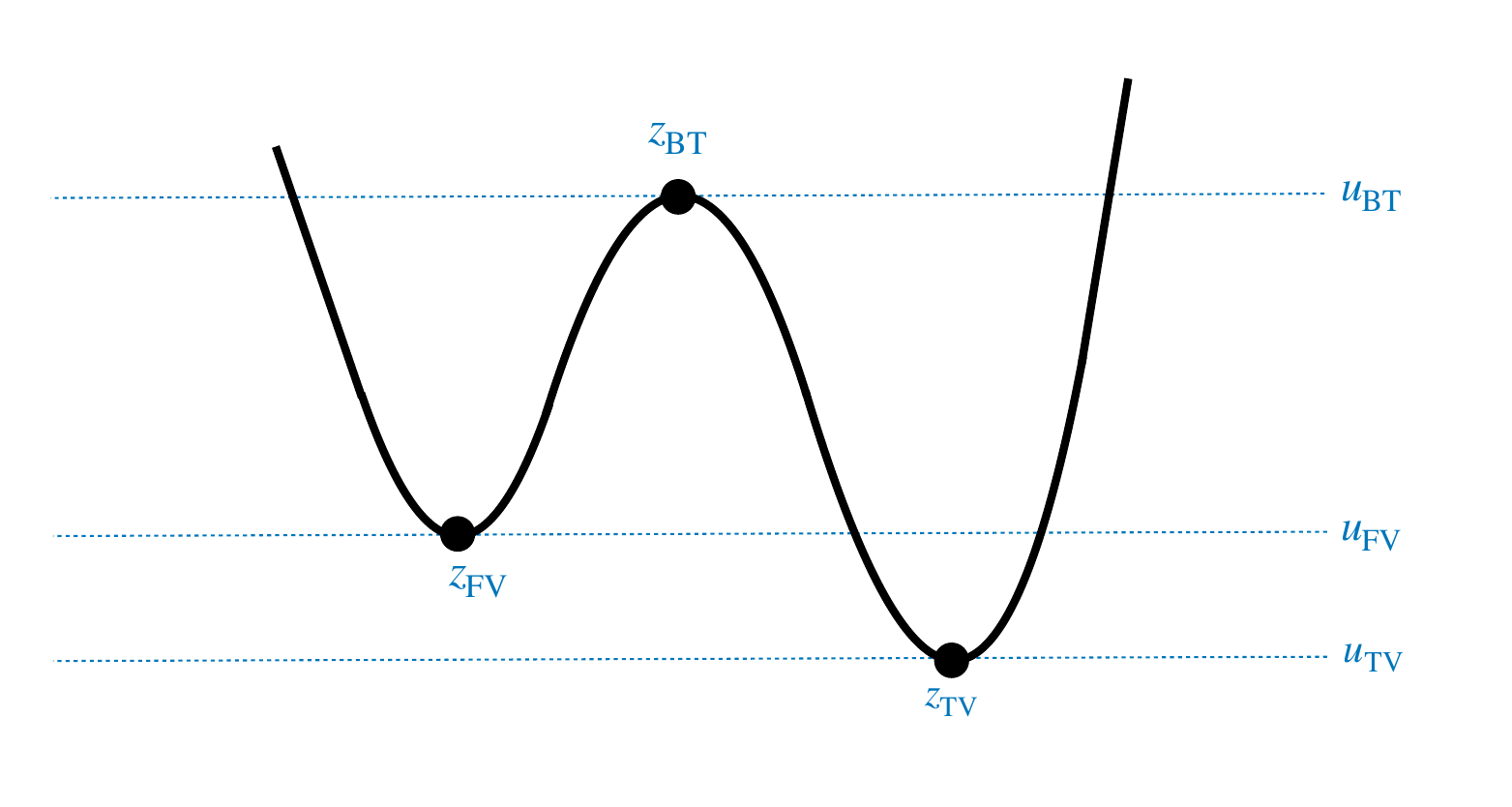}
	\caption{Tilted double-well potential for $\gamma>0$.  The locations and classical energies of the false vacuum (FV), barrier top (BT) and true vacuum (TV) are indicated, providing the reference saddles and energy levels.} \label{Figure:TDW_potential}
\end{figure}

To show this invariance, we first focus on the bounce actions $S_\mcalB^{(i)}$ of $V_\mrmTDW$ since, apart from the frequencies $\o_i$, it is the only quantity we need in order to compute $f_2^{(i)}(\g)$ and $f_3^{(i)}(\g)$. In general, there are three independent bounce configurations, which are associated with the energy levels $u^\mrmTV < u^\mrmFV < u^\mrmBT$, as depicted in Fig~\ref{Figure:TDW_potential}. 

Note that although we are interested in the specific form of the potential in \eqref{TDW_generic}, we found that starting with the generic curve is more instructive and even easier at various steps. Therefore, using the translation invariance of the problem, around each level, we bring the bounce actions into the following form:
\begin{align}
	S^{(i)}_\mcalB(t^{(i)}_1,t^{(i)}_2) &= 2\int_0^{t^{(i)}_1} \mrmd x\, \sqrt{2 \tV^{(i)}(x)}\, , \quad \tV^{(i)}(x) = V(x) - u_i \,, \label{GenericInstanton_Action} 
\end{align}
where 
\begin{equation}
	\tV^{(i)}(x) = \frac{1}{2}x^2 (x-t_1^{(i)})(x-t_2^{(i)}) \, , \label{scaledPotential_TDW}
\end{equation}
and we assume $t_{1,2}^{(i)}$ are implicitly dependent on the deformation parameter $\g$. Note that for each $u_i$, we set the double turning point to be at $x=0$, the simple turning points where the instanton bounces from to be at $t^{(i)}_1$ and the other turning point at the corresponding energy level $u_i$ to be at $t^{(i)}_2$.
 
For the generic form of the scaled potential in \eqref{scaledPotential_TDW}, we get the same form for all three bounce actions:
\begin{equation}\label{BounceAction_TDW_General}
	S_\mcalB^{(i)}(t_1^{(i)},t_2^{(i)}) = \frac{8}{15}\sqrt{t_1^{(i)}t_2^{(i)}} \left(t_1^{(i)}\right)^2 {}_2F_1\left(-\frac{1}{2},2,\frac{7}{2},\frac{t_1^{(i)}}{t_2^{(i)}}\right)\, .
\end{equation}
Since all the bounce actions for TDW has the same form, their discontinuities, so that the associated residue functions, are in the same form as well: 
\begin{equation}
	R^{(i)}(t_1^{(i)},t_2^{(i)}) = \frac{1}{16}(t_1^{(i)} - t_2^{(i)})^2 (t_1^{(i)}+ t_2^{(i)})\, . \label{residueFunction_TDW_Generic}
\end{equation}
Finally, using \eqref{BounceAction_TDW_General} and \eqref{residueFunction_TDW_Generic} for each $u^{(i)}$, we find
\begin{equation}
	\det M^{(i)} = \o_i \det \tM^{(i)} \,,\label{determinant_factorized}
\end{equation}
where 
\begin{equation}
	\det \tM^{(i)} = \frac{t_1^{(i)} t_2^{(i)}}{12}\left(t_1^{(i)}-t_2^{(i)}\right)  \left(t_1^{(i)} \dee_\g t_2^{(i)}-t_2^{(i)} \dee_\g t_1^{(i)}\right) \, , \label{determinantScaled}
\end{equation}
and we used  $\left(t_1^{(i)} t_2^{(i)}\right)^{1/2} = \o_i$ to get the final expression. Note that bringing $\det M^{(i)}$ to the form in \eqref{determinant_factorized} reveals its frequency dependence on the perturbative well around $u=u_i$, which cancels against $\o_i(\g)$ in the numerators in \eqref{PNP_parameters}.

\paragraph{\underline{Remark}:} Note that the bounce actions, $S_\mcalB^\mrmFV$ and $S_\mcalB^\mrmTV$, at $u_\mrmFV$ and $u_\mrmTV$ are singular when $t^{\mrmFV}_1 = t^{\mrmFV}_2$ and $t^\mrmTV_1 = t^\mrmTV_2$, respectively. This corresponds to the symmetric double well\footnote{The meaning of the singular behaviour of $S_\mcalB^{\mrmFV}$ and $S_\mcalB^\mrmTV$ in this limit is straightforward: The bounce of the TDW for $\g\neq 0$ becomes a bion, which is a combination of instanton and anti-instanton events, and it is only well-defined after a proper analytic continuation \cite{Zinn-Justin:1981qzi}.} limit $\g\rightarrow 0$ and it induces the discontinuity for the action as described in  \cite{Cavusoglu:2024usn}. At $u_\mrmBT$, on the other hand, the limit $\g\rightarrow 0$ leads to $t_1^\mrmBT = -t_2^\mrmBT$, which does not correspond to a singularity of the hypergeometric function in \eqref{BounceAction_TDW_General}. However, this does not eliminate such a discontinuity to exist. Indeed, we observe that for the TDW in \eqref{TDW_generic}, the action $S_\mcalB^{\mrmBT}$ is singular at $\g=-2$ and the deformed P-NP relation \eqref{P_NP_deformed_general} holds around $u_\mrmBT$. We verify this numerically in Section~\ref{Section: TDW}. \\

For the generic TDW, the explicit expressions for $R^{(i)}$ and $\det \tM^{(i)}$ for different $u_i$ cannot be obtained. Therefore, we now use the specific form of TDW in \eqref{TDW_generic}. Then, we find the turning points $t^{(i)}_{1,2}$ at each level as follows:
\begin{itemize}
	\item \underline{At $u=u_\mrmFV$}: 
	\begin{align}
		t^{\mrmFV}_1 = 1 + \g - \sqrt{2 \g + \g^2}\, ,  \qquad t^{\mrmFV}_2 =  1+ \g+\sqrt{2\g +\g^2}  \, . \label{TurningPoints_FV}
	\end{align}
	\item \underline{At $u=u_\mrmTV$}: 
	\begin{align}
		t^{\mrmTV}_1 &= \frac{1}{2} \left(-1-\g- \D + \sqrt{1+ 2\g+ \g^2-\D-\g\D}\right) \, , \nonumber \\ 
		t^{\mrmTV}_2 &= \frac{1}{2} \left(-1-\g- \D - \sqrt{1+ 2\g+ \g^2-\D-\g\D}\right)\, . \label{TurningPoints_TV}
	\end{align}	
	\item \underline{At $u=u_\mrmBT$}: 
	\begin{align}
		t^{\mrmBT}_1 &= \frac{1}{2} \left(-1-\g- \D + \sqrt{1+ 2\g+ \g^2+\D+\g\D}\right) \, , \nonumber\\ 
		t^{\mrmBT}_2 &= \frac{1}{2} \left(-1-\g- \D - \sqrt{1+ 2\g+ \g^2+\D+\g\D}\right) \, . \label{TurningPoints_BT}
	\end{align}	
\end{itemize}
where $\D = \sqrt{1+ 18 \g +9 \g^2}$. Inserting these expressions in \eqref{residueFunction_TDW_Generic} and \eqref{determinantScaled}, we obtain
\begin{align}
	R^{(i)}(\g) = \frac{1}{2}\g(\g+1)(\g+2)\, ,\qquad \det \tM^{(i)} = -\frac{1}{3}\, ,
\end{align}
for all three cases. Then, finally $f_{2,3}(\g)$ in the P-NP relation become
\begin{equation}
	f_2^{(i)}(\g) = -\frac{3}{2}\left[3\g(\g+2) + 2\right] \qquad , \qquad f_3^{(i)}(\g) = -\frac{3}{2}\g(\g+1)(\g+2) \, , \label{TDW_PNP_parameters}
\end{equation} 
for all cases, showing the invariance of $f_2^{(i)}(\g)$ and $f_3^{(i)}(\g)$ terms in P-NP relation for TDW potential.

This argument shows that despite the loss of simple scaling dependence of the classical perturbative and non-perturbative actions, the parameters $f_2(\g)$ and $f_3(\g)$ are invariant at least for the TDW potential. This sets the invariance of the P-NP relation up to $f_1(\g)$ term. Contrary to the symmetric cases, $f_1(\g)$ does not scale with the frequencies associated with the perturbative saddles. However, we remind that the only role of $f_1(\g)$ is to cancel the $O(g^0)$ term of the non-perturbative quantum actions, $a^\mrmD(u,g)$ or $a(u,g)$ depending on the sector of interest. Therefore, the way $f_1(\g)$ term changes does not affect the instanton function $\mcalG(\tu,g)$. This is particularly apparent in \eqref{HigherOrders_Homog} and \eqref{HigherOrders_Inhomog}, where $\frac{\dee \tu^{(i)}}{\dee \mcalF}$ is the only factor depending on the perturbative saddle of interest.

\section{Asymmetric triple-well}\label{Section: ATW}

\begin{figure}
	\centering
	\includegraphics[width=0.7\textwidth]{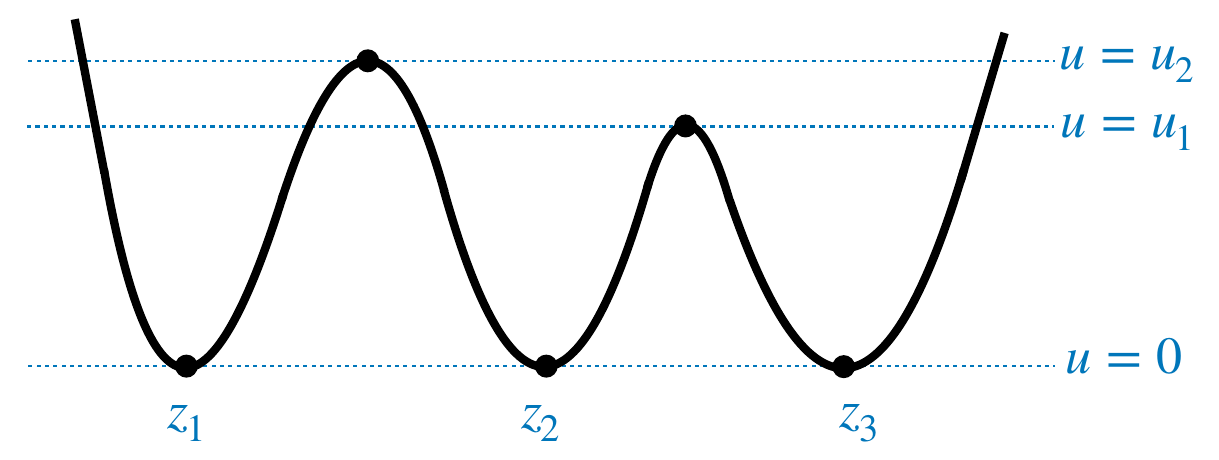}
	\caption{Asymmetric triple-well potential with non-degenerate minima at $z_i$, which lie at $u=0$. Other critical points at $u=u_1$ and $u=u_2$ are indicated, separating the spectrum into 3 continuously connected sectors.}	\label{Figure: ATW_Potential}
\end{figure}

As a generalization of our previous work \cite{Misumi:2024gtf} to the non-degenerate saddles, we firstly consider the asymmetric triple-well (ATW) potential. In its most general form, the triple well corresponds to a sextic curve with five real saddle points, i.e 3 minima and 2 maxima. In this section, we set all minima to $u=0$ level and consider the potential of the following form
\begin{equation}\label{ATW_potential}
	V_\mrmATW(x) = (x-t_1)^2 (x-t_2)^2 (x-t_3)^2\, ,
\end{equation} 
where we assume $t_1<t_2<t_3$ without loss of generality. Note that in general, the barrier tops of the potential are not at the same level unless $t_2-t_1 = t_3-t_2$, i.e. the symmetric limit.  In the following, we consider the case with a higher left barrier than the right one for concreteness in our discussion. (See Fig~\ref{Figure: ATW_Potential}.)

We first briefly discuss the exact quantization of the ATW in all sectors. Then, we solve the exact quantization condition around each minimum. The latter will be the first precise demonstration of the trans-series construction for a potential without any symmetry between its perturbative and non-perturbative cycles. We illuminate the roles of bion configurations in the trans-series in connection with the symmetric limit discussed in \cite{Dunne:2020gtk}. 

Later, we take the symmetric triple-well (STW) limit and focus on the corresponding Chebyshev potential as a non-trivial example on the transformations of P-NP relations, which we discussed in Section~\ref{Section: P_NP_revisit}. In the symmetric limit, the dual potential with appropriate boundary conditions is associated with a PT-symmetric system which is known to have a real spectrum\footnote{In fact, a classically $PT$-symmetric Hamiltonian, which commutes with $PT$ operator, could have a complex eigenvalues if the eigenstates of the Hamiltonian were not eigenstates of the $PT$-operator, which is possible since $PT$ is an anti-linear operator. Such a case us called \textit{broken} $PT$-symmetric phase. While it would be interesting to study in EWKB formalism, we only consider unbroken phase here.} \cite{Bender:2023cem}. At the end of this section, we study its exact quantization and show the reality of the energy trans-series in EWKB formalism. We also verify the ambiguity cancellations at the leading order using the large-order low-order resurgence properties between the perturbation series and the non-perturbative configurations, which was previously studied only in Hermitian systems.

\subsection{Transition between different sectors}\label{Section: ATW_Transition}
\begin{figure}
	\centering
	\caption*{$\bm{\underline{0 < u < u_1}}$}
	\vspace{6pt}
	\begin{subfigure}[h]{0.48\textwidth}
		\caption{\underline{$\t_{\l_1} = 0^-$}}	\label{Figure: ATW_StokesDiagram1_Minus}
		\includegraphics[width=\textwidth]{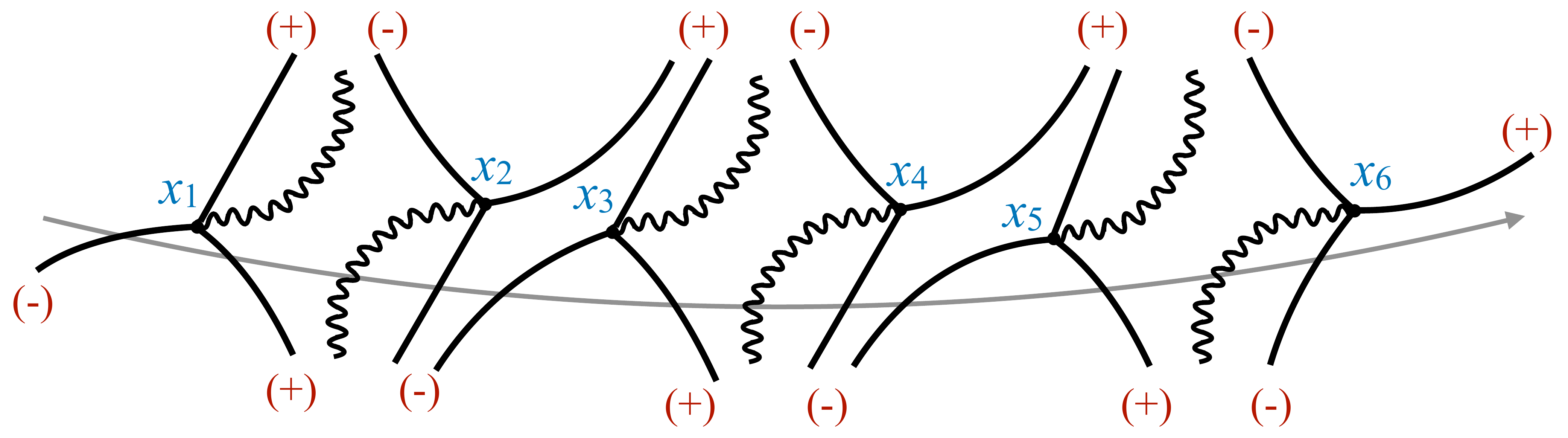}
	\end{subfigure}
	~\hfill 
	\begin{subfigure}[h]{0.48\textwidth}
		\caption{\underline{$\t_{\l_1} = 0^+$}}	\label{Figure: ATW_StokesDiagram1_Plus}
		\includegraphics[width=\textwidth]{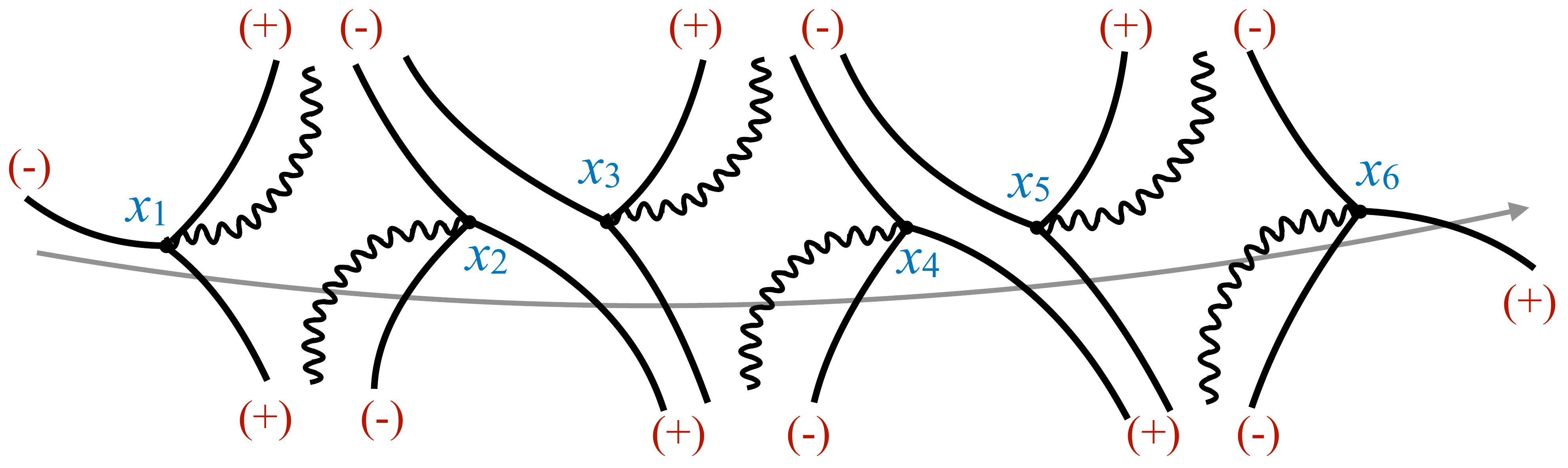}
	\end{subfigure}
	
	\vspace{15pt}
	\caption*{$\bm{\underline{u_1 <  u < u_2}}$}
	\vspace{6pt}
	\begin{subfigure}[h]{0.48\textwidth}
		\caption{\underline{$\t_{\l_1} = -\pi^+$ $(\t_{\l_2} = 0^-)$}}	\label{Figure: ATW_StokesDiagram2_Minus}
		\includegraphics[width=\textwidth]{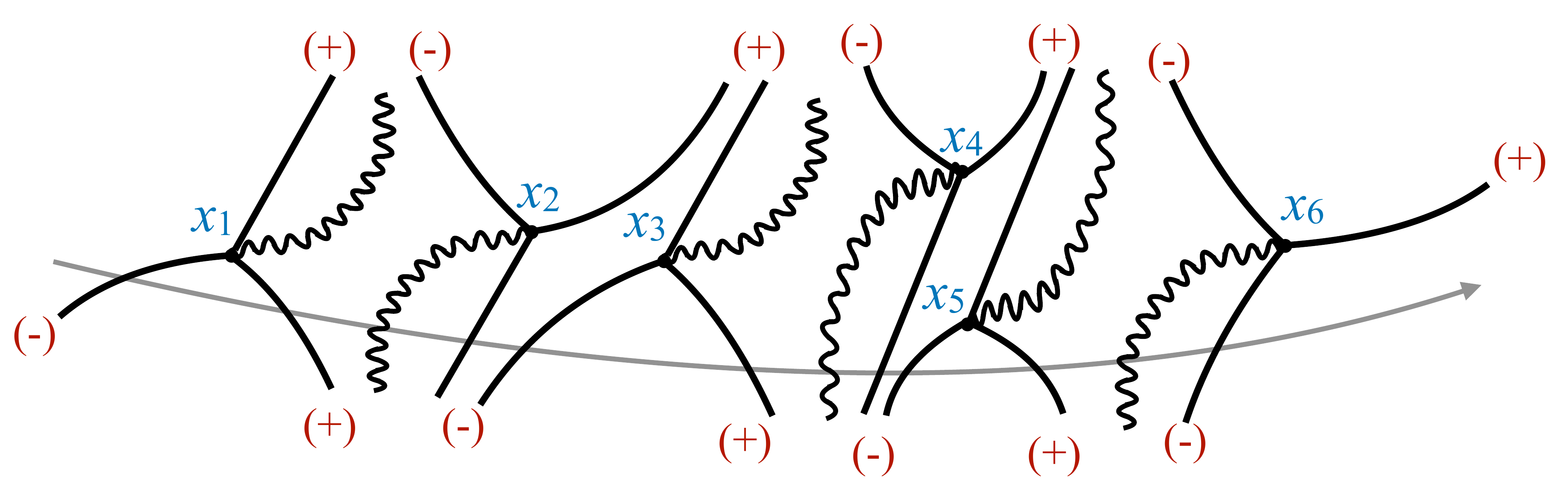}
	\end{subfigure}
	~\hfill 
	\begin{subfigure}[h]{0.48\textwidth}
		\caption{\underline{$\t_{\l_1} = \pi^-$ $(\t_{\l_2} = 0^+)$}}	\label{Figure: ATW_StokesDiagram2_Plus}
		\includegraphics[width=\textwidth]{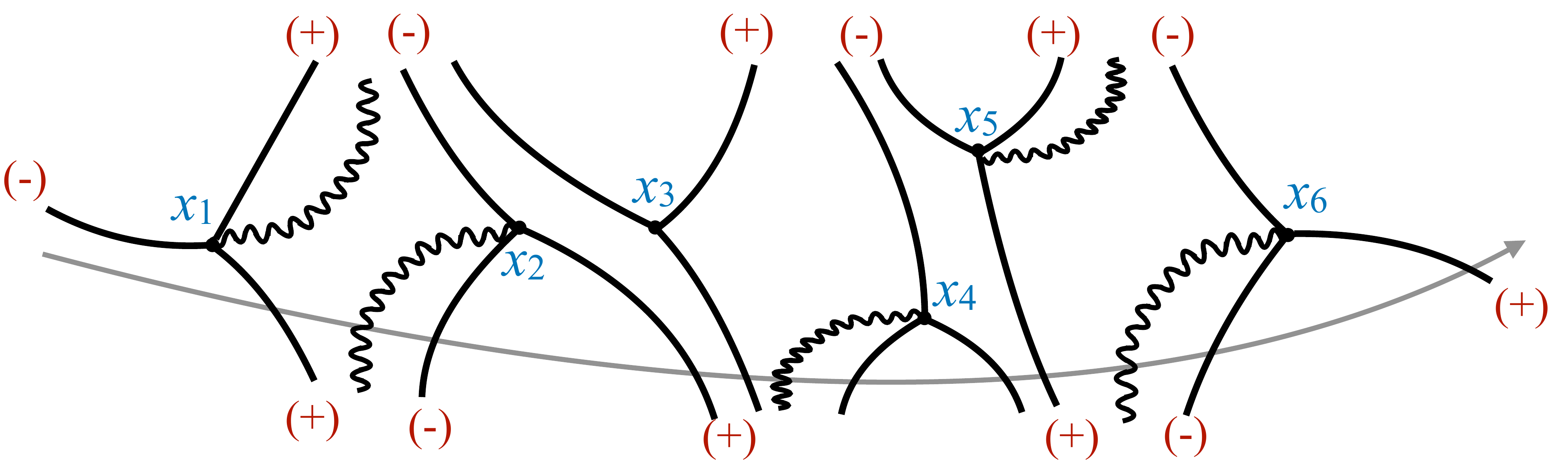}
	\end{subfigure}
	
	\vspace{15pt}
	\caption*{$\underline{\bm{u_2 < u}} $}
	\vspace{6pt}
	\begin{subfigure}[h]{0.48\textwidth}
		\caption{\underline{$\t_{\l_2} = -\pi^+$}}	\label{Figure: ATW_StokesDiagram3_Minus}
		\includegraphics[width=\textwidth]{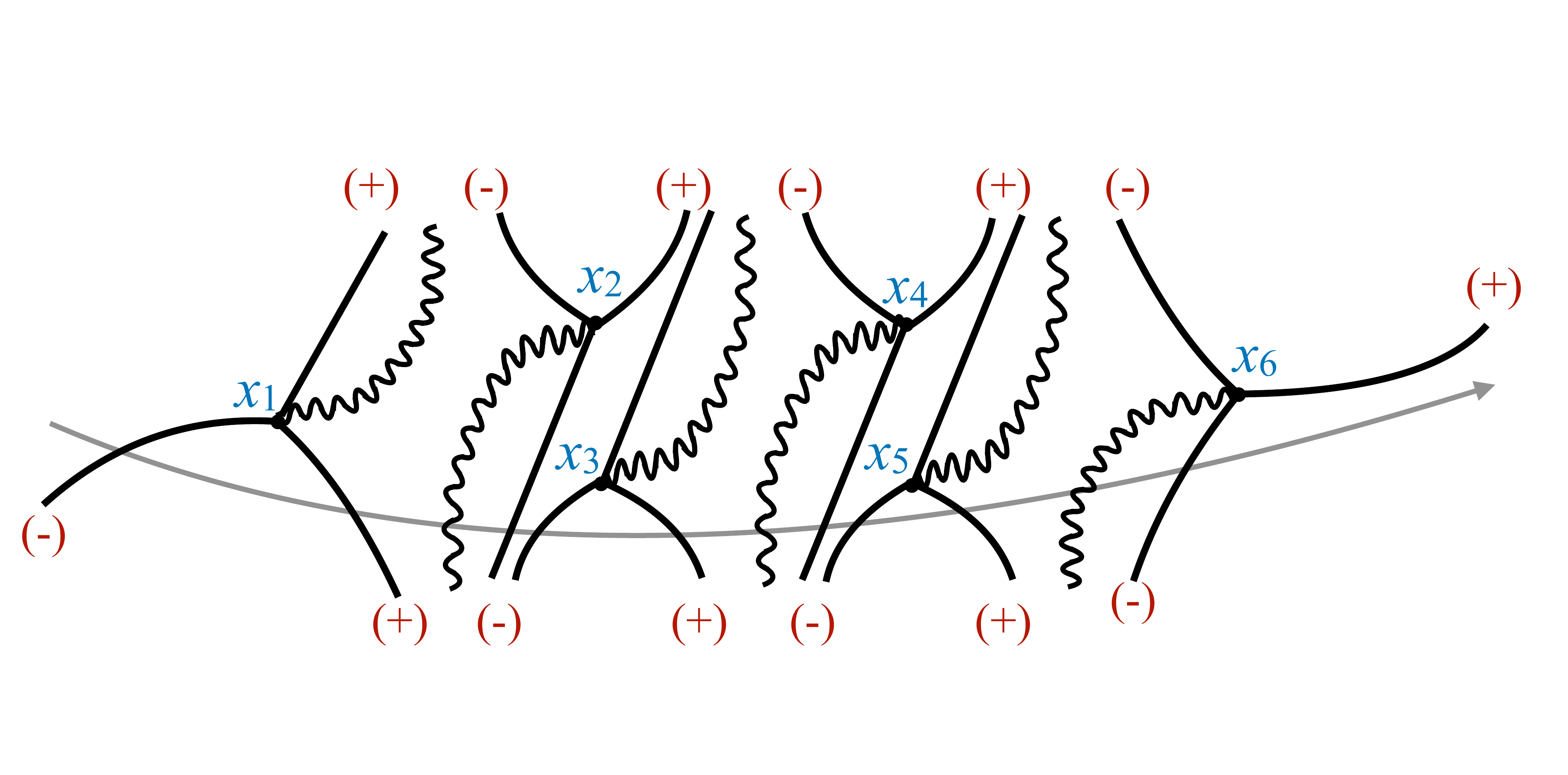}
	\end{subfigure}
	~\hfill 
	\begin{subfigure}[h]{0.48\textwidth}
		\caption{\underline{$\t_{\l_2} = \pi^-$}}	\label{Figure: ATW_StokesDiagram3_Plus}
		\includegraphics[width=\textwidth]{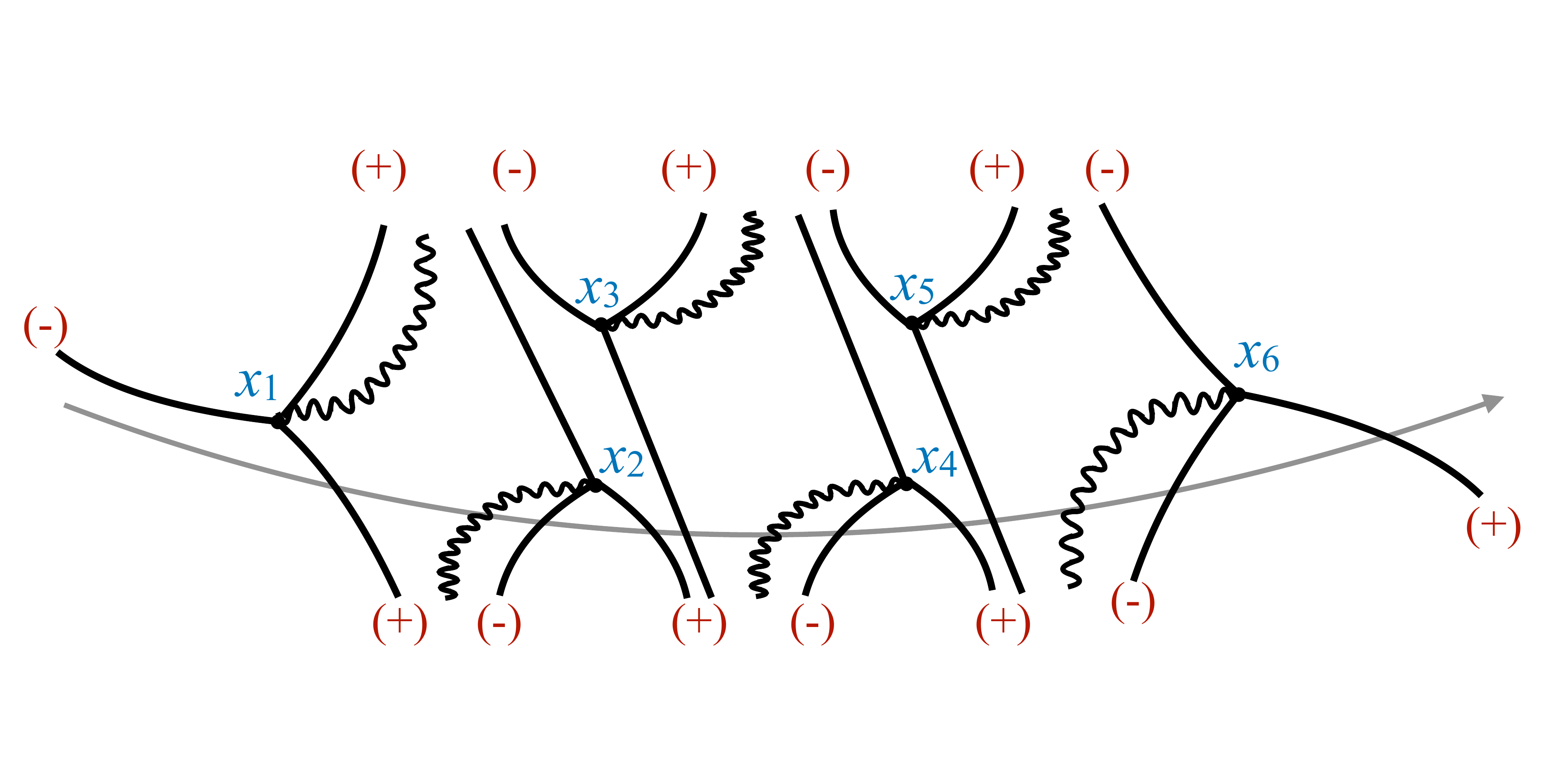}
	\end{subfigure}
	\vspace{15pt}
	\caption{Stokes diagrams for the ATW potential in the three spectral sectors, i.e. $0< u < u_1$, $u_1< u < u_2$ and $u_2< u$. 
	} \label{Figure: ATW_Diagrams}
\end{figure}

For the potential \eqref{ATW_potential}, there are two transition points, i.e.~the levels of barrier tops at $u_1$ and  $u_2$. The Stokes diagrams in each sector are illustrated in Fig.~\ref{Figure: ATW_Diagrams}. Through the analytic continuation of the energy parameter, the Stokes diagrams are continuously connected to each other in the light of Fig.~\ref{Figure: Transition_aboveBarrierTop}. 

Let us describe the two-step transition: Around $u=u_1$, we first parametrize the Airy-type algebraic curve as
\begin{equation}
	P_\mrmATW = 2\left(V_\mrmATW - u_1 + |\l_1|e^{\t_{\l_1}} \right) \, .  \label{ATW_curve1}
\end{equation}
Then, $\t_{\l_1}= 0, \pm \pi$ correspond to $u<u_1$ and $u>u_1$ sectors, respectively. The analytic continuations of the Stokes diagrams concerning the change in $\t_{\l_1}$ are given as
\begin{itemize}
	\item $\t_{\l_1}: 0^- \rightarrow -\pi^+ \quad  \Longrightarrow \quad $  Fig.~\ref{Figure: ATW_StokesDiagram1_Minus} $\rightarrow$  Fig.~\ref{Figure: ATW_StokesDiagram2_Minus} ,
	\item $\t_{\l_1}: 0^+ \rightarrow +\pi^- \quad \Longrightarrow \quad $  Fig.~\ref{Figure: ATW_StokesDiagram1_Plus} $\rightarrow$  Fig.~\ref{Figure: ATW_StokesDiagram2_Plus}  .
\end{itemize}
Since the transitions are continuous, they do not change the quantization conditions:
\begin{align}
	D_{\t_{\l_1}=0^-} = D_{\t_{\l_1}=-\pi^+}  &= \O_\mrmATW\, \Bigg\{ \left(1+\Pi_A^{(1)}\right)\left(1+\Pi_A^{(2)}\right)\left(1+\Pi_A^{(3)}\right) +  \Pi_B^{(1)} \Pi_B^{(2)}  \nonumber \\
	&\quad \qquad   + \Pi_B^{(1)}\left(1+\Pi_A^{(3)}\right)   + \Pi_B^{(2)}\left(1+\Pi_A^{(1)}\right)\Bigg\}	\,,\label{ATW_EQC_Minus1} \\ 
	D_{\t_{\l_1}=0^+} = D_{\t_{\l_1}=\pi^-}  &= \O_\mrmATW \, \Bigg\{ \left(1+\Pi_A^{(1)}\right)\left(1+\Pi_A^{(2)}\right)\left(1+\Pi_A^{(3)}\right) + \Pi_A^{(1)}\Pi_A^{(2)}\Pi_A^{(3)} \Pi_B^{(1)} \Pi_B^{(2)}  \nonumber \\
	&\quad \quad   + \Pi_A^{(1)}\Pi_A^{(2)}\Pi_B^{(1)}\left(1+\Pi_A^{(3)}\right) + \Pi_A^{(2)} \Pi_A^{(3)}\Pi_B^{(2)}\left(1+\Pi_A^{(1)}\right)\Bigg\} \, ,  \label{ATW_EQC_Plus1}
\end{align}
where $\O_\mrmATW = \left(\Pi_A^{(1)}\Pi_A^{(2)}\Pi_A^{(3)}\Pi_B^{(1)}\Pi_B^{(2)}\right)^{-1/2}$.

Similarly, around $u=u_2$, we parametrize the curve as
\begin{equation}
	P_\mrmATW = 2\left(V_\mrmATW - u_2 + |\l_2|e^{\t_{\l_2}} \right) \, .  \label{ATW_curve2}
\end{equation}
Note that $\t_{\l_2}=0$ in \eqref{ATW_curve2} is the same as $\t_{\l_1}=\pm \pi$ in \eqref{ATW_curve1} for suitable values of $\l_1$ and $\l_2$, i.e. $\l_1 + \l_2 = u_2 - u_2$. Then, it is straightforward to see that $T_{\t_{\l_1}=\mp \pi^\pm} = T_{\t_{\l_2} = 0^\mp}$ and therefore, 
\begin{equation}
	D_{\t_{\l_1}=\mp \pi^\pm} = D_{\t_{\l_2} = 0^\mp}\, , \label{u1u2_relation}
\end{equation}
which is nothing but the manifestation of unchanged Stokes geometry between $u=u_1$ and $u=u_2$.

The transition across $u=u_2$ is carried out by the following analytic continuations of the Stokes diagrams: 
\begin{itemize}
	\item $\t_{\l_2}: 0^- \rightarrow -\pi^+$  Fig.~\ref{Figure: ATW_StokesDiagram2_Minus} $\rightarrow$  Fig.~\ref{Figure: ATW_StokesDiagram3_Minus} \, ,
	\item $\t_{\l_2}: 0^+ \rightarrow +\pi^-$  Fig.~\ref{Figure: ATW_StokesDiagram2_Plus} $\rightarrow$  Fig.~\ref{Figure: ATW_StokesDiagram3_Plus} \, .
\end{itemize}
Again, it is manifest that the exact quantization conditions in both sectors are
\begin{align}
	D_{\t_{\l_2}=0^-} &= D_{\t_{\l_2}=-\pi^+}     \label{ATW_EQC_Minus2} \\
	D_{\t_{\l_2}=0^+} &= D_{\t_{\l_2}=\pi^-}  \, , \label{ATW_EQC_Plus2}
\end{align}
which are equal to \eqref{ATW_EQC_Minus1} and \eqref{ATW_EQC_Plus1} respectively.
Therefore, we observe that the form of the exact quantization conditions stays intact throughout all sectors, as expected.\\

\noindent {\bf \underline{Smooth transition of the spectrum}:}~While the smooth connection of the Stokes geometries is the manifestation of having the same exact quantization conditions in different energy sectors, it does not directly mean the smooth transition of the spectrum or the associated trans-series. Note that, solving the quantization conditions leads to the energy trans-series, which represents the ``exact'' spectrum of the quantum system. The procedure of solving them is well-established around the bottom of wells, but solving the quantization conditions around barrier tops is a much harder task. It is conjectured by some numerical computations for certain examples \cite{Basar:2015xna,Dunne:2016qix} that the trans-series structure is preserved although the perturbative/non-perturbative properties of cycles change in different regions of the spectrum.

A conclusive (formal) observation on the smooth transition of the spectrum, on the other hand, can be achieved by considering the median summation of the quantization conditions via Stokes automorphism \cite{Misumi:2024gtf}. This is nothing but the Borel summation procedure applied at the quantization condition level \cite{DDP2}. (See also \cite{Kamata:2023opn,Kamata:2024tyb} for an accessible review.) Then, as in \cite{Misumi:2024gtf}, we remove the ambiguities from the quantization conditions for each sector and show that the resulting quantization conditions, which we call ``median QC'', are manifestly real and are in the same form for all energy sectors. This establishes the reality of the spectrum and its smooth transition across different sectors.

\begin{figure}
	\centering
	\includegraphics[width=1.0\textwidth]{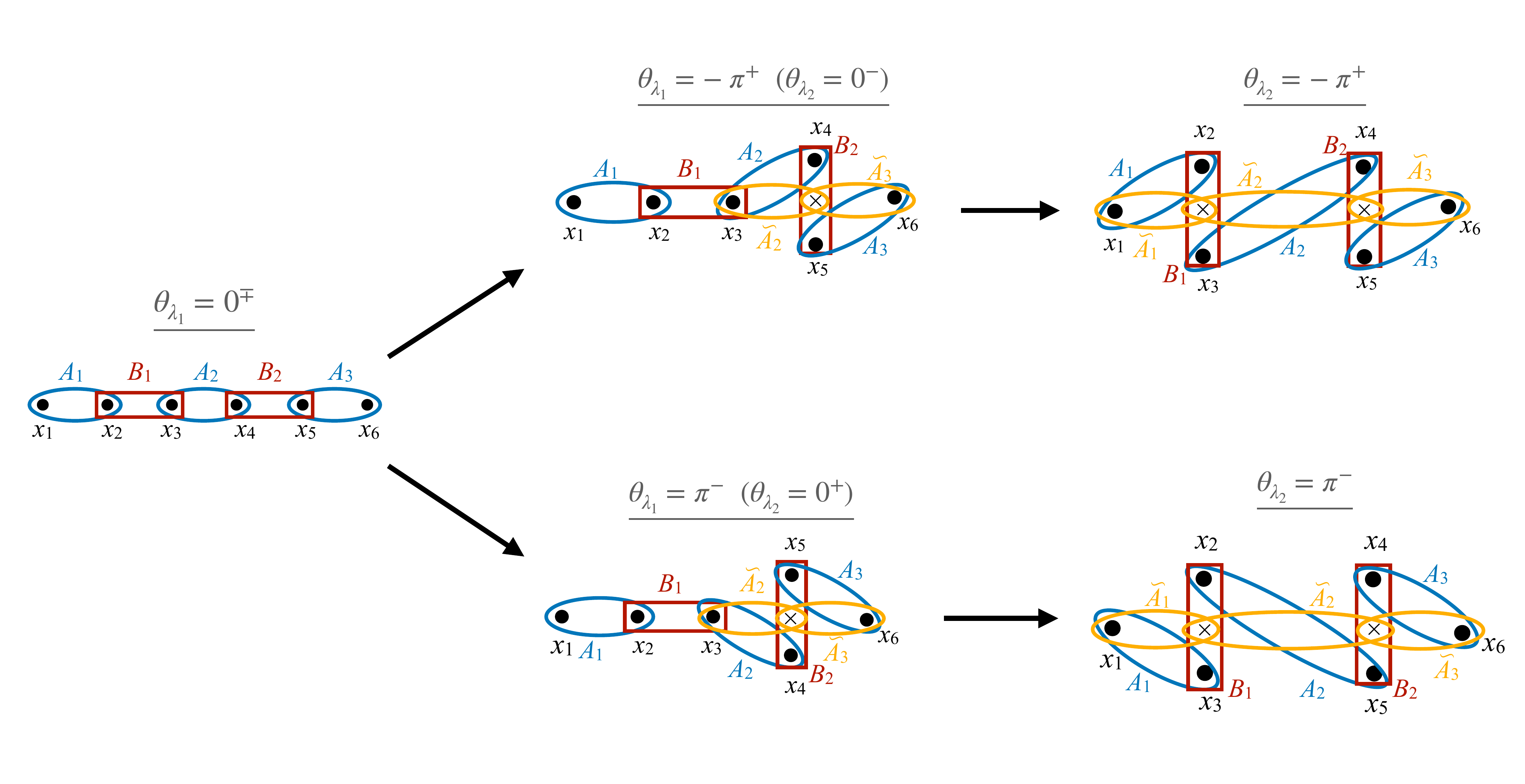}
	\caption{WKB cycles in all sectors of the ATW potential. Same coloured cycle in different sectors are direct analytic continuations of each other. In $u_1<u<u_2$ and $u_2<u$, the $A_i$-cycles associated with the barrier top sectors in respective regions are redefined as $\tA_i$-cycles, which are part of the homology basis of the cycles. These re-definitions also keep the median QCs in all sectors intact.
    }	\label{Figure: ATW_Cycles}
\end{figure}

Before precise computations, we recall that while transitioning to above-barrier-top sectors, the perturbative actions hit a singularity and require an analytic continuation. In \cite{Misumi:2024gtf}, for genus-1 Chebyshev potentials, we adapt the analytic properties of the (Gauss) hypergeometric function, which represents the quantum actions, to analyze the analytic continuations across the barrier top. For higher genus potentials, the actions are not simply given by the hypergeometric functions and in fact it may not be possible to obtain a closed form for all cases. However, in the following, we use the prescription introduced in \cite{Misumi:2024gtf} by using the Stokes geometry and imposing linear independence between WKB cycles in each sector, and show that all the median QCs in the different energy sectors become identical. 

Note that the WKB cycles are originally introduced to the quantization conditions via connection matrices. However, it is not always the case that they lead to a linearly independent set which forms a homology basis of cycles. For the ATW, we illustrate the WKB cycles in all sectors in Fig.~\ref{Figure: ATW_Cycles}. For $0<u<u_1$ sector, $A_i$ and $B_i$ are given directly by the connection matrices and are linearly independent. However, this is not the case for the sectors $u_1<u<u_2$ and $u_2<u$. For example, in $u_1<u<u_2$ sector, for both analytic continuations, we observe that the $A_{2,3}$-cycles can be rewritten as $A= \tA_2 + B^{1/2}_2$. This shows their dependence on the $B_2$ cycle, while $\tA_2$ is independent from the $B_2$-cycle. Similarly, in $u_2<u$ sector, all the original $A_{1,2,3}$-cycles can be re-written in terms of $B_{1,2}$ cycles as $A_{1,3} = \tA_{1,3}+B^{1/2}_{1,2}$ and $A_2 = \tA_2 + B_1^{1/2} B_2^{1/2}$.

At this point, we signify that the above relations are purely geometric and simply inferred from the orientations of the cycles in Fig.~\ref{Figure: ATW_Cycles}. However, as we showed in \cite{Misumi:2024gtf}, re-writing $A_i = \tA_i + B_j$, in fact, reveals the analytic structures of the associated quantum actions, where $B_j$ represents the discontinuities of the branch cuts on $u$-plane. Then, we observe that removing the discontinuity leads to the linearly independent cycles which should be taken into account in the quantization procedure\footnote{We note that this is in equal footing with the constructing a Lefschetz thimble decomposition using an appropriate homology basis in path integral formalism. In \cite{Sueishi:2020rug,Sueishi:2021xti}, the connection between the thimbles and the WKB-cycles has been verified in certain examples around corresponding potentials' minima. The precise extension of this relationship to all sectors is an open problem.}. The last point was verified in \cite{Misumi:2024gtf} and we verify it again in the following:\\


\noindent{\bf{\underline{${\bm{D_\med}}$ for $\bm{0<u<u_1}$}:}} Let us start with removing the discontinuities in the three-well region ($0<u<u_1$) in Fig~\ref{Figure: ATW_Potential}. We first note that in this region the quantization conditions $D_{\t_{\l_1}=0^\pm} $ are related via a Stokes automorphism, i.e. $D_{\t_{\l_1}=0^+} = \mfrS_0 D_{\t_{\l_1}=0^-}$, 
where the Stokes automorphism induces 
\begin{align}
	\mfrS_0:& \Pi^{(1,3)}_{A_0^-} \mapsto \Pi^{(1,3)}_{A_0^+} \left(1 + \Pi^{(1,2)}_{B_0}\right) \, , \label{stokesAuto_ATW_1} \\ 
	\mfrS_0:& \Pi^{(2)}_{A_0^-} \mapsto \Pi^{(2)}_{A_0^+} \left(1 + \Pi^{(1)}_{B_0}\right) \left(1 + \Pi^{(2)}_{B_0}\right) \, . \label{stokesAuto_ATW_2}
\end{align} 
Then, the median QC is given by 
\begin{equation}
	D^\med_1 = \mfrS_0^{1/2} D_{\t_{\l_1}=0^-}= \mfrS_0^{-1/2} D_{\t_{\l_1}=0^+} \label{TDW_medianQC_1}
\end{equation}
and after a straightforward but tedious calculation, for both $D_{\t_{\l_1}=0^\mp}$, we get 
\begin{align}
	D_1^\med &= \O_\mrmATW \Bigg\{\sqrt{1+\Pi_B^{(1)}}\sqrt{1+ \Pi_B^{(2)}} + \Pi_{A}^{(1)}\sqrt{1 + \Pi_B^{(2)}} + \Pi_{A}^{(3)}\sqrt{1 + \Pi_B^{(1)}} + \Pi_{A}^{(2)} \Pi_{A}^{(3)}\sqrt{1 + \Pi_B^{(2)}} \nonumber \\
	& \quad  +\Pi_{\tA}^{(1)}\Pi_{A}^{(2)}\sqrt{1 + \Pi_B^{(1)}}  + \Pi_{A}^{(1)}\Pi_{A}^{(3)}+\Pi_{A}^{(1)}\Pi_{A}^{(2)}\Pi_{A}^{(3)}\sqrt{1+\Pi_B^{(1)}}\sqrt{1+ \Pi_B^{(2)}}\Bigg\} \, ,  \label{ATW_medianQC_Result1}
\end{align}	
where the prefactor $\O$ is given below \eqref{ATW_EQC_Plus1}. \\

\noindent{\bf{\underline{{$\bm {D_\med}$ for $\bm {u_1<u<u_2}$}:}}}	Upon transitioning to $u_1<u<u_2$ sector in Fig~\ref{Figure: ATW_Potential}, the geometry of the Stokes diagram and the WKB cycles change. As we discussed above, proper $A$-cycles in this sector are $\tA_{2,3}$, rather than $A_{2,3}$, which are originally indicated via the Stokes diagrams. Therefore, after the Stokes automorphisms are applied to the quantization conditions, the actions associated with $A$-cycles should be transformed to ones associated with $\tA$-cycles using 
\begin{equation}
	\Pi^{(2,3)}_{A_0} \rightarrow \Pi^{(2,3)}_{\tA_{\pm \pi}} = \left(\Pi^{(2)}_{B_{\pm \pi}}\right)^{\pm 1/2}\, \Pi^{(2,3)}_{A_{\pm \pi}}\, . \label{ATW_actionAC}
\end{equation}
We note that the difference between the power of $\Pi^{(2)}_{B_{\pm \pi}}$ is simply due to the relative differences between the branch cut orientations at $\t_{\l_1} = \pm \pi$. This results in the actions $\Pi_A^{(2,3)}$ are mapped to different Riemann sheets during the analytic continuations $\t_{\l_1}: 0^\mp \rightarrow \pi^\pm$. We refer to Appendix A of \cite{Misumi:2024gtf} for a detailed discussion on this property of the analytic continuations of the Stokes diagrams.

Using these guidelines, we now compute the median QCs at $\l_1=\pm \pi$: Contrary to the sector $0<u<u_1$, the resulting Stokes diagrams and associated quantization conditions at $\t_{\l_1}=\pm\pi^\mp$ are not related to each other via Stokes automorphism, which is again due to the differences in their branch cut structures. Despite this disconnection, however, their median QCs are still the same. 

For convenience in the computations, we start with inverting $B_2$ cycles in $D_{\l_1=\mp\pi^{\pm}}$. With this inversion, the Stokes automeorphisms become
\begin{align}
	\mfrS_{\mp \pi^\pm}:& \; \Pi^{(1)}_{A_{\mp \pi^\mp}} \mapsto \Pi^{(1)}_{A_{\mp \pi^\pm}} \left(1 + \Pi^{(1)}_{B_{\mp \pi}}\right) \, , \label{stokesAuto_ATW_2_1} \\ 
	\mfrS_{\mp \pi^\pm }:& \;\Pi^{(2)}_{A_{\mp \pi^\mp}} \mapsto \Pi^{(2)}_{A_{\mp \pi^\mp}} \left(1 + \Pi^{(1)}_{B_{\mp \pi}}\right) \left(1 + \left(\Pi^{(2)}_{B_{\mp \pi}}\right)^{-1}\right) \, , \\ \label{stokesAuto_ATW_2_2}
	\mfrS_{\mp \pi^\pm}:& \; \Pi^{(3)}_{A_{\mp \pi^\mp}} \mapsto \Pi^{(3)}_{A_{\mp \pi^\pm}} \left(1 + \left(\Pi^{(2)}_{B_{\mp \pi}}\right)^{-1}\, \right)\,,
\end{align}
and we compute the respective median quantization via
\begin{align}
	D^\med_2 = \mfrS^{ \pm 1/2}_{\t_{\l_1} =\mp \pi}\; D_{\t_{\l_1}= \mp \pi^\pm }\, .  \label{ATW_EQC_Transition1_Median}
\end{align}	
Finally, inverting $B_2$-cycle back to its original direction and imposing \eqref{ATW_actionAC}, we get the same form of the median QC as in \eqref{ATW_medianQC_Result1}:
\begin{align}
	D_2^\med &= \O_\mrmATW \Bigg\{\sqrt{1+\Pi_B^{(1)}}\sqrt{1+ \Pi_B^{(2)}} + \Pi_{A}^{(1)}\sqrt{1 + \Pi_B^{(2)}} + \Pi_{\tA}^{(3)}\sqrt{1 + \Pi_B^{(1)}} + \Pi_{\tA}^{(2)} \Pi_{\tA}^{(3)}\sqrt{1 + \Pi_B^{(2)}} \nonumber \\
	& \quad  +\Pi_{A}^{(1)}\Pi_{\tA}^{(2)}\sqrt{1 + \Pi_B^{(1)}}  + \Pi_{A}^{(1)}\Pi_{\tA}^{(3)}+\Pi_{A}^{(1)}\Pi_{\tA}^{(2)}\Pi_{\tA}^{(3)}\sqrt{1+\Pi_B^{(1)}}\sqrt{1+ \Pi_B^{(2)}} \Bigg\}\, , \label{ATW_medianQC_Result2}
\end{align}
where we removed the subscripts $\pm\pi$ in the notation for simplicity. Note also that with the transformation \eqref{ATW_actionAC}, we get $\Pi^{(2,3)}_{\tA} =\Pi^{(2,3)}_{\tA_\pi} =\Pi^{(2,3)}_{\tA_{-\pi}}$, indicating a single action associated with the $\tA$-cycles in this sector. 
\\

\noindent{\bf{\underline{{$\bm {D_\med}$ for $\bm {u_2<u}$}:}}} Finally, we show that the median QC in $u>u_2$ sector in Fig~\ref{Figure: ATW_Potential} exhibits the same form as those in the other sectors by following the same procedure. We note that the Stokes automorphisms acts on the actions of original $A$-cycles, i.e. $\Pi^{(i)}_{A}$. Therefore, to obtain $D^\med$ in this sector (which we denote $D^\med_3$), we need to start with the exact quantization conditions $D_{\t_{\l_2}=\mp\pi^\pm}$ in \eqref{ATW_EQC_Minus2} and \eqref{ATW_EQC_Plus2}.

Upon transition to $u_2<u$ sector from $u_1<u<u_2$, the geometry of $A_1$ and $B_2$ cycles change. This means that there is no change in the transformations of the actions $\Pi_{A_{\pm\pi}}^{(3)}$; they are still given by \eqref{ATW_actionAC}. However, due to the change in the orientation of $B_1$ cycle, $\Pi^{(2)}_{A_{\pm\pi^\mp}}$ obtain extra contributions. Then, at $\t_{\l_2}=\pm \pi$, we have
\begin{equation}
	\Pi^{(2)}_{A_0} \rightarrow \Pi^{(2)}_{\tA_{\pm \pi}} = \left(\Pi^{(1)}_{B_{\pm \pi}}\right)^{\pm 1/2} \left(\Pi^{(2)}_{B_{\pm \pi}}\right)^{\pm 1/2}\, \Pi^{(2)}_{A_{\pm \pi}}\,  \label{ATW_actionAC_2}
\end{equation}
The action of $A_{1}$ cycle, on the other hand, is transformed by the addition of only the $B_{1}$ cycle as
\begin{equation}
	\Pi^{(1)}_{A_0} \rightarrow \Pi^{(1)}_{\tA_{\pm \pi}} = \left(\Pi^{(1)}_{B_{\pm \pi}}\right)^{\pm 1/2} \Pi^{(1)}_{A_{\pm \pi}}\, . \label{ATW_actionAC_3}
\end{equation}

To compute the median QC, we invert both $B_1$ and $B_2$ cycles. Then, the associated Stokes automorphisms read as
\begin{align}
	\mfrS_{\pm\pi}:& \Pi^{(1,3)}_{A_{\pm\pi^\mp}} \mapsto \Pi^{(1,3)}_{A_{\pm\pi^\pm}} \left(1 + \left(\Pi^{(1,2)}_{B_{\pm\pi}}\right)^{-1}\right) \, , \label{stokesAuto_ATW_3_1} \\ 
	\mfrS_{\pm \pi}:& \Pi^{(2)}_{A_{\pm\pi^\mp}} \mapsto \Pi^{(2)}_{A_{\pm\pi^\pm}} \left(1 + \left(\Pi^{(1)}_{B_{\pm\pi}}\right)^{-1}\right) \left(1 + \left(\Pi^{(2)}_{B_{\pm\pi}}\right)^{-1}\right) \, , \label{stokesAuto_ATW_3_2}
\end{align}
where the angles in subscripts refer $\t_{\l_2} = \pm \pi$. Then, we obtain the both median QC via
\begin{equation}
	D^\med_3 = \mfrS^{\pm 1/2}_{\t_{\l_2}=\mp \pi}\,  D_{\t_{\l_2}= \mp \pi^\pm } \,.\label{ATW_EQC_Transition2_Median}
\end{equation}
Finally, performing appropriate transformations to each $A$-cycles, we recover the same form of the median QC:

\begin{align}
	D_3^\med &=\O_\mrmATW \Bigg\{ \sqrt{1+\Pi_B^{(1)}}\sqrt{1+ \Pi_B^{(2)}} + \Pi_{\tA}^{(1)}\sqrt{1 + \Pi_B^{(2)}} + \Pi_{\tA}^{(3)}\sqrt{1 + \Pi_B^{(1)}} + \Pi_{\tA}^{(2)} \Pi_{\tA}^{(3)}\sqrt{1 + \Pi_B^{(2)}} \nonumber \\
	& \quad  +\Pi_{\tA}^{(1)}\Pi_{\tA}^{(2)}\sqrt{1 + \Pi_B^{(1)}}  + \Pi_{\tA}^{(1)}\Pi_{\tA}^{(3)}+\Pi_{\tA}^{(1)}\Pi_{\tA}^{(2)}\Pi_{\tA}^{(3)}\sqrt{1+\Pi_B^{(1)}}\sqrt{1+ \Pi_B^{(2)}} \Bigg\} \, .  \label{ATW_medianQC_Result3}
\end{align}

Our argument shows that for a potential with all minima at the same classical energy level, the median QC in each sector stays intact, even when the symmetry of the potential is broken. This indicates the smooth transition of the trans-series of the spectrum. In addition to that the spectrum should be real as the potential is bounded. In EWKB formalism, this is provided by the median quantizations to be invariant under complex conjugation, i.e. $\mathcal{C}\left[D^\med\right] = D^\med$ \cite{DDP2, Kamata:2023opn}, which is performed using $\mcalC\left[\Pi_A\right] = \Pi_A^{-1}$ and $\mcalC\left[\Pi_B\right] = \Pi_B$. 

In all sectors of the ATW potential, the complex conjugation of $D^\med$ leads to
\begin{align}
	\mcalC\left[D^\med\right] &= \O_\mrmATW \Bigg\{ \sqrt{1+\Pi_B^{(1)}}\sqrt{1+ \Pi_B^{(2)}} + \Pi_A^{(1)}\sqrt{1 + \Pi_B^{(2)}} + \Pi_A^{(3)}\sqrt{1 + \Pi_B^{(1)}} \nonumber \\
	& \quad  + \Pi_A^{(2)} \Pi_A^{(3)}\sqrt{1 + \Pi_B^{(2)}}+\Pi_A^{(1)}\Pi_A^{(2)}\sqrt{1 + \Pi_B^{(1)}}  + \Pi_A^{(2)}\Bigg\} \, .\label{ATW_medianQC_conjugate}
\end{align}
Note that we use $\Pi_A^{(i)}$ rather than $\Pi_{\tA}^{(i)}$ for all sectors to simplify the notation and we assume the appropriate transformations are carried out for $A$-cycles in the above-barrier regions. Comparing \eqref{ATW_medianQC_conjugate} with \eqref{ATW_medianQC_Result1},\eqref{ATW_medianQC_Result2} and \eqref{ATW_medianQC_Result3}, we observe that they differ just by the last term. This means that the spectrum is real only if 
\begin{equation}
	\Pi_A^{(1)} \Pi_A^{(3)} = \Pi_A^{(2)}\, . \label{ATW_realityCondition}
\end{equation} 

It is well-known that the condition \eqref{ATW_realityCondition} is satisfied in the symmetric limit \cite{Basar:2017hpr}, which is shown to have a real spectrum \cite{Dunne:2020gtk}. When the symmetry is broken, on the other hand, \eqref{ATW_realityCondition} indicates that the contributions of the $A_1$ and $A_3$ cycles change in a way that the overall change in $\Pi_A^{(1)}\Pi_A^{(3)}$ is always equal to the change in $\Pi_A^{(2)}$. For the generic ATW potential a general verification of \eqref{ATW_realityCondition} might not be possible. Instead, below, for specific examples, we demonstrate how the constraint \eqref{ATW_realityCondition} works together with the bion contributions to the spectrum via the large-order low-order resurgence relations. As \eqref{ATW_medianQC_conjugate} and \eqref{ATW_realityCondition} suggests this coherence becomes the key in reality of the spectrum when the symmetry is broken.

\subsection{Trans-series around non-degenerate minima}\label{Section: ATW_trans-series} Now, we solve the exact quantization conditions around $u=0$ and analyze trans-series stucture, which involve multiple independent bion contributions. We utilize large-order low-order resurgence relations between the perturbative and non-perturbative sectors to numerically verify the reality of the spectrum.

Let us start with the perturbative quantization first: For each well around $z_i$ for $i=1,2,3$ in Fig~\ref{Figure: ATW_Potential}, setting the non-perturbative actions to zero, the exact quantization conditions \eqref{ATW_EQC_Minus1} and \eqref{ATW_EQC_Plus1} simply read
\begin{equation}
	1+\Pi_A^{(i)}= 0\;   \Longrightarrow \; \mcalF_i(\tu,g) = N_i + \frac{1}{2} \equiv \n_i \label{ATW_PerturbativeQC}\, ,\qquad  i=1,2,3 \, .
\end{equation}
Note that for the generic ATW, none of the perturbative levels are degenerate. Then, the energy of perturbative states, namely the actions $\mcalF_i$, are related by a difference. For future use, we define the differences as
\begin{equation}
	\mcalF_2 = \mcalF_1 + \D_{1,2} \; , \qquad \mcalF_3 = \mcalF_1 + \D_{1,3} \; , \qquad \mcalF_3 = \mcalF_2 + \D_{2,3}\; . \label{ATW_differences}
\end{equation}

When the non-perturbative effects are included, the quantization conditions \eqref{ATW_PerturbativeQC} alters accordingly. In the following, we solve the exact quantization conditions \eqref{ATW_EQC_Minus1} and \eqref{ATW_EQC_Plus1} to get the non-perturbative corrections around the bottom of wells.

\paragraph{\underline{Around $\bm {z_1}$}:} We first start with the outer-most left well in Fig~\ref{Figure: ATW_Potential}. With the corresponding non-perturbative shift, the actions $\mcalF_i$ are written as
\begin{equation}
	\mcalF_1 = \n_1 + \d_1\; , \qquad  \mcalF_2 = \n_1 + \D_{1,2}+ \d_1 \; ,\qquad \mcalF_3 = \n_1 + \D_{1,3} + \d_1 \, . \label{NPshift_ATW_1}
\end{equation}
To determine $\d_1$, we first isolate $1+\Pi_A^{(1)}$ and using $\Pi_A^{(i)} = e^{-2\pi i \mcalF_i}$, and we rewrite the quantization conditions as 
\begin{equation}
	2\cos\pi\mcalF_1 = - \frac{\Pi_B^{(1)} \, e^{\mp \pi i \left(\mcalF_1 + \mcalF_2\right)}}{2\cos\pi\mcalF_2} \, \frac{1+\mcalM_1}{1+\mcalM_2}\, , \label{ATW_EQC_around_z1}
\end{equation}
where 
\begin{align}
	\mcalM_1 &= \frac{\Pi_B^{(2)}\, e^{\mp \pi i \mcalF_3}}{2\cos\pi\mcalF_3} \, , \qquad 
	\mcalM_2  = \frac{\Pi_B^{(2)}\, e^{\mp \pi i \left(\mcalF_2 + \mcalF_3 \right)}}{2\cos\pi\mcalF_2 \;\;  2\cos\pi\mcalF_3}\,. \label{shorthand_1}
\end{align}
The $(\pm)$ signs in \eqref{ATW_EQC_around_z1} indicate the ambiguities which are canceled in the full trans-series expression. 

Bringing the quantization condition in the form of \eqref{ATW_EQC_around_z1} makes clear that the $\Pi_B^{(1)}$ is the only source of the first non-perturbative correction at the leading order. The terms $\mcalM_1$, $\mcalM_2$ and $\cos\pi\mcalF_2$ in the first term, on the other hand, starts contributing at next to leading non-perturbative order when \eqref{ATW_EQC_around_z1} is expanded regarding $\d_1\ll 1$. This shows that the divergence of the perturbative expansion around $x=z_1$ is governed by the $B_1$-cycle which is also expected from the Stokes geometry and Stokes automorphisms in \eqref{stokesAuto_ATW_1}.

In the following, we perform the solution up to second non-perturbative corrections. For convenience, we first expand $\frac{1}{1+\mcalM_2}$ up to $\Pi_B^{(2)}$ order and express \eqref{ATW_EQC_around_z1} as
\begin{equation}
	2\cos\pi\mcalF_1 \simeq - \frac{\Pi_B^{(1)} \, e^{\mp \pi i \left(\mcalF_1 + \mcalF_2\right)}}{2\cos\pi\mcalF_2} - \frac{\Pi_B^{(1)}\Pi_B^{(2)}\, e^{\mp \pi i \left(\mcalF_1+\mcalF_2+\mcalF_3\right)}}{2\cos\mcalF_2\;2\cos\pi\mcalF_3} +\frac{\Pi_B^{(1)}\Pi_B^{(2)}\, e^{\mp \pi i \left(\mcalF_1+2\mcalF_2 + \mcalF_3 \right)}}{\left(2\cos\pi\mcalF_2\right)^2 \;\;  2\cos\pi\mcalF_3} \, . \label{ATW_EQC_approx}
\end{equation}
We recall the Weber-type representations for the $B_1$ and $B_2$ cycles in \eqref{Dictionary_NP_Well_Bion}. For ATW potential around $u\sim 0$, they are expressed as
\begin{equation}
	\Pi_B^{(1)} =\frac{2\pi e^{\mcalG_1}\left(\frac{g}{c_1}\right)^{-\mcalF_1} \left(\frac{g}{c_2}\right)^{-\mcalF_2}}{\Gamma\left(\frac{1}{2}+\mcalF_1\right)\Gamma\left(\frac{1}{2}+\mcalF_2\right)} \, , \qquad 	\Pi_B^{(2)} =\frac{2\pi e^{\mcalG_2}\left(\frac{g}{d_2}\right)^{-\mcalF_2} \left(\frac{g}{d_3}\right)^{-\mcalF_3}}{\Gamma\left(\frac{1}{2}+\mcalF_2\right)\Gamma\left(\frac{1}{2}+\mcalF_3\right)} \,.\label{weberType_Bcycles}
\end{equation}
Note that except the symmetric limit, the instanton functions $\mcalG_{1,2}$ and constants $c_{1,2}$ and $d_{2,3}$ are no longer identical. As we will show below, however, these are merely quantitative effects and the roles of the $B_{1,2}$ cycles remain intact, even if one of them becomes more dominant quantitatively.

We now use \eqref{NPshift_ATW_1} in \eqref{ATW_EQC_approx} and expand it with respect to $\d_1 \ll 1$. Recall that $\mcalG_{1,2}$ are also functions of $\mcalF_{1,2}$ so they are expanded with respect to $\d_1$ as well. We express them as
\begin{equation}
	e^{\mcalG_{i}} = e^{-\frac{S^{(i)}_\mcalB}{g}}\sum_{k=0}^\infty H_k^{(i)} \d_1^{k} \, , \qquad i=1,2\, , \label{expansion_bionATW}
\end{equation}
where $S_\mcalB^{(i)}$ is the bion action associated with $\mcalB_i$ cycle, and bring \eqref{ATW_EQC_approx} to the following form:
\begin{align}
	\sum_{n=0} b_n \d_1^{n+1}& = K_1^\mp \sum_{k=0}^\infty J_k^{(1)} \d_1^k \; \pm i (-1)^{N_1} K_1^\mp K_2^\mp \sum_{k=0}^\infty J_k^{(2)} \d_1^k \; \nonumber \\
	& \quad  + e^{\mp i \pi \D_{1,2}}K_1^\mp K_2^\mp \sum_{k=0}^\infty J_k^{(3)} \d_1^k \, , \label{ATW_arranged_expansion}
\end{align}
where 
\begin{align}
	K_1^\mp  &= e^{-\frac{S_\mcalB^{(1)}}{g}} e^{\mp i \pi \D_{1,2}} \left(\frac{c_1 c_2}{g^2}\right)^{N_1+\frac{1}{2}} \left(\frac{c_2}{g}\right)^{\D_{1,2} } \, , \label{ATW_bion_leading1}\\ 
	K_2^\mp &= e^{-\frac{S_\mcalB^{(2)}}{g}} e^{\mp i \pi \D_{1,3}} \left(\frac{d_2 d_3}{g^2}\right)^{N_1+\frac{1}{2}} \left(\frac{d_2}{g}\right)^{\D_{1,2}}\left(\frac{d_3}{g}\right)^{\D_{1,3}} \label{ATW_bion_leading2}
\end{align}
are leading order non-perturbative contributions. They are sometimes called \textit{fugacity} in relation to the dilute instanton gas picture \cite{Callan:1977gz}. Moreover, each series in \eqref{ATW_arranged_expansion} is defined as follows:
\begin{align}
	\sum_{k=0}^\infty J_k^{(1)} \d_1^k & := \Gamma\left(-N_1 - \D_{1,2} - \d_1 \right) e^{-2\d_1\log\left(\frac{g^2}{c_1 c_2}\right)}  \sum_{k=0}^\infty H_k^{(1)} \d_1^{k}\, ,  \label{expansion_ATW_1} \\ \nonumber \\ 
	\sum_{k=0}^\infty J_k^{(2)} \d_1^k & := \frac{\Gamma\left(-N_1 - \D_{1,2} - \d_1 \right)\Gamma\left(-N_1 - \D_{1,3} - \d_1 \right)}{\Gamma\left(1+N_1 + \D_{1,2} + \d_1 \right)} \nonumber \\ 
	& \quad e^{\mp \pi i \d_1} e^{-2\d_1 \left(\log\frac{g^2}{c_1 c_2} + \log\frac{g^2}{d_1 d_2}\right)} \sum_{k=0}^\infty H_k^{(1)} \d_1^{k}\; \sum_{k=0}^\infty H_k^{(2)} \d_1^{k} \, , \label{expansion_ATW_2}\\
	\sum_{k=0}^\infty J_k^{(3)} \d_1^k & := \big[\Gamma\left(-N_1 - \D_{1,2} - \d_1 \right)\big]^2\;  \Gamma\left(-N_1 - \D_{1,3} - \d_1 \right)\nonumber \\ 
	& \quad e^{\mp 2\pi i \d_1} e^{-2\d_1 \left(\log\frac{g^2}{c_1 c_2} + \log\frac{g^2}{d_1 d_2}\right)} \sum_{k=0}^\infty H_k^{(1)} \d_1^{k}\; \sum_{k=0}^\infty H_k^{(2)} \d_1^{k} \, , \label{expansion_ATW_3} \\
	\sum_{n=0} b_n \d_1^{n+1} &:= \frac{e^{\pm 2\pi i \d_1}}{\Gamma\left(-N_1 -\d_1 \right)} \, , \label{expansion_ATW_4}
\end{align}

We can solve \eqref{ATW_arranged_expansion} for $\d_1$ iteratively up to $O(e^{-\frac{S_\mcalB^{(1)} + 2S_\mcalB^{(2)}}{g}})$. Note that the limit on the order of the solution is forced by our decision to expand \eqref{ATW_EQC_around_z1} up to $\Pi_\mrmB^{(2)}$ order in \eqref{ATW_EQC_approx}. Given the form of the expression in \eqref{ATW_arranged_expansion}, the appropriate ansatz for the iterative solution is
\begin{equation}
	\d_1^\pm = \sum_{n=0} \b^{(1)}_n \left(K_1^\mp\right)^n + \sum_{n=0}^\infty \b^{(2)}_n \left(K_1^\mp K_2^\mp \right)^n\, , \label{ATW_ansatz_z1}
\end{equation}
and using it we get the first two non-perturbative corrections for $\mcalF_1$ as
\begin{align}
	\d_1^\pm &\simeq \frac{J_0^{(1)}}{b_0}K_1^\mp + \left[\pm i \frac{(-1)^{N_1}}{b_0}J_0^{(2)} + \frac{e^{\mp i \pi \D_{1,2}}}{b_0} J_0^{(3)}\,\right] K_1^\mp K_2^\mp \nonumber \\ & \; + \frac{1}{b_0^2}\left[\frac{J_0^{(1)}\, J_1^{(1)} }{b_0^2} - \frac{b_1}{b_0}\left(J_0^{(1)}\right)^2\right] \left(K_1^\mp\right)^2 \, . \label{NP_correction_z1}
\end{align}
Finally, the non-perturbative corrections to the perturbative energy can be obtained by using \eqref{NP_correction_z1}, which we use to verify the reality of the spectrum via large-order low-order resurgence later in this section. 

\paragraph{\underline{Around $\bm{z_3}$}:} The quantization conditions \eqref{ATW_EQC_Minus1} and \eqref{ATW_EQC_Plus1} are symmetric under the exchange of the actions of $A_1$ and $A_3$ cycles. Therefore, solving them for the well around $z=z_3$ is qualitatively the same as the procedure we just discussed for the well around $z=z_1$. Main differences are due to the changes in the roles of cycles. In this case, $A_3$ cycle is chosen as the base and all the perturbative shifts $\D_{3,i}$ are determined accordingly. Similarly, the leading order non-perturbative sector is governed by $B_2$-cycle which exchanges the roles with $B_1$-cycle and the latter contributes to the higher orders in the trans-series. The remaining specific expressions for the trans-series solution can be obtained as we explain above; thus we don't repeat the same discussion here.

\paragraph{\underline{Around $\bm{z_2}$}:} Finally, we shift our focus to the inner well around $x=z_2$ in Fig~\ref{Figure: ATW_Potential}. Isolating $1+\Pi_A^{(2)}$ term and using $\Pi_\mrmA^{(i)} = e^{-2\pi i \mcalF_i}$, we get
\begin{equation}
	2\cos\pi \mcalF_2 = - \frac{\Pi_\mrmB^{(1)}\, e^{\mp \pi i \left(\mcalF_1 + \mcalF_2\right)}}{2\cos\pi\mcalF_1} - \frac{\Pi_\mrmB^{(2)}\, e^{\mp \pi i \left(\mcalF_2 + \mcalF_3\right)}}{2\cos\pi\mcalF_3} - \frac{\Pi_\mrmB^{(1)}\Pi_\mrmB^{(2)}\, e^{\mp \pi i \left(\mcalF_1 + \mcalF_2+\mcalF_3\right)}}{2\cos\pi\mcalF_1\; \; 2\cos\pi\mcalF_3} \, . \label{ATW_EQC_around_z2}
\end{equation}
In this case, we relate the actions as
\begin{equation}
	\mcalF_2 = \n_2 + \d_2\; , \qquad  \mcalF_1 = \n_2 - \D_{1,2}+ \d_2 \; ,\qquad \mcalF_3 = \n_2 + \D_{2,3} + \d_2 \, .  \label{NPshift_ATW_3}
\end{equation}
Then, similarly to the case around $z_1$, using the expressions in \eqref{weberType_Bcycles}, \eqref{expansion_bionATW} and the expansion in \eqref{expansion_ATW_4}, we rewrite the exact quantization condition as
\begin{align}
	\sum_{n=0}\tb_n \d_2^{n+1}= \tK_1^\pm \sum_{k=0}^\infty  \tJ_k^{(1)} \d_2^k + \tK_2^\mp \sum_{k=0}^\infty \tJ_k^{(2)} \d_2^k \pm (-1)^{N_2} \tK_1^\pm \tK_2^\mp \sum_{k=0}^\infty \tJ_k^{(3)} \d_2^{(3)}\, , \label{ATW_arranged_expansion2}
\end{align}
where the bion fugacities become
\begin{align}
	\tK_1^\pm &= e^{-\frac{S_\mcalB^{(1)}}{g}} e^{\pm i \pi \D_{1,2}} \left(\frac{c_1 c_2}{g^2}\right)^{N_1+\frac{1}{2}} \left(\frac{c_1}{g}\right)^{-\D_{1,2} } \, , \label{ATW_bion_leading3}\\ 
	\tK_2^\mp &= e^{-\frac{S_\mcalB^{(2)}}{g}} e^{\mp i \pi \D_{2,3}} \left(\frac{d_2 d_3}{g^2}\right)^{N_1+\frac{1}{2}} \left(\frac{d_3}{g}\right)^{\D_{2,3}} \, .\label{ATW_bion_leading4}
\end{align}
Note that these expressions are very similar to the ones appear for the non-perturbative corrections around $z_1$. The main distinctions are due to the relative differences of $\mcalF_i$ in \eqref{NPshift_ATW_1} and \eqref{NPshift_ATW_3}. These lead to the sign difference of the term $\left(\frac{c_1}{g}\right)^{\D_{1,2}}$ in $\tK_1^\pm$ with respect to $K_1^\pm$ as well as the sign difference in the quantization conditions, arising from $e^{\pm i \pi \D_{1,2}}$. Similarly, the disappearance of $\left(\frac{d_2}{g}\right)^{\D_{1,2}}$ in $\tK_2^\mp$ in comparison to $K_2^\pm$ is due to these relative differences. Similar changes appear in the expansion coefficients $\tJ_k^{(i)}$, which are now written as
\begin{align}
	\sum_{k=0}^\infty \tJ_k^{(1)} \d_2^k & := \Gamma\left(-N_2 + \D_{1,2} - \d_2 \right) e^{-\d_2\log\left(\frac{g^2}{c_1 c_2}\right)}  \sum_{k=0}^\infty H_k^{(1)} \d_2^{k}\, ,  \label{expansion_ATW_11} \\ \nonumber \\ 
	\sum_{k=0}^\infty \tJ_k^{(2)} \d_2^k & := \Gamma\left(-N_2 - \D_{2,3} - \d_2 \right) e^{-\d_2\log\left(\frac{g^2}{d_1 d_2}\right)}  \sum_{k=0}^\infty H_k^{(2)} \d_2^{k}\, ,\label{expansion_ATW_22}\\
	\sum_{k=0}^\infty \tJ_k^{(3)} \d_2^k & := \frac{\Gamma\left(-N_2 + \D_{1,2} - \d_2 \right)\Gamma\left(-N_2 - \D_{2,3} - \d_2 \right)}{\Gamma\left(1+ N_2 + \d_2\right)}\nonumber \\ 
	& \quad  e^{-2\d_2 \left(\log\frac{g^2}{c_1 c_2} + \log\frac{g^2}{d_1 d_2}\right)}  \sum_{k=0}^\infty H_k^{(1)} \d_2^{k}\; \sum_{k=0}^\infty H_k^{(2)} \d_2^{k} \, . \label{expansion_ATW_33}
\end{align}

The series expansion of the quantization condition in \eqref{ATW_arranged_expansion2} suggests the following ansatz: 
\begin{equation}
	\d_2^\pm  = \sum_{n=0}^\infty \a_n^{(1)} \left(K_1^\pm\right)^{n+1} + \sum_{n=0}^\infty \a_n^{(2)} \left(K_2^\pm\right)^{n+1} + \sum_{n=0}^\infty \a_n^{(3)} \left(K_1^\pm K_2^\mp\right)^{n+1}\, .\label{ATW_ansatz_z2}
\end{equation}
Then, solving it for $\d_2^\pm$ iteratively, we get first two orders of the non-perturbative corrections to $\mcalF_3$ as
\begin{align}
	\d_2^\pm &\simeq \frac{\tJ_0^{(1)}}{\tb_0} \tK^\pm_1  + \frac{\tJ_0^{(2)}}{\tb_0} \tK^\mp_2 + \left(\pm i \frac{\tJ_0^{(3)}}{\tb_0} + \frac{\tJ_0^{(1)} \tJ_1^{(1)}}{\tb_0^2} + \frac{\tJ_0^{(2)} \tJ_1^{(2)}}{\tb_0^2} - 2\frac{\tb_1}{t_b^3} \tJ_0^{(1)}\tJ_0^{(2)} \right) \tK_1^\pm\, \tK_2^\mp  \nonumber \\ 
	& \quad 
	+ \left(\frac{\tJ_0^{(1)} \tJ_1^{(1)}}{\tb_0^2} - \frac{\tb_1}{\tb_0^3} \left(\tJ_0^{(1)}\right)^2 \right) \left(\tK_1^\pm \right)^2 + \left(\frac{\tJ_0^{(2)} \tJ_1^{(2)}}{\tb_0^2} - \frac{\tb_1}{\tb_0^3}\left(\tJ_0^{(2)}\right)^2\right) \left(\tK_2^\mp\right)^2 \, . \label{NP_correction_z2}
\end{align}
\vspace{0.5cm}

Given the first two orders in the non-perturbative sectors of the trans-series in \eqref{NP_correction_z1} and \eqref{NP_correction_z2}, we now relate these expressions to the bion configurations of path integrals and underlying resurgence structure:

\begin{itemize}[wide]
	\item The organization of the fugacity terms in \eqref{NP_correction_z1} and \eqref{NP_correction_z2} reveals the instanton configurations that organize the non-perturbative sector of the path integral of the partition function. When $K_{1,2}$ (or $\tK_{1,2}$) terms appear alone, it means that it is a contribution of a single bion molecule in the diluted instanton gas description and appearance of $\left(K_{1,2}\right)^n$ (and $\left(\tK_{1,2}\right)^n$) indicates their proliferation. Note that they are the same as the \textit{neutral} bion, which were discussed in the symmetric triple-well \cite{Dunne:2020gtk} as they don't carry any topological charge and they are analogue of the interacting instanton anti-instanton molecules which are famously known from the symmetric double-well potential \cite{Zinn-Justin:1981qzi}.
	
	Combinations of $K_1$ and $K_2$ terms, on the other hand, correspond to correlated events of individual bion molecules in the cluster expansion. Considering the topological charges, they are still \textit{neutral} molecules. For the inner well (around $z_2$), this is in parallel with the symmetric case \cite{Dunne:2020gtk}. For the outer wells (around $z_1$ or $z_3$), on the other hand, the \textit{topological bion} configurations, which contribute as $\left(K_1\right)^{m+\frac{1}{2}}\left(K_2\right)^{n+\frac{1}{2}}$ ($m,n\in \mbbR$) in the symmetric limit, disappear from the spectrum. This is simply due to the asymmetry or, more technically, the lack of degeneracy in the perturbative quantization, which leads to $\D_{k,l}\neq 0$ terms in \eqref{NPshift_ATW_1} and \eqref{NPshift_ATW_3}. In the path integral formalism, the disappearance implies the vanishing contributions of the topological configurations to the spectrum \cite{Dunne:2020gtk,Ture:2024nbi}. Note that the quantization of ATW is also different from the tilted triple well case (both SUSY and non-SUSY) discussed in \cite{Behtash:2017rqj,Kamata:2021jrs}, where the perturbative degeneracy between outer wells is preserved and is broken non-perturbatively. 
	
	\item Note that the correlated and proliferated organization of the bion molecules\footnote{In fact, at a deeper level, bions are also correlated events arising from the instanton and anti-instanton interaction. However, since they disappear from the spectrum of asymmetric potentials, we keep the bion configurations as the most basic building block of the non-perturbative sector.} indicates a cluster expansion at the level of partition function. This is in parallel with the previous discussions on the symmetric potentials in quantum mechanics \cite{Behtash:2018voa}, and also ones induced by a quantum mechanical reduction of a QFT \cite{Pazarbasi:2021ifb,Pazarbasi:2021fey}. Here, we provide an example on its generalization when the symmetry is lost, strengthening the link between the dilute instanton gas picture and the resurgent trans-series.   
	
	\item The leading order non-perturbative contributions to each well comes with an imaginary ambiguity. For outer-wells, this is similar to the symmetric case, where a real contribution of the topological bion is a higher order term in the non-perturbative sector, while there is a complex ambiguous shift at the leading order. In \eqref{NP_correction_z1}, the ambiguity is simply due to the $e^{\mp i \pi \D_{1,2}}$ term in the bion fugacity where $\D_{1,2}$ is analogue of the \textit{hidden topological angle}, which is first found in the context of SUSY in relation to the complex bions \cite{Behtash:2015zha, Behtash:2015loa,Behtash:2017rqj}. In our case, the source of $\D_{1,2}$ is the asymmetry between the neighbouring wells, which is in the same line as the symmetric triple-well \cite{Dunne:2020gtk}. 
	
	\item  The asymmetry between the outer-wells has a more profound outcome for the trans-series associated with the inner well: The leading order non-perturbative terms in \eqref{NP_correction_z2}, i.e.~one bion order, come from both bion configurations. In the symmetric limit, their contributions are the same and appear as a multiplicative factor in the non-perturbative corrections, which is crucial to have a real spectrum. When the symmetry is broken, although the contributions of $B_1$ and $B_2$ cycles are not the same, both of them continue contributing to the spectrum with imaginary ambiguities and the overall contribution should be canceled against the Borel ambiguity of the perturbative sector to get a real spectrum. 
	
	All these are in line with the constraint we found in \eqref{ATW_realityCondition}. Note that the individual bions of $B_1$ and $B_2$ cycles are linked to the perturbative spectrum around $z_1$ and $z_3$, which is crucial for real and unambiguous spectrum. Then, it is natural to expect that the combination of the perturbative sectors around $z_1$ and $z_3$ to be linked to the total contribution coming from both bions ($z_{1}\leftrightarrow z_{2} $ and $z_{3}\leftrightarrow z_{2} $). Then, the latter, namely the bion contributions, would be naturally linked to the perturbative sector around $z_2$, which again leads to the real and unambiguous spectrum. Next, we verify these points quantitatively in specific examples of ATW.
	
\end{itemize}

\subsubsection*{Large-order low-order Resurgence:}
To understand the link between the perturbative expansions and the associated non-perturbative sectors, we utilize the Borel summation. Let us consider a non-alternating divergent series for the perturbative energy around $x=z_i$ in Fig~\ref{Figure: ATW_Potential}:
\begin{equation}
	\ve^{(i)}(g) = \sum_{k=0}^\infty \ve^{(i)}_k g^k \, , \qquad \ve_k \sim \mcalA^{-k-\a} \Gamma\left(k+\b +1\right)\, .  \label{genericDivergentSeries}
\end{equation}
The leading order Borel summation to \eqref{genericDivergentSeries} leads to an imaginary contribution
\begin{equation}
	\Im \ve^{(i)} \sim \mp \pi \mcalA^{-\a+1+\b} g^{-\b - 1} e^{-\frac{\mcalA}{g}}\, , \label{imaginaryBorel}
\end{equation} 
where the $(\pm)$ sign is determined by the choice of the analytic continuation direction, thus it is ambiguous. It is now well-known that for stable systems, such an ambiguity is canceled by an ambiguous imaginary contribution from the non-perturbative sector \cite{Zinn-Justin:1981qzi,Dunne:2014bca}, which is due to one of the bions in ATW case. Then, extending these cancellations to the entire trans-series ensures the reality of the spectrum.

Up to the first non-perturbative correction, we express the trans-series for ATW associated with the perturbative sector around $x=z_i$ 
\begin{equation}
	u_i \simeq \ve^{(i)}(N_i) + \d_i \frac{\dee \ve^{(i)}}{\dee \mcalF_i}\Big|_{\mcalF_i=N_i+ \frac{1}{2}} + O(\d_i^2) \, , \label{ATW_energy_TransSeries}
\end{equation}
where the specific expression of $\d_i$ for $i=1,2$ are given in \eqref{NP_correction_z1} and \eqref{NP_correction_z2}. To observe the cancellations at the leading order, it is enough to include only the fugacity terms $K_i$ or $\tK_i$ in $\Im \d_i \frac{\dee \ve^{(i)}}{\dee \mcalF_i}$. Then, the numerical matching with the perturbative part is satisfied via \eqref{genericDivergentSeries} and \eqref{imaginaryBorel}; and the matching becomes better as the higher order terms are taken into account. It is also possible to include the fluctuations $J_k$ or $\tJ_k$ to get a quicker matching. However, we only use the leading order fugacity terms, as it is enough for our purposes in the following numerical analysis.

\begin{figure}
	\centering
	\begin{subfigure}[h]{0.48\textwidth}
		\caption{\underline{Outer wells vs. neighbouring bions}}	\label{Figure: ATW_Ratio_LeftRight1}
		\vspace{10pt}
		\includegraphics[width=\textwidth]{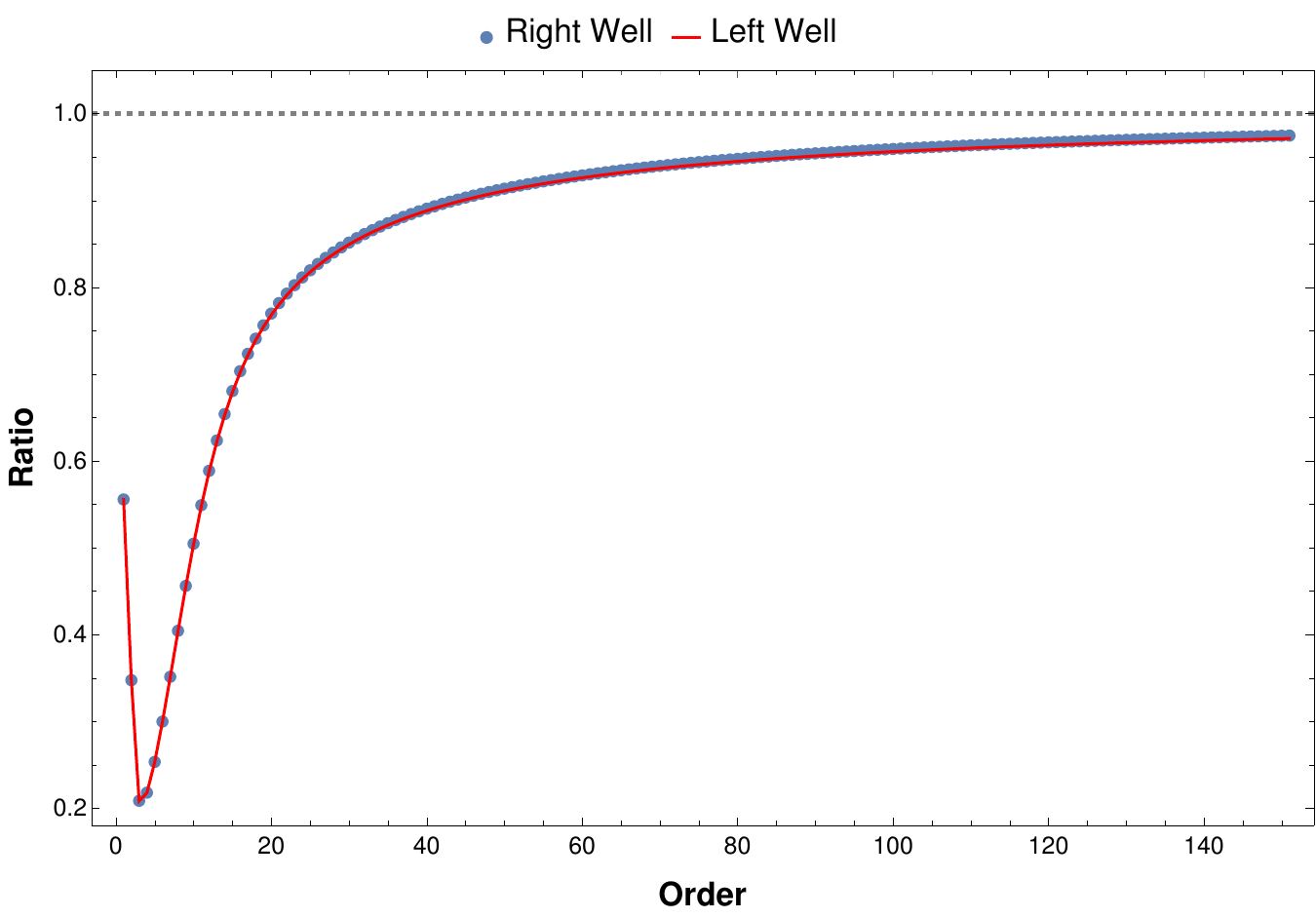}
		\end{subfigure}
	~\hfill 
	\begin{subfigure}[h]{0.48\textwidth}
		\caption{\underline{Inner well vs bions}}	\label{Figure: ATW_Ratio_Inner1}
		\vspace{10pt}
		\includegraphics[width=\textwidth]{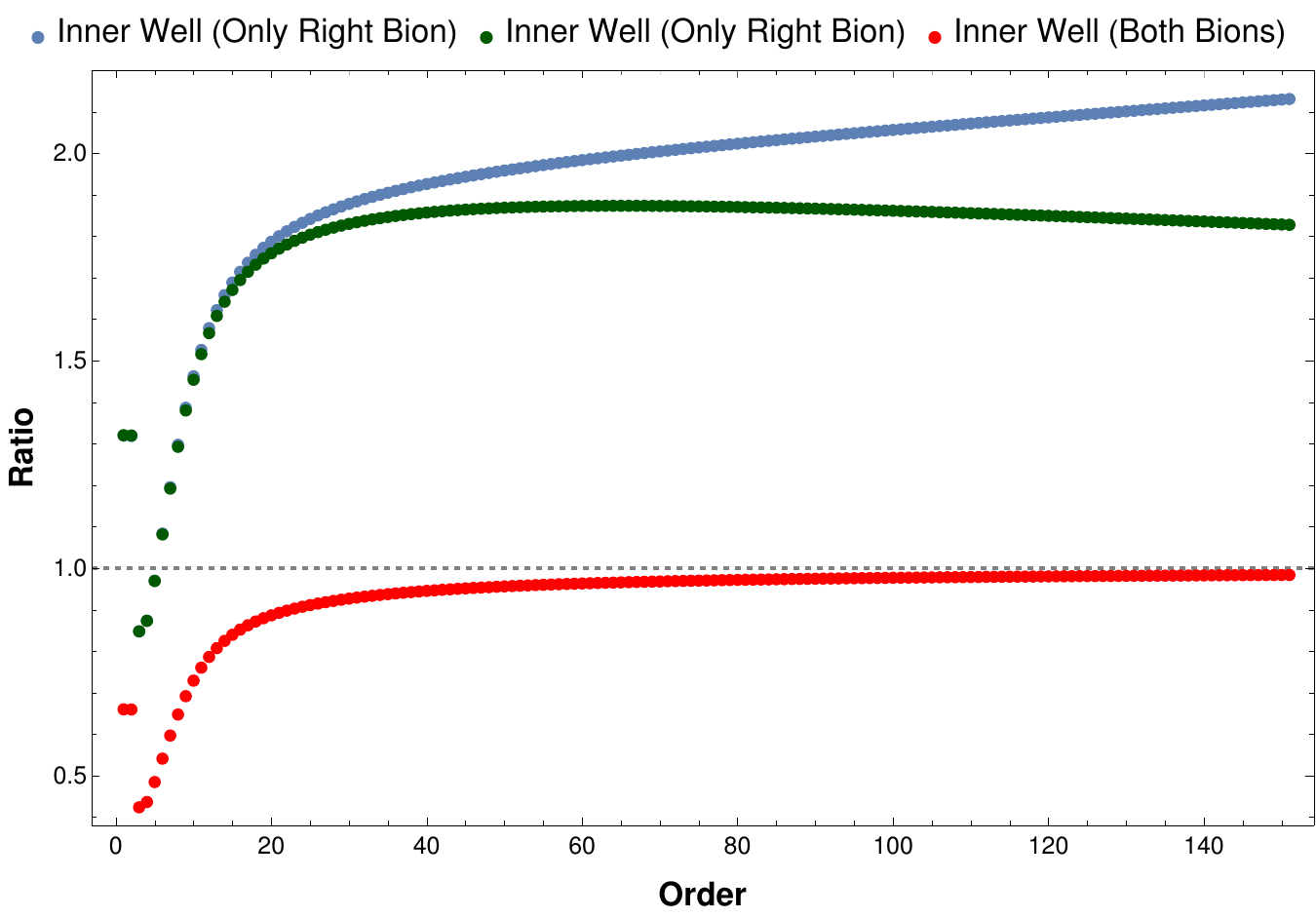}
	\end{subfigure}
	\vspace{6pt}
	\caption{Numerical comparisons between exact perturbative series and divergent series from imaginary bion contributions for $V_\mrmATW^{(1)}$. {\bf (a)} Both ratios for outer wells approach unity in similar ways. These behaviours don't differ from the case of $V_\mrmATW^{(2)}$ plotted in Fig.~\ref{Figure: ATW_Ratio_LeftRight2}. \\
	{\bf (b)} For the inner well, a single contribution from any bion does not match with the exact series. Instead, the ratio between the overall contribution arising from both bions and the exact series approaches unity quickly. This numerical behaviour ceases to be apparent once the asymmetric gets larger as plotted in Fig.~\ref{Figure: ATW_Ratio_Inner2}.}
	\vspace{30pt} \label{Figure: ATW_Ratio1}
\end{figure}
\begin{figure}[h]
	\centering
	\begin{subfigure}[h]{0.48\textwidth}
		\caption{\underline{Outer wells vs. neighbouring bions}} \label{Figure: ATW_Ratio_LeftRight2}
		\vspace{10pt}
		\includegraphics[width=\textwidth]{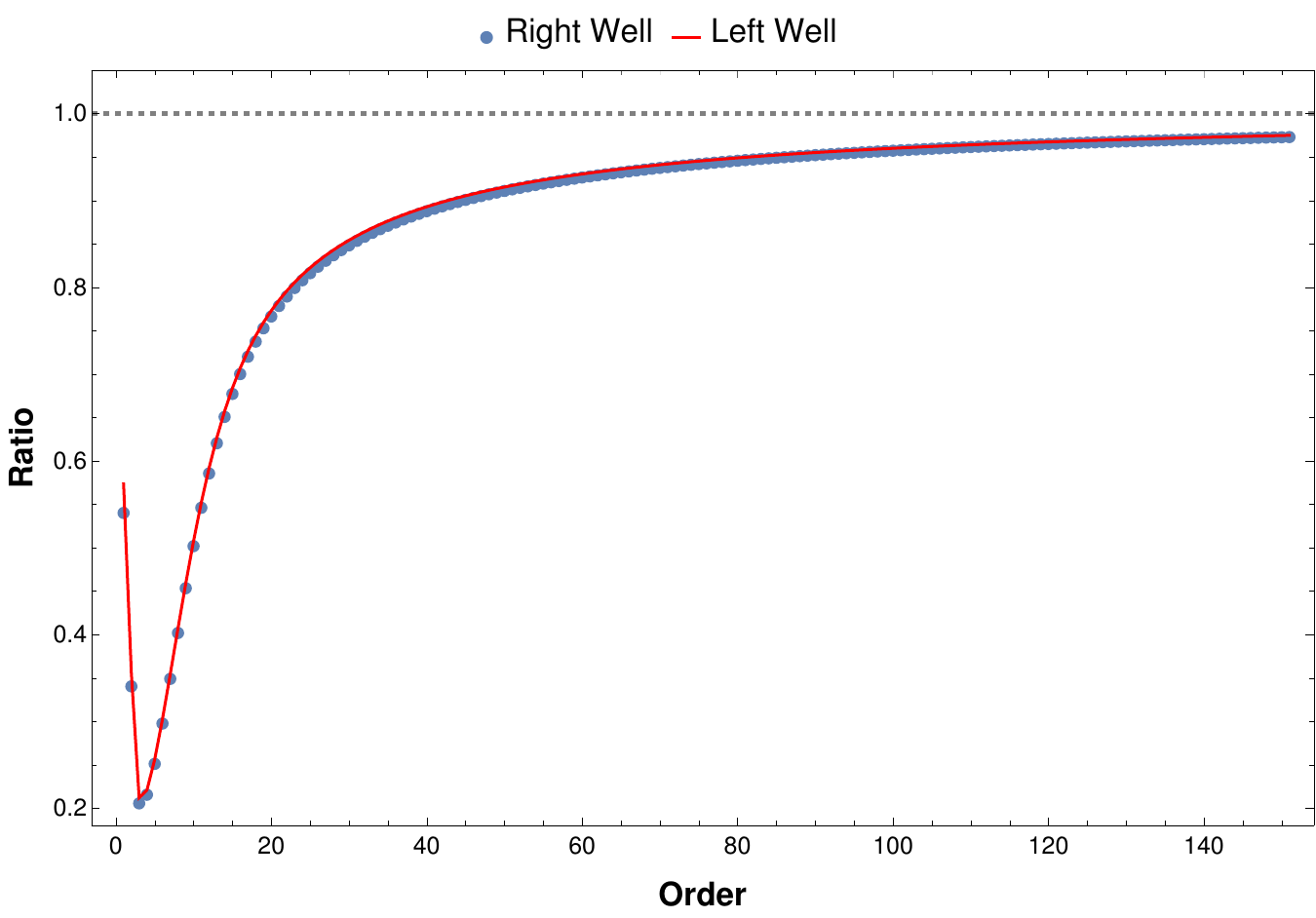}
	\end{subfigure}
	~\hfill 
	\begin{subfigure}[h]{0.48\textwidth}
		\caption{\underline{Inner well vs bions}}	\label{Figure: ATW_Ratio_Inner2}
		\vspace{10pt}
		\includegraphics[width=\textwidth]{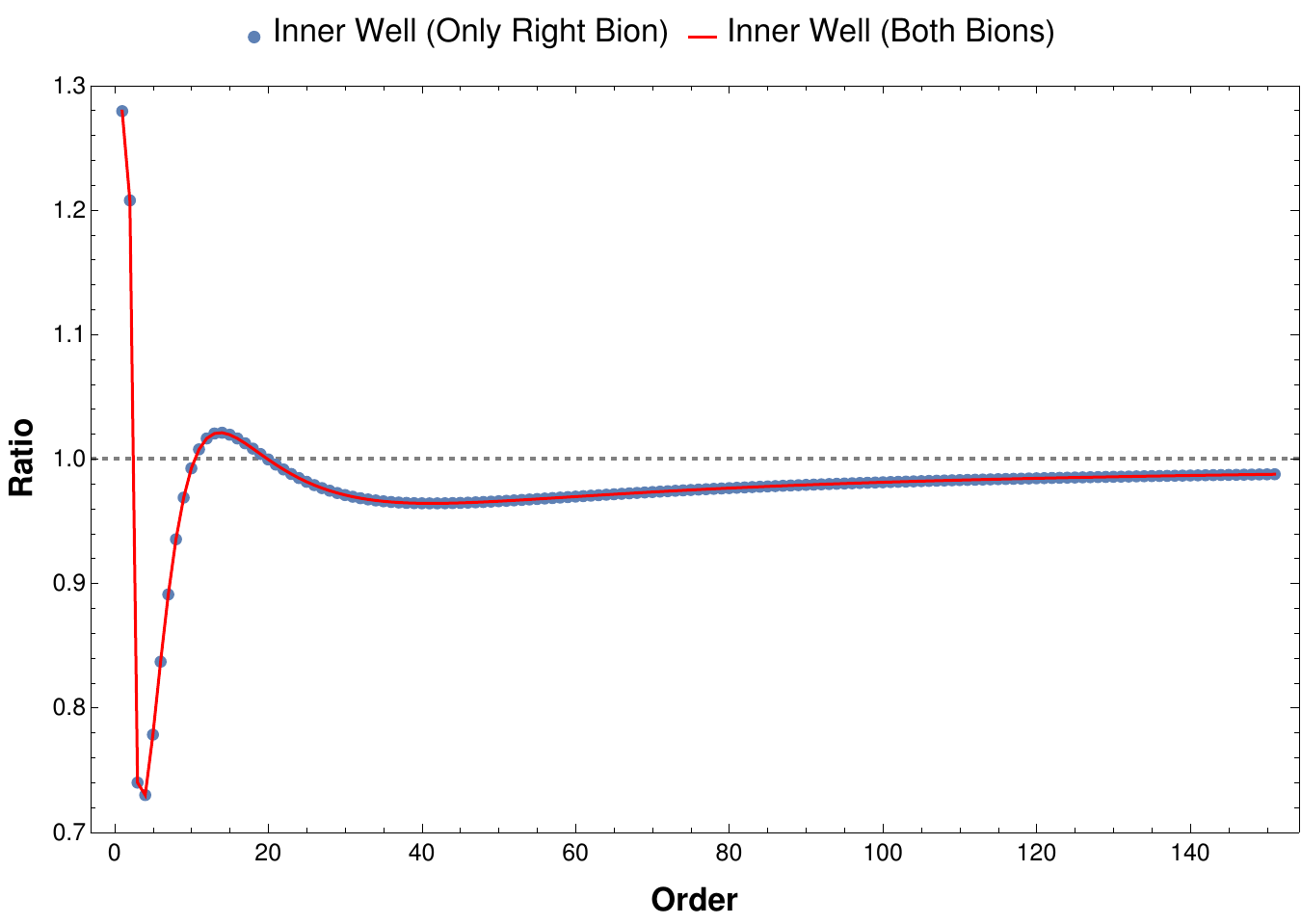}
	\end{subfigure}
	\vspace{6pt}
	\caption{Numerical comparisons between exact perturbative series and divergent series from imaginary bion contributions for $V_\mrmATW^{(2)}$. {\bf (a)} For outer wells, both ratios approach unity as for $V^{(1)}_\mrmATW$ in Fig~\ref{Figure: ATW_Ratio_LeftRight1}. {\bf (b)} The series for inner well matches with the imaginary contribution of the right bion. But, this is misleading since the trans-series structure is not different from the case $V^{(1)}_\mrmATW$ plotted in Fig~\ref{Figure: ATW_Ratio_Inner1}. The ratio associated with the overall imaginary contribution of both bions shows left bion's contribution is negligible numerically. } \label{Figure: ATW_Ratio2}
\end{figure}

To understand the roles of both non-perturbative saddles better, we pick the following forms for ATW:
\begin{equation}
	V^{(1)}_\mrmATW = \frac{1}{2} x^2 (x-1)^2 \left(x-\frac{5001}{10000}\right)^2 \, , \qquad V^{(2)}_\mrmATW = \frac{1}{2} x^2 (x-1)^2 \left(x-\frac{51}{100}\right)^2\, . \label{ATW_PotentialsSpecific}
\end{equation}
Note that for both potentials, left barrier is higher than the right one and all the qualitative properties of these two systems are the same. As we observe below, however, straightforward computations without considering the correct trans-series structure around $x=z_2$ in \eqref{NP_correction_z2} seem to be misleading for $V_\mrmATW^{(2)}$. 

For both $V^{(1)}_\mrmATW$ and $V^{(2)}_\mrmATW$, by computing the constants in the bion fugacities using the formulas provided in the Weber-type EWKB in Section~\ref{Section: EWKB_review}, we plot the ratios between the exact series, which we obtain using the Bender-Wu package \cite{Sulejmanpasic:2016fwr}, and the estimations via \eqref{genericDivergentSeries} and \eqref{imaginaryBorel} in Fig.~\ref{Figure: ATW_Ratio1} and Fig.~\ref{Figure: ATW_Ratio2}, respectively. 

They are aligned with the solutions of the exact quantization conditions in \eqref{NP_correction_z1} and \eqref{NP_correction_z2}: For outer-wells, the perturbation series of right and left wells are controlled by a single bion associated with right and left barriers respectively. For the inner-well, on the other hand, the overall contribution of both bions control the divergent series. This is particularly apparent in Fig.~\ref{Figure: ATW_Ratio_Inner1}, where we show the individual bion contributions do not match the behaviour of the perturbations series, while their combination via \eqref{NP_correction_z2} generates the correct divergent expansion which matches the exact perturbation series at high orders.

This last point is not clear for $V^{(2)}_\mrmATW$ potential. In Fig.~\ref{Figure: ATW_Ratio_Inner2}, we show that the contribution of the left bion (i.e. $B_2$-cycle) is numerically negligible, compared with the case of $V^{(1)}_\mrmATW$. However, as there is no qualitative difference between $V^{(1)}_\mrmATW$ and $V^{(2)}_\mrmATW$, this naive numerical observation is misleading. Note that the associated resurgence structures of them, which yield the real spectrum, are the same.

\subsection{Symmetric Limit, P-NP Relations and PT-symmetric case}\label{Section: ATW_SymmetricLimit}

\begin{figure}
	\centering
	\includegraphics[width=\textwidth]{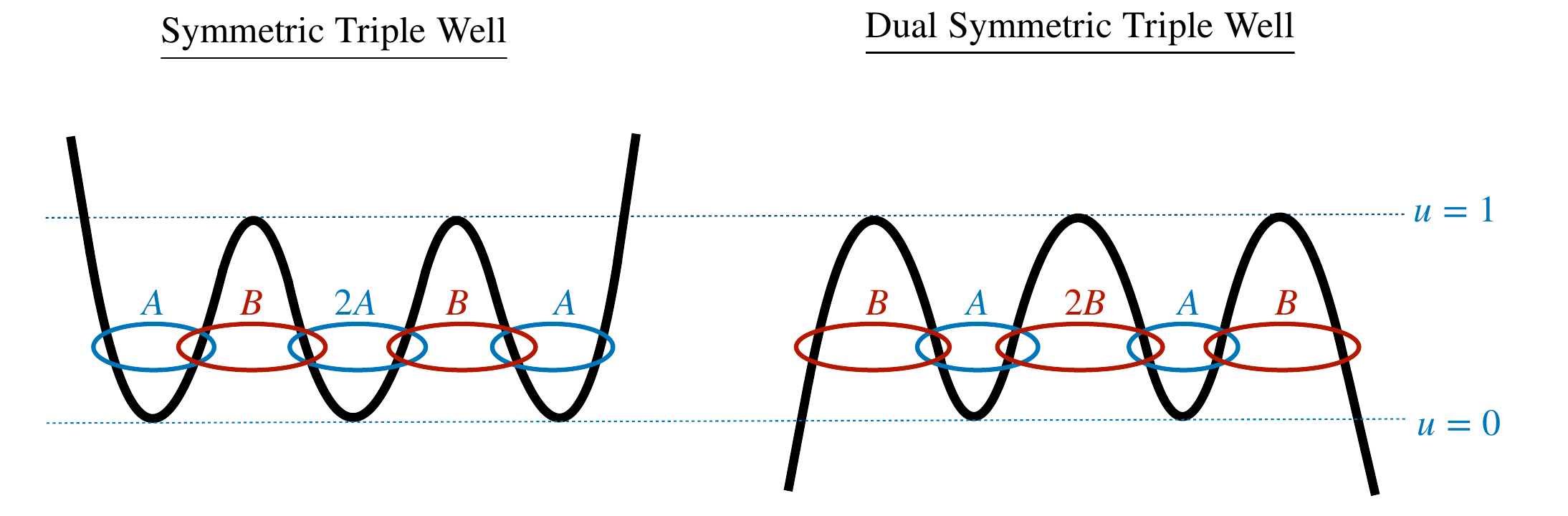}
	\caption{Chebyshev type STW and DSTW potentials with corresponding $A$ and $B$ cycles, which come with only one independent pair.}	\label{Figure: STW_PT}
\end{figure}

Before finishing this section, we discuss the symmetric limit of ATW, i.e. $u_1=u_2$, and focus on the P-NP relations of the symmetric triple well potential (STW) and its dual potential (DSTW). For clarity in the discussions, we focus on the corresponding Chebyshev potential:
\begin{equation}
	V_\mathrm{STW} = T_3^2(x) = x^2 (3- 4 x^2)^2 \, , \label{Chebyshev_TW}
\end{equation}
for which we have $u_1= u_2=1$ in Fig~\ref{Figure: ATW_Potential}. Then, the dual theory (DSTW) is given by the potential
\begin{equation}
	V_{\mrmDSTW} = -\left(1-4 x^2\right)^2 \left(x^2-1\right)\, .
\end{equation}
We illustrate both potentials with the corresponding WKB cycles, which we use below as shown in Fig.\ref{Figure: STW_PT}.

In the STW case, the frequencies of the inner and outer wells are $\o_{\iin} = 3\sqrt{2}$ and $\o_\out = 6\sqrt{2}$. The perturbative sector around the inner-well is linked to the outer ones via the proportionality of their frequencies. In light of Section~\ref{Section: P_NP_revisit}, this relation also extends to the P-NP relations associated with these wells. For DSTW case, both wells have the same frequency, $\o_\dual=2\sqrt{6}$. Its perturbative sector, however, is not related to the STW via the ratio of the frequencies. Instead, the resurgence structures or more specifically associated P-NP relations of these two dual systems are related by the duality transformations.

In the following, we discuss the P-NP relations of each neighbouring well-barrier pairs for STW and DSTW potentials\footnote{The properties of the barrier top region of STW can be obtained by inserting imaginary factors as discussed in \cite{Misumi:2024gtf}.} and illuminate their dependence of the frequencies. Note that with appropriate boundary conditions, the dual potential can also be seen as a PT-symmetric theory, which has a real spectrum. We end the section with a brief discussion on its resurgence structure and the reality of the associated spectrum.


\paragraph{\underline{P-NP Relations}:} In the symmetric limit, the perturbative sectors of STW are proportional to each other. Around the inner-well, the perturbative action is 
\begin{align}
	\mcalF^{u\sim 0}_\iin &= \frac{\tu}{3 \sqrt{2}}+\left(\frac{1}{6 \sqrt{2}}+\frac{\tu^2}{27 \sqrt{2}}\right)
	g +\left(\frac{5 \sqrt{2} \tu}{81}+\frac{5 \sqrt{2} \tu^3}{729}\right) g^2 \nonumber \\	
	& \quad +\left(\frac{25}{162 \sqrt{2}}+\frac{245 \tu^2}{2187 \sqrt{2}}+\frac{70
		\sqrt{2} \tu^4}{19683}\right) g^3 +\dots  \, .\label{STW_ActionPert_Inner}
\end{align}
Then, inverting the series yields
\begin{align}
	\tu_\iin &= 3 \sqrt{2} \mcalF+\left(-\frac{1}{2}-2 \mcalF^2\right) g +\left(-\frac{1}{9} \left(7
	\sqrt{2} \mcalF\right)-\frac{8 \sqrt{2} \mcalF^3}{9}\right) g^2 \nonumber \\ 
	& \quad +\left(-\frac{11}{36}-\frac{274 \mcalF^2}{81}-\frac{140 \mcalF^4}{81}\right) g^3 +\dots \, , \label{STW_EnergyPert_Inner}
\end{align}
which matches an independent check with the Bender-Wu package \cite{Sulejmanpasic:2016fwr}. Around outer-wells, the perturbative action and associated energy series become
\begin{align}
	\mcalF_\out^{u\sim 0} &= \frac{\tu}{6 \sqrt{2}}+\left(\frac{1}{12 \sqrt{2}}+\frac{\tu^2}{54 \sqrt{2}}\right) g +\left(\frac{5 \tu}{81 \sqrt{2}}+\frac{5 \tu^3}{729 \sqrt{2}}\right) g^2 \nonumber \\
	& \quad +\left(\frac{25}{324 \sqrt{2}}+\frac{245 \tu^2}{4374 \sqrt{2}}+\frac{35 \sqrt{2} \tu^4}{19683}\right) g^3 +\dots \, , \label{STW_ActionPert_Outer} 
\end{align} 
and 
\begin{align}
	\tu_\out &= 6 \sqrt{2} \mcalF+\left(-\frac{1}{2}-8 \mcalF^2\right) g +\left(-\frac{1}{9} \left(14 \sqrt{2} \mcalF\right)-\frac{64 \sqrt{2} \mcalF^3}{9}\right) g^2 \nonumber \\ 
	& \quad +\left(-\frac{11}{36}-\frac{1096 \mcalF^2}{81}-\frac{2240 \mcalF^4}{81}\right) g^3 \dots \, . \label{STW_EnergyPert_Outer}
\end{align}
Note that as it is well-known \cite{Basar:2017hpr,Dunne:2020gtk, Kamata:2021jrs}, the actions are simply proportional to each other by $\mcalF_\out^{u\sim0} = \frac{\o_\iin}{\o_\out} \mcalF_\iin^{u\sim 0}$, and the energy series around inner and outer wells are linked via the same proportionality of the actions: $u_\out(\mcalF,g) = u_\iin\left(2\mcalF,g\right)$.

In the dual theory, on the other hand, the perturbative sectors are exactly the same. The corresponding action and the energy series are written as
\begin{align}
	\mcalF^{\dual} &= \frac{\tu}{2 \sqrt{6}}+\left(\frac{11}{36 \sqrt{6}}+\frac{\tu^2}{18 \sqrt{6}}\right) g +\left(\frac{10}{81} \sqrt{\frac{2}{3}} \tu+\frac{5 \tu^3}{243 \sqrt{6}}\right) g^2 \nonumber \\
	& \quad +\left(\frac{4835}{8748 \sqrt{6}}+\frac{1015 \tu^2}{4374 \sqrt{6}}+\frac{35 \sqrt{\frac{2}{3}} \tu^4}{6561}\right) g^3 \dots \, , \label{STW_ActionPert_Dual}
\end{align}
and 
\begin{align}
	\tu^\dual &= 2 \sqrt{6} \mcalF+\left(-\frac{11}{18}-\frac{8 \mcalF^2}{3}\right) g +\left(-\frac{58}{27} \sqrt{\frac{2}{3}} \mcalF-\frac{64}{27} \sqrt{\frac{2}{3}} \mcalF^3\right) g^2 \nonumber \\
	& \quad +\left(-\frac{7393}{8748}-\frac{4712 \mcalF^2}{729}-\frac{2240 \mcalF^4}{729}\right) g^3 \dots \, . \label{STW_EnergyPert_Dual}
\end{align}

The expansion in \eqref{STW_ActionPert_Dual} is not related to \eqref{STW_ActionPert_Inner} or \eqref{STW_ActionPert_Outer} by a simple multiplication. However, it does not prevent the P-NP relations being related by simple transformations. This is because of the fact that the $f_1(0)$ and $f_2(0)$ terms in P-NP relations depends solely on a few parameters, which govern the relationship between perturbative and non-perturbative sectors entirely. Indeed, following guidelines of Section~\ref{Section: P_NP_revisit}, we obtain the P-NP relations for STW and its dual as 
\begin{equation}
	-\frac{\dee \tu_\iin}{\dee \mcalF} = g \left(4\mcalF_\iin^{u\sim0} + \frac{8}{3} g\frac{\dee \mcalG^{u\sim0}}{\dee g} \right) \; , \quad 	-\frac{\dee \tu_\out}{\dee \mcalF} = g \left(16\mcalF_\out^{u\sim0} + \frac{16}{3}g\frac{\dee \mcalG^{u\sim 0}}{\dee g} \right) \; , \label{PNP_STW}
\end{equation}
and 
\begin{equation}
	-\frac{\dee \tu^\dual}{\dee \mcalF} = g \left(\frac{16}{3}\mcalF^{\dual} + \frac{8}{3} g\frac{\dee \mcalG^\dual_\iin}{\dee g} \right) \; , \quad -\frac{\dee \tu^\dual}{\dee \mcalF} = g \left(\frac{16}{3}\mcalF^{\dual} + \frac{16}{3} g\frac{\dee \mcalG^\dual_\out}{\dee g} \right)\, , \label{PNP_DSTW}
\end{equation}
where we use \eqref{PNP_VanishingOrder} and \eqref{f2_NonDeformed} to fix the constants. 

Comparing these equations, we can now easily verify that they are related to each other by
\begin{equation}
	f_1^{\out} = \frac{\o^2_{\out}}{\o^2_{\iin}}f_1^{\iin}\; , \qquad f_1^{\dual} = \frac{\o^2_\dual}{\o^2_\iin}f_1^\iin \, , \label{transformationSTW_f1}
\end{equation} 
as predicted in \eqref{f1_FreqDependence}. Similarly, the $f_2$ terms in \eqref{PNP_STW} and \eqref{PNP_DSTW} are related via \eqref{f2_NonDeformed_DifferentPerturbative} and \eqref{f2_NonDeformed_DifferentNP_1}, respectively:
\begin{equation}
	f_2^{\out} = \frac{\o_{\out}}{\o_{\iin}}f_2^{\iin}\; , \qquad f_{2,\out}^{\dual} = \frac{S_\iin^{\dual}}{S_\out^\dual}f_{2,\dual}^\iin \, , \label{transformationSTW_f2}
\end{equation}
Finally, the direct duality transformation can be used to map the inner(outer) perturbative cycle(s) to the inner(outer) non-perturbative one(s). This maps the first(second) equation in \eqref{PNP_STW} to the first(second) one in \eqref{PNP_DSTW}. In this way, we verify that it leaves the $f_2(0)$ term invariant, as predicted in \eqref{f2_NonDeformed_Dual}.

Finally, using the P-NP equations \eqref{PNP_STW} and \eqref{PNP_DSTW}, it is straightforward to compute the instanton functions $\mcalG$ for each barrier for both STW and DSTW. Due to the parity symmetry, there is a single instanton function for STW: 
\begin{align}
	\mcalG^{u\sim 0} &=  \frac{9}{4 \sqrt{2} g }+\left(\frac{7}{12 \sqrt{2}}+\frac{\tu^2}{9 \sqrt{2}}\right) g +\frac{\left(1557 \sqrt{2} \tu+142 \sqrt{2} \tu^3\right) g ^2}{5832} \nonumber \\
	& \quad +\frac{\left(138753 \sqrt{2}+79272 \sqrt{2} \tu^2+4256 \sqrt{2} \tu^4\right) g ^3}{314928} + \dots\, .
\end{align}
For the dual case, we have two instanton functions which are proportional to each other by a factor of $2$:
\begin{align}
	\mcalG_\iin^\dual & = \frac{3 \sqrt{\frac{3}{2}}}{2 g }+\left(\frac{29}{18 \sqrt{6}}+\frac{1}{9} \sqrt{\frac{2}{3}} \tu^2\right) g +\left(\frac{721 \tu}{486\sqrt{6}}+\frac{71 \tu^3}{729 \sqrt{6}}\right) g ^2 \nonumber \\
	& \quad +\left(\frac{126281}{26244 \sqrt{6}}+\frac{518}{729} \sqrt{\frac{2}{3}} \tu^2+\frac{532 \sqrt{\frac{2}{3}} \tu^4}{19683}\right) g ^3 + \dots\, ,  \label{NPfunction_STWdual_inner}\\ \nonumber \\ 
	\mcalG_\out^\dual & = \frac{3 \sqrt{\frac{3}{2}}}{4 g }+\left(\frac{29}{36 \sqrt{6}}+\frac{\tu^2}{9 \sqrt{6}}\right) g +\left(\frac{721 \tu}{972 \sqrt{6}}+\frac{71 \tu^3}{1458 \sqrt{6}}\right) g ^2 \nonumber \\ 
	& \quad + \left(\frac{126281}{52488 \sqrt{6}}+\frac{259}{729} \sqrt{\frac{2}{3}}	\tu^2+\frac{266 \sqrt{\frac{2}{3}} \tu^4}{19683}\right) g ^3 + \dots\,.\label{NPfunction_STWdual_outer}
\end{align}
Note that this proportionality is reminiscent of the one for the perturbative cycles of STW, and the factor 2 arises simply from the ratio of the instanton actions in \eqref{transformationSTW_f2} and its effect on \eqref{PNP_DSTW}.

\noindent {\bf \underline{Spectrum for PT-symmetric case}:} The duality transformations between STW and DSTW only relate the quantum actions; but they also help to understand the resurgence structure of the quantized spectrum. While the P-NP relations are linked to the underlying quantum geometry, the spectrum depends on the boundary conditions, which determine the quantization conditions for a specific problem at hand. Therefore, although they share the same P-NP relations, the resurgence structure of the spectra of STW and DSTW should be investigated individually.

In Section~\ref{Section: ATW_trans-series}, we have already discussed the resurgence structure of the energy trans-series for ATW case, which assumes vanishing solutions at real infinities. In the symmetric limit, the Stokes diagram in Figs.~\ref{Figure: ATW_StokesDiagram1_Minus} and \ref{Figure: ATW_StokesDiagram1_Plus}, so that the quantization conditions of them, are the same for the $u<1$ region, with the additional property $\Pi_{A}^{(1)} = \Pi_A^{(3)}$. This equality indicates the perturbative degeneracy of the outer wells which is broken by the non-perturbative corrections, i.e. \textit{topological bions}. Here, we don't discuss the solution to this case and refer to the related discussions in the literature \cite{Dunne:2020gtk}.

In the dual case, on the other hand, the effect of boundary conditions on the spectrum is more apparent as they decide if the spectrum shows the resonance or PT-symmetric behaviors. For the latter case, although the potential is unbounded, the quantum system is stable\footnote{More specifically, it is stable unless the system is in broken PT-symmetric phase. We do not discuss such cases here and refer \cite{Bender:2023cem} and references in that a general discussion.} with a real spectrum \cite{Bender:2023cem}. In the following, we analyze the reality of the spectrum of PT-symmetric case of DTSW potential in EWKB formalism and show the leading order ambiguity cancellations, yielding to a real spectrum.\\

\begin{figure}
	\centering
	\begin{subfigure}[h]{0.48\textwidth}
		\caption{\underline{$\t_{\l} = 0^-$}}	\label{Figure: STW_PT_diagram1}		\includegraphics[width=\textwidth]{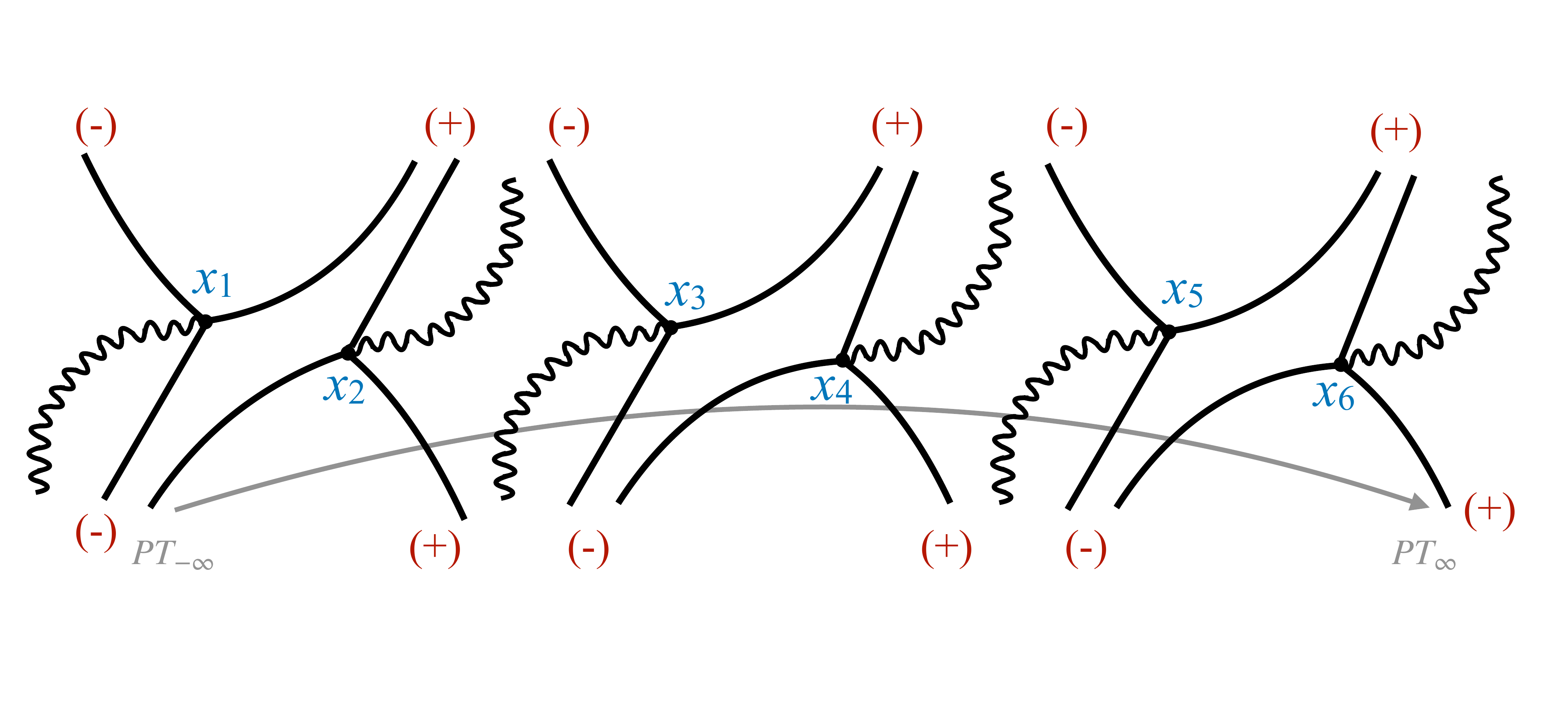}
	\end{subfigure}
	~\hfill 
	\begin{subfigure}[h]{0.48\textwidth}
		\caption{\underline{$\t_{\l} = 0^+$}}	\label{Figure: STW_PT_diagram2}
		\includegraphics[width=\textwidth]{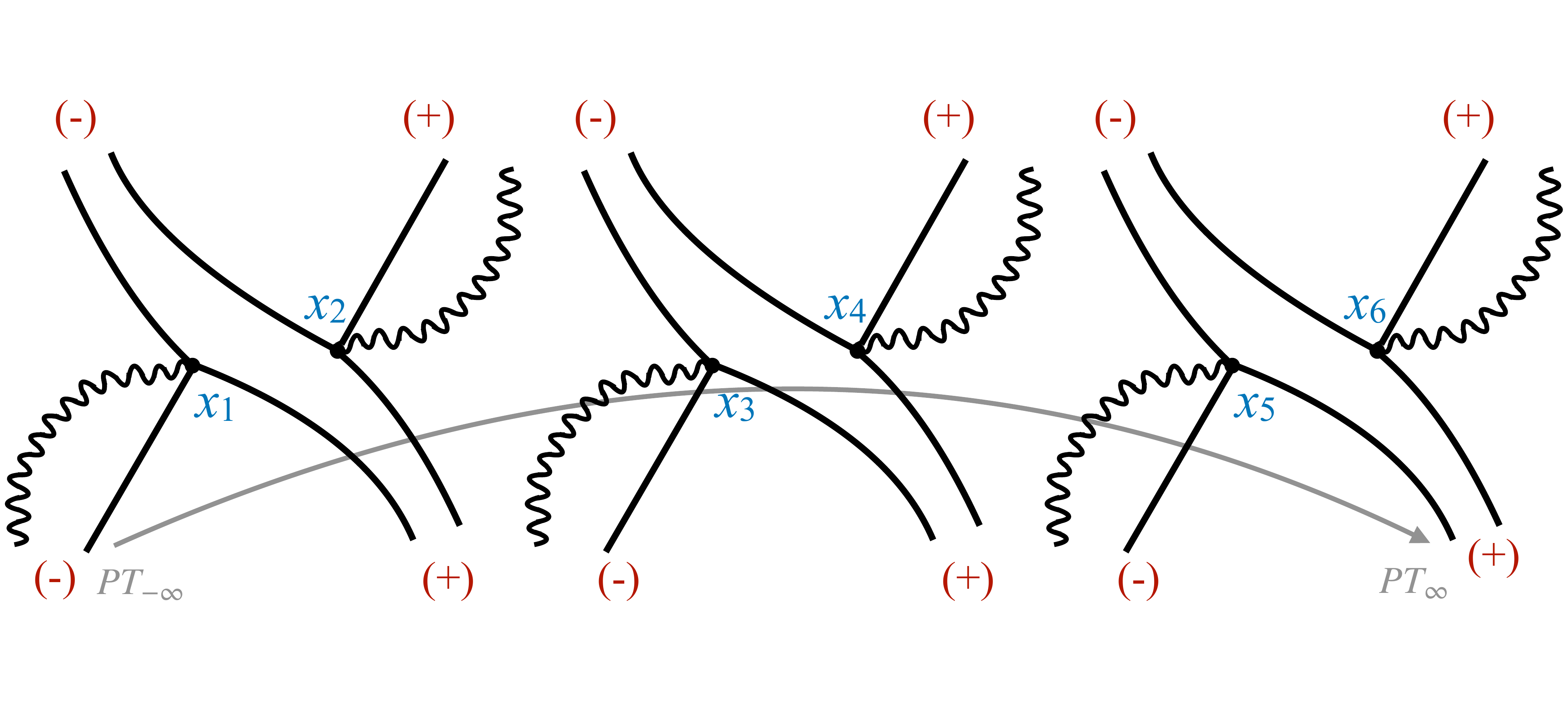}
	\end{subfigure}
	\caption{Stokes geometry of $V_{\mrmDSTW}$ around $u\sim 0$ for $\t_\l = 0^\mp$. $\mcalPT_{\pm\infty}$ indicates the directions of complex infinities where the exponential dumping requirement of the solutions imposed as boundary conditions.} \label{Figure: STW_PT_StokesDiagrams}
\end{figure}

Let us first start with the Stokes geometry of $V_{\mrmDSTW}$.~Around $u\sim 0$ of $V_{\mrmDSTW}$, the diagrams for $\t_\l = 0^\mp$ are given in Fig~\ref{Figure: STW_PT_StokesDiagrams}. To quantize the PT-symmetric system, we connect the infinities labeled as $\mcalPT_{\pm\infty}$, leading to the following quantization conditions:
\begin{align}
	D_{\t_\l = 0^-}^\mcalPT &= \sqrt{\Pi_A^{-2}\Pi_B^{-3}}\left[\left(1+\Pi_A\right)^2 + \Pi_B \left(1+\Pi_A\right) + \Pi_B^2 \left(1+\Pi_B\right) \right] = 0 \, , \label{STW_PT_QC1}\\
	D_{\t_\l = 0^+}^\mcalPT &= \sqrt{\Pi_A^{-2}\Pi_B^{-3}}\left[\left(1+\Pi_A\right)^2 + \Pi_A\Pi_B \left(1+\Pi_A\right) + \Pi_A^2 \Pi_B^2 \left(1+\Pi_B\right) \right] = 0 \, , \label{STW_PT_QC2}
\end{align}
where 
\begin{equation}
	\Pi_B = \frac{\sqrt{2 \pi} e^{-G_{\out}^\dual(\tu,g)}}{\Gamma\left[\frac{1}{2}+\mcalF(\tu,g)\right]} \left(\frac{g}{c_\dual}\right)^{-\mcalF(\tu,g)}\, , \quad c_\dual = \frac{9}{2}\sqrt{\frac{3}{2}} \, .  \label{STW_PT_bounce}
\end{equation}
The quantization conditions \eqref{STW_PT_QC1} and \eqref{STW_PT_QC2} are related by  a Stokes automorphism given as \[\mfrS_0: \Pi_A\mapsto \Pi_A \left(1+\Pi_B\right)\left(1+\Pi_B^2\right)\, , \] and we find the associated median quantization $D^\mcalPT_\med = \mfrS_0^{\pm 1/2} D_{\t_\l = 0^\mp}^\mcalPT $ as 
\begin{equation}
	D^\mcalPT_\med  = \sqrt{\Pi_A^{-2} \Pi_B^{-3}}\left[\sqrt{1 + \Pi_B}\sqrt{1+\Pi^2_B} \left(1+ \Pi_A^2\right) + 2\Pi_A + \Pi_A \Pi_B \right]\, .
\end{equation}

To understand the quantitative aspect of the real spectrum, we solve \eqref{STW_PT_QC1} and \eqref{STW_PT_QC2} and obtain the non-perturbative corrections for the energy trans-series. Note that both quantization conditions are quadratic equations for $(1 + \Pi_A)$. Then, we express the formal solutions as
\begin{equation}
	2\cos \pi \mcalF = - \frac{\Pi_B\, e^{\mp i \pi \mcalF}}{2} + \ve i\, e^{\mp i \pi \mcalF}\,  \frac{\sqrt{3}\, \Pi_B}{2}\, \sqrt{1+\frac{4}{3}\, \Pi_B} \, , \label{STW_PT_QC}
\end{equation}
where $\ve=\pm 1$ and we used $\Pi_A = e^{- i \pi \mcalF}$. Since $\Pi_B$ is small for $u\sim 0$, we expand the square root terms in $\Pi_B$ up to $O\left(\Pi_B^2\right)$. Then, using $\eqref{STW_PT_bounce}$ and 
\begin{equation}
	e^{\mcalG_\out^\dual} = e^{-\frac{S^\dual_\mcalB}{g}}\sum_{k=0}^\infty H^\dual_k \d^{k} \, , \label{expansion_bouncePT} 
\end{equation}
we rewrite \eqref{STW_PT_QC} up to $e^{-\frac{2S_\mcalB^\dual}{g}}$ as 
\begin{align}
	\sum_{n=1}^\infty h_n \d^n &= \frac{\ve \sqrt{3} \mp i}{2} \mcalK_{\mcalB}^\dual \sum_k \tH_k^{(1)}\, \d^k  + \frac{2\pi \ve\, \left(\mcalK_\mcalB^\dual\right)^2}{\sqrt{3}}  \, \sum_{k=0} \tH_k^{(2)} \d^k \, , \label{STW_PT_arranged_expansion}
\end{align}
where the bounce fugacity is
\begin{equation}
	\mcalK_\mcalB^\dual = (-1)^N \, \frac{e^{-S_\mcalB^{\dual}}}{2\sqrt{2\pi}} \left(\frac{c_\dual}{g}\right)^{\frac{1}{2}+N} \, ,
\end{equation}
and the series expansions are defined as
\begin{align}
	\sum_{k=0}^\infty \tH_k^{(1)} \d^k &:= e^{\d \left[\log\frac{c_\dual}{g} \mp i \pi\right]}\, \sum_{k=0} H_k^\dual \d^k \, ,\\ 
	\sum_{k=0}^\infty \tH_k^{(2)} \d^k &:= (-1)^N\frac{e^{\d \left[2\log\frac{c_\dual}{g} \mp i \pi\right]}}{\Gamma\left(1+ N +\d\right)} \left(\sum_{k=0} H_k^\dual \d^k \right)^2\, , \\ 
	\sum_{n=1}^\infty h_n \d^n &:= \frac{1}{\Gamma\left(-N-\d\right)}   \, ,\qquad h_1 = -1\, .
\end{align}
Finally, using the ansatz
\begin{equation}
	\d^\pm = \sum_n^\infty \x^\pm_n \left(\mcalK_\mcalB^\dual\right)^n\, ,
\end{equation}
we solve \eqref{STW_PT_arranged_expansion} order by order; and obtain the first two non-perturbative corrections as
\begin{align}
	\d^\pm  & \simeq  \mp \frac{i}{2 h_1} \tH_0^{(1)}\, \mcalK_\mcalB^\dual  + \ve \frac{\sqrt{3}}{2 h_1}  \tH_0^{(1)} \mcalK_\mcalB^\dual +  \frac{2\pi \ve}{\sqrt{3}}\tH_0^{(2)} \left(\mcalK_{\mcalB}^\dual\right)^2 \nonumber \\
	& \quad + \left(\frac{\sqrt{3}\ve \mp i }{2 h_1}\right)^2 \left[\tH_0^{(1)}\, \tH_1^{(1)}\, - h_1 \left(\tH_0^{(1)}\right)^2\right]\left(\mcalK_\mcalB^\dual\right)^2   \, .  \label{NP_correction_PT}
\end{align}

As expected, the non-perturbative corrections arise with degeneracy breaking and ambiguous terms. Note that at the leading order in \eqref{NP_correction_PT}, 
we express the non-perturbative corrections in terms of a single fugacity term, i.e. $\mcalK_\mcalB^\dual$. However, the sources of those two terms are not the same: The ambiguous term comes from the bounce configuration associated with $B_3$ cycle, while the splitting term is due to a single instanton contribution governing the tunnelling between perturbative states.

\begin{figure}
	\centering
	\includegraphics[width=0.5\textwidth]{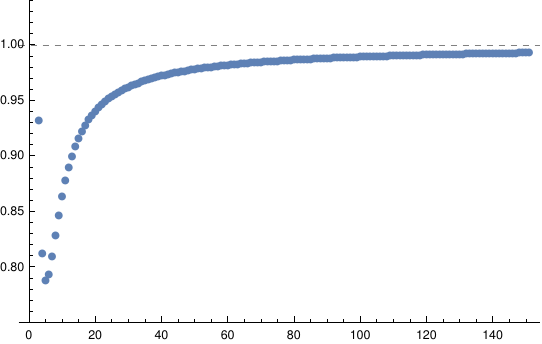}
	\caption{Ratio between the divergent series associated with $K_{B}^{\mathrm{dual}}/2$ and the exact perturbation series for the ground-state energy of the PT-symmetric STW system.}	\label{Figure: STW_PT_Ratio}
\end{figure}

As the PT-symmetric system is stable, it has real eigenvalues which indicate the ambiguous imaginary non-perturbative contributions should be canceled against perturbative Borel ambiguities. For the ground state, we compare the divergent series associated with $\frac{\mcalK_\mcalB^\dual}{2}$ with the exact perturbation series and  plot the corresponding ratio in Fig.~\ref{Figure: STW_PT_Ratio}, which verifies the cancellations at the leading order. 

We finally note that in the PT-symmetric DSTW, the divergence of the perturbative sector seems to be controlled by a single non-perturbative sector, i.e.~first term in \eqref{NP_correction_PT}, arising from the bounce configuration associated with the $B_3$ cycle. The bion associated with $B_2$ cycle, on the other hand, contributes at the next order, i.e.~$\left(\mcalK_\mcalB^\dual\right)^2$. This trans-series structure is in the same form with the other symmetric potentials which have only one independent pair of perturbative and non-perturbative sectors \cite{Zinn-Justin:2004vcw, Dunne:2013ada, Sueishi:2020rug,Sueishi:2021xti} as well as the STW potential \cite{Dunne:2020gtk}. However, once the symmetry is broken, this would not be the case and the trans-series would be governed by multiple independent non-perturbative sectors as in the case of ATW.  It would be interesting to see the change in the roles of the non-perturbative cycles in such cases. We leave this problem for a future work.


\section{Tilted double-well}\label{Section: TDW}

In this section, we discuss tilted double-well (TDW) potential as another example of potential with non-degenerate saddles. For concreteness, we focus on the specific form we introduced in \eqref{TDW_generic}, which we rewrite here again: 
\begin{equation}
	V_\mrmTDW(x,\g) = \frac{1}{2}x^2(1-x)^2 - \g x^3\, , \qquad  \g>0\, . \label{TDW_example}
\end{equation}  
Its saddle points are at 
\begin{align}
	z_\mrmFV = 0\, ,\; z_\mrmBT = \frac{1}{4}\left(3 + 3 \g - \sqrt{1+ 18\g + 9 \g^2}\right)\, , \;  z_\mrmTV = \frac{1}{4}\left(3 + 3 \g + \sqrt{1+ 18\g + 9 \g^2}\right)\, ,
\end{align}
which are at the classical energy levels labeled as $u_\mrmFV$, $u_\mrmBT$ and $u_\mrmTV$ in Fig.~\ref{Figure:TDW_potential}.

In Section~\ref{Section: P_NP_revisit}, we have discussed the deformed P-NP relation of this system and show that except the $f_1(\g)$ term, it has the identical form at all saddle points. Despite this, we show in this section that the spectrum has multiple disconnected sectors due to the Stokes phenomena occurring during the transition across the $u_\mrmFV$ level. This behavior is different from that of ATW in Section~\ref{Section: ATW} and other examples discussed in \cite{Misumi:2024gtf} where all sectors are smoothly connected. Note that the TDW system we will see also demonstrates the difference between the resurgence relation of perturbative and non-perturbative cycles and the resurgence structure of the exact spectrum.

The P-NP relations of TDW potential has recently been discussed in \cite{Cavusoglu:2023bai,Cavusoglu:2024usn}. Their discussion, however, was only focused on $x=z_\mrmFV$ for $\g>0$ and $\g<0$ cases, which correspond to false-vacuum and true vacuum cases, respectively. Instead of focusing on one saddle point, by using EWKB techniques, we verify the P-NP relations around all saddle points, which have different frequencies. This verifies that the connection between the P-NP relations around different levels are given by the prescription shown in Section~\ref{Section: P_NP_revisit}. 

Moreover, the analysis around $x=z_\mrmBT$, which wasn't discussed in \cite{Cavusoglu:2023bai}, provides a neat example of resurgence relations when there is one perturbative saddle and two non-perturbative saddles (bounces) for a deformed genus-1 potential. We show that although there is no parity symmetry in the system, the corresponding P-NP relations are the same for different bounces, which is in sync with the invariance of $f_{2,3}(\g)$ terms that we have discussed in Section~\ref{Section: P_NP_revisit}. 

Finally, we note that the supersymmetric version of the TDW potential was previously discussed in EWKB setting in  \cite{Kamata:2021jrs}, where a linear \textit{quantum} deformation parameter was used. This differs from our cubic deformation in \eqref{TDW_example}, which is a classical one. We compare these two systems after performing exact quantization of \eqref{TDW_example} in all sectors.
\vspace{-1pt}

\subsection{Transition between different sectors}
\begin{figure}
	\centering
	\caption*{$\underline{\bm{u_\mrmBT<u}}$}
	\vspace{6pt}
	\begin{subfigure}[h]{0.48\textwidth}
		\caption{\underline{$\t_\mrmBT = 0^-$ $(\t_\mrmFV = 0^-)$}}	\label{Figure: TDW_StokesDiagram1_Minus}
		\includegraphics[width=\textwidth]{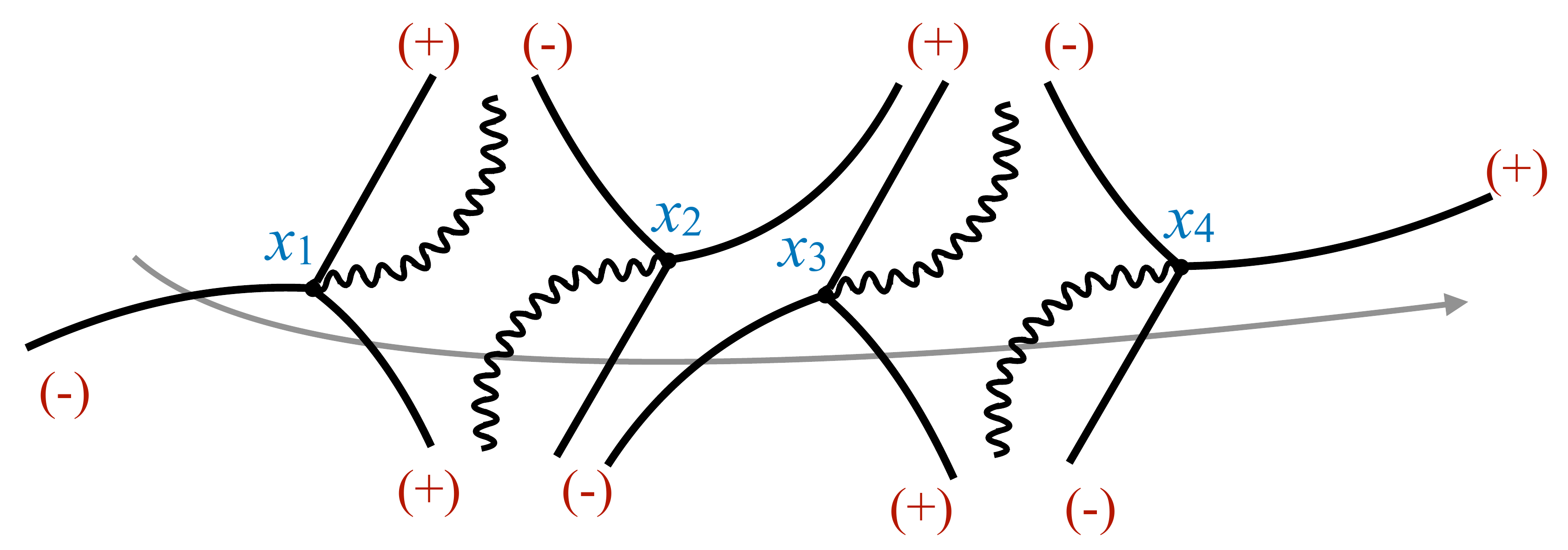}
	\end{subfigure}
	~\hfill 
	\begin{subfigure}[h]{0.48\textwidth}
		\caption{\underline{$\t_\mrmBT = 0^+$ $(\t_\mrmFV = 0^+)$}}	\label{Figure: TDW_StokesDiagram1_Plus}
		\includegraphics[width=\textwidth]{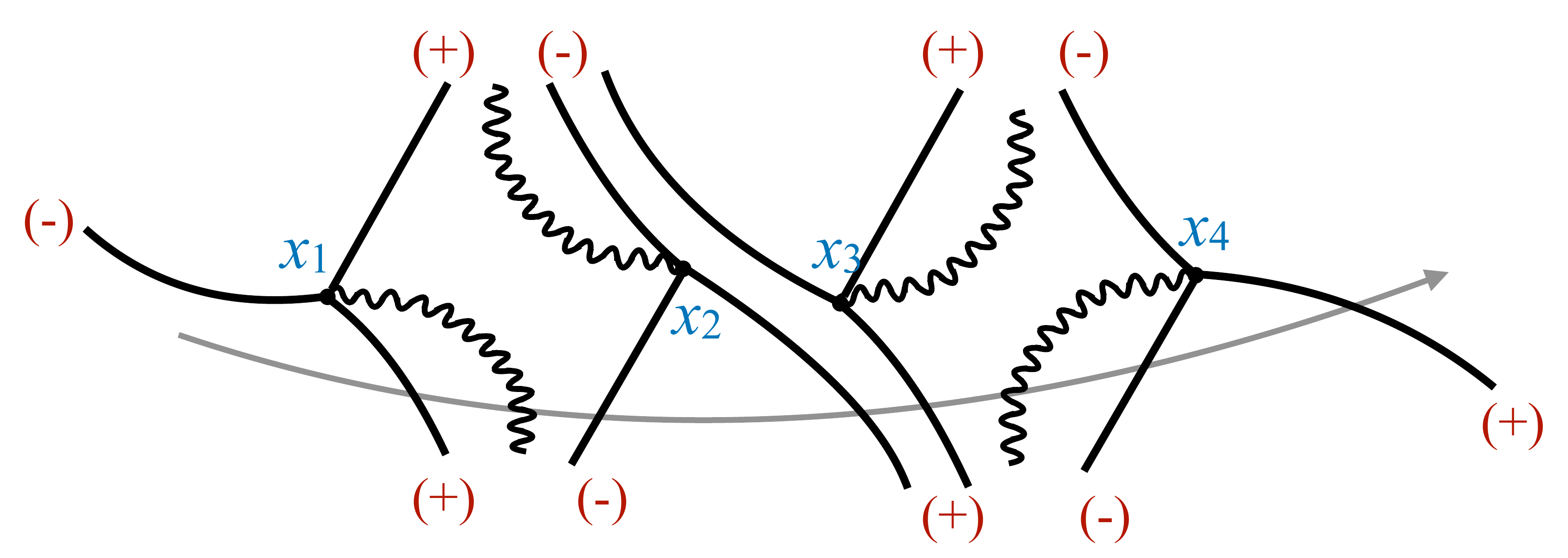}
	\end{subfigure}
	\vspace{10pt}
	\caption*{$\underline{\bm{u_\mrmFV<u < u_\mrmBT}}$}
	\vspace{6pt}
	\begin{subfigure}[h]{0.48\textwidth}
		\caption{\underline{$\t_\mrmBT = -\pi^+$}}	\label{Figure: TDW_StokesDiagram2_Minus}
		\includegraphics[width=\textwidth]{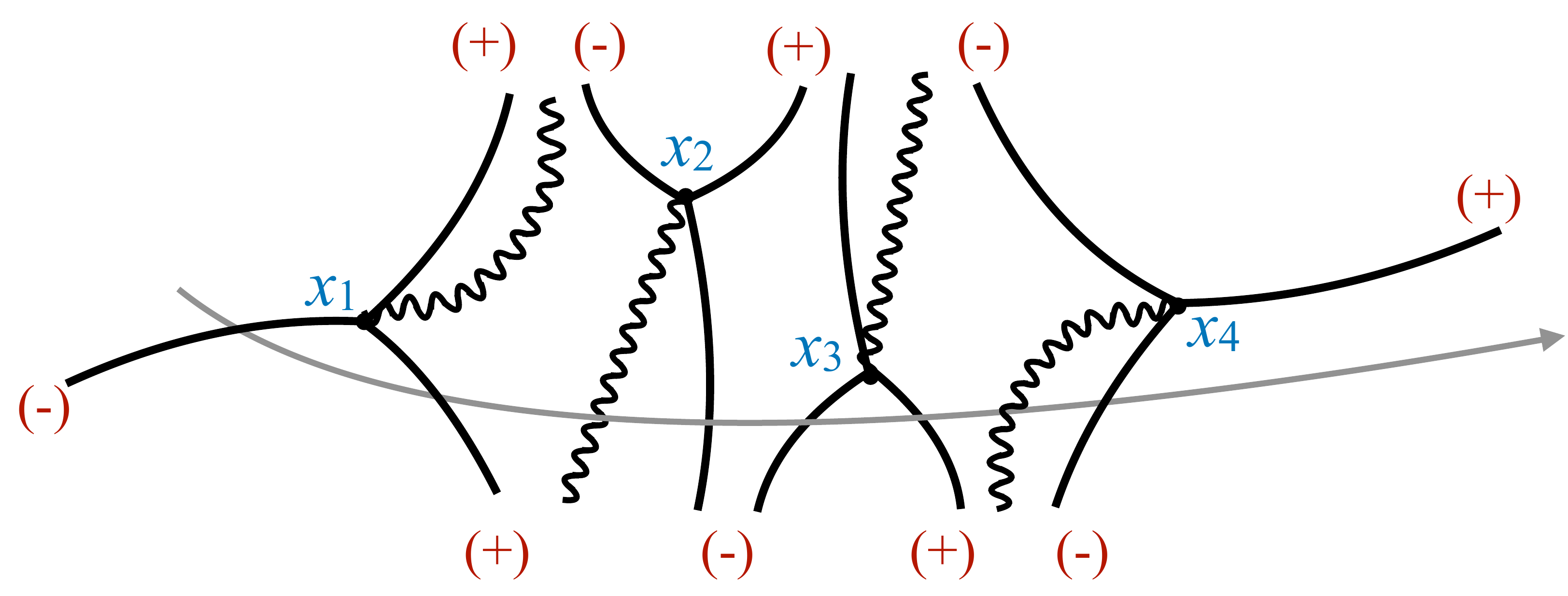}
	\end{subfigure}
	~\hfill 
	\begin{subfigure}[h]{0.48\textwidth}
		\caption{\underline{$\t_\mrmBT = \pi^-$}}	\label{Figure: TDW_StokesDiagram2_Plus}
		\includegraphics[width=\textwidth]{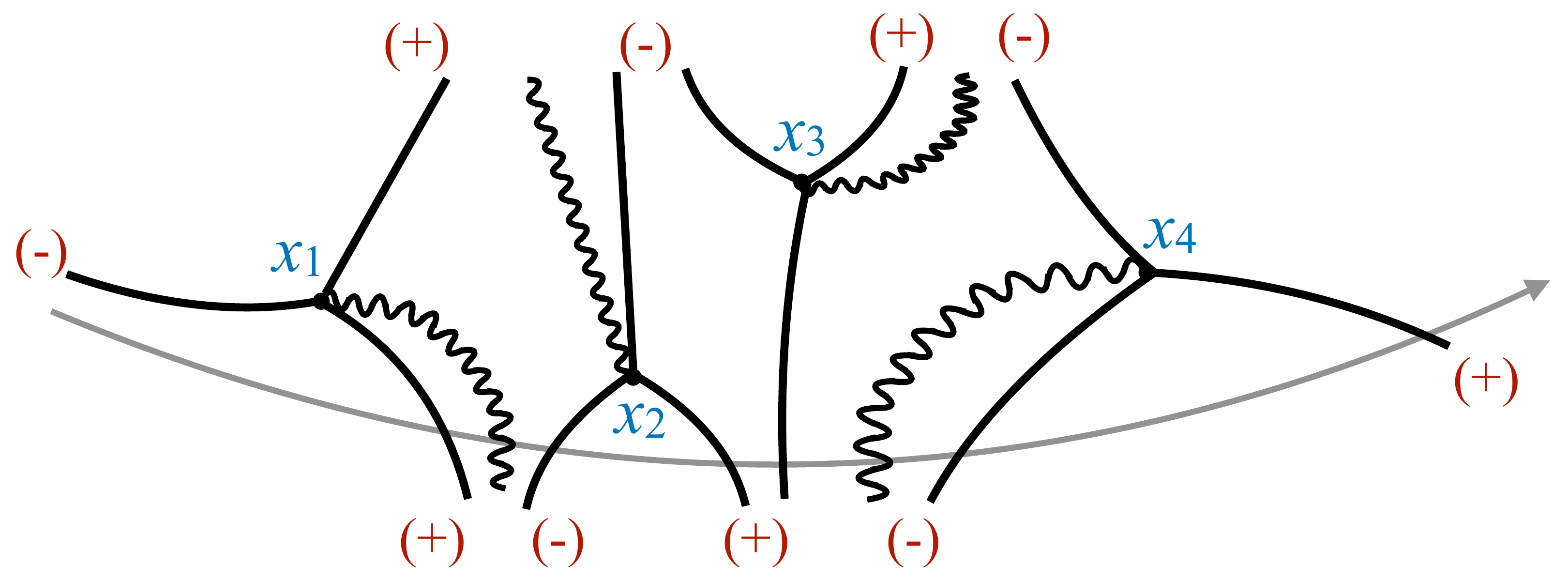}
	\end{subfigure}
	
	\vspace{10pt}
	\caption*{$\underline{\bm{u_\mrmTV<u < u_\mrmFV}}$}
	\vspace{6pt}
	\begin{subfigure}[h]{0.45\textwidth}
		\caption{\underline{$\t_\mrmFV = -\pi^+$}}	\label{Figure: TDW_StokesDiagram3_Minus}
		\includegraphics[width=\textwidth]{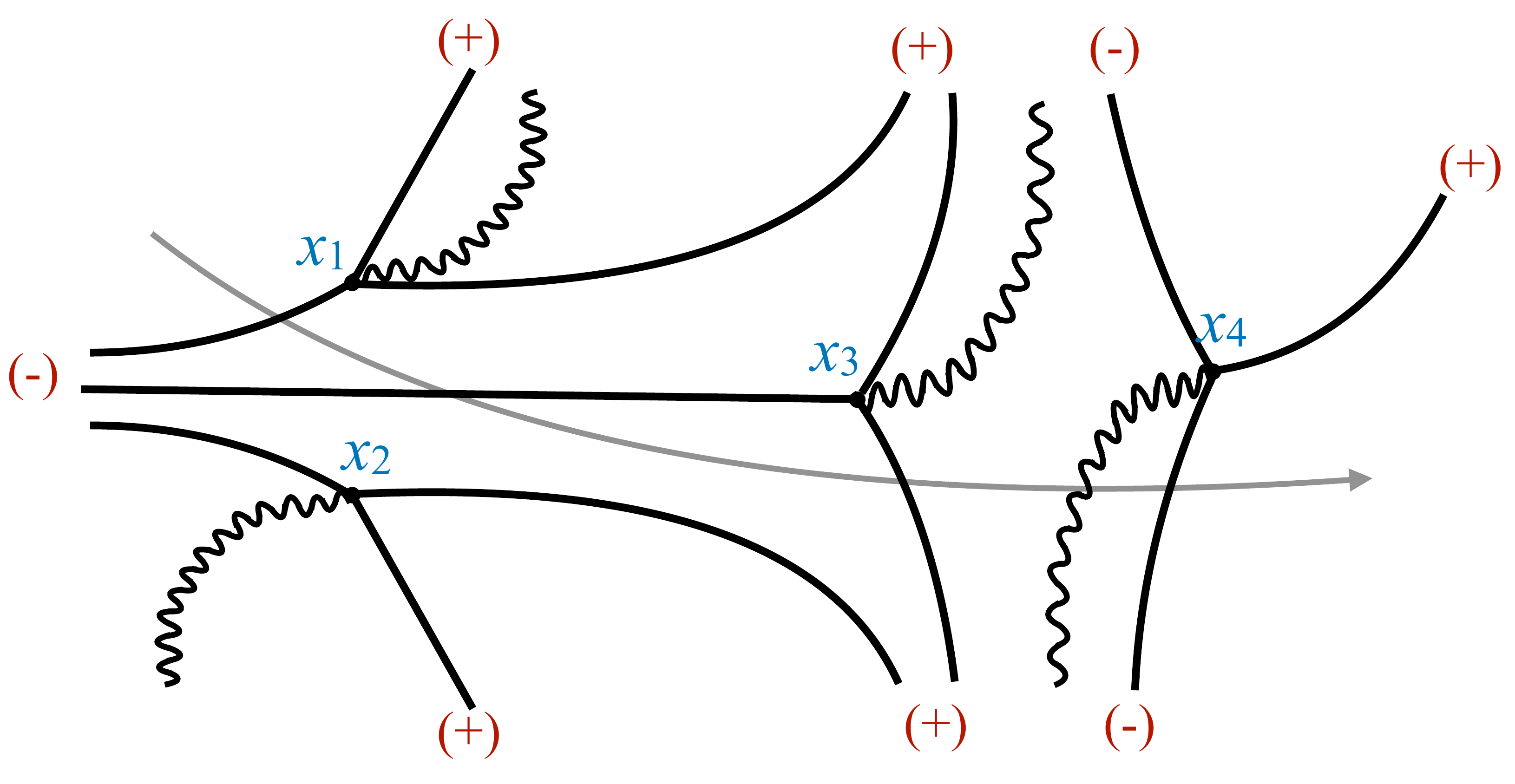}
	\end{subfigure}
	~\hfill 
	\begin{subfigure}[h]{0.45\textwidth}
		\caption{\underline{$\t_\mrmFV = \pi^-$}}	\label{Figure: TDW_StokesDiagram3_Plus}
		\includegraphics[width=\textwidth]{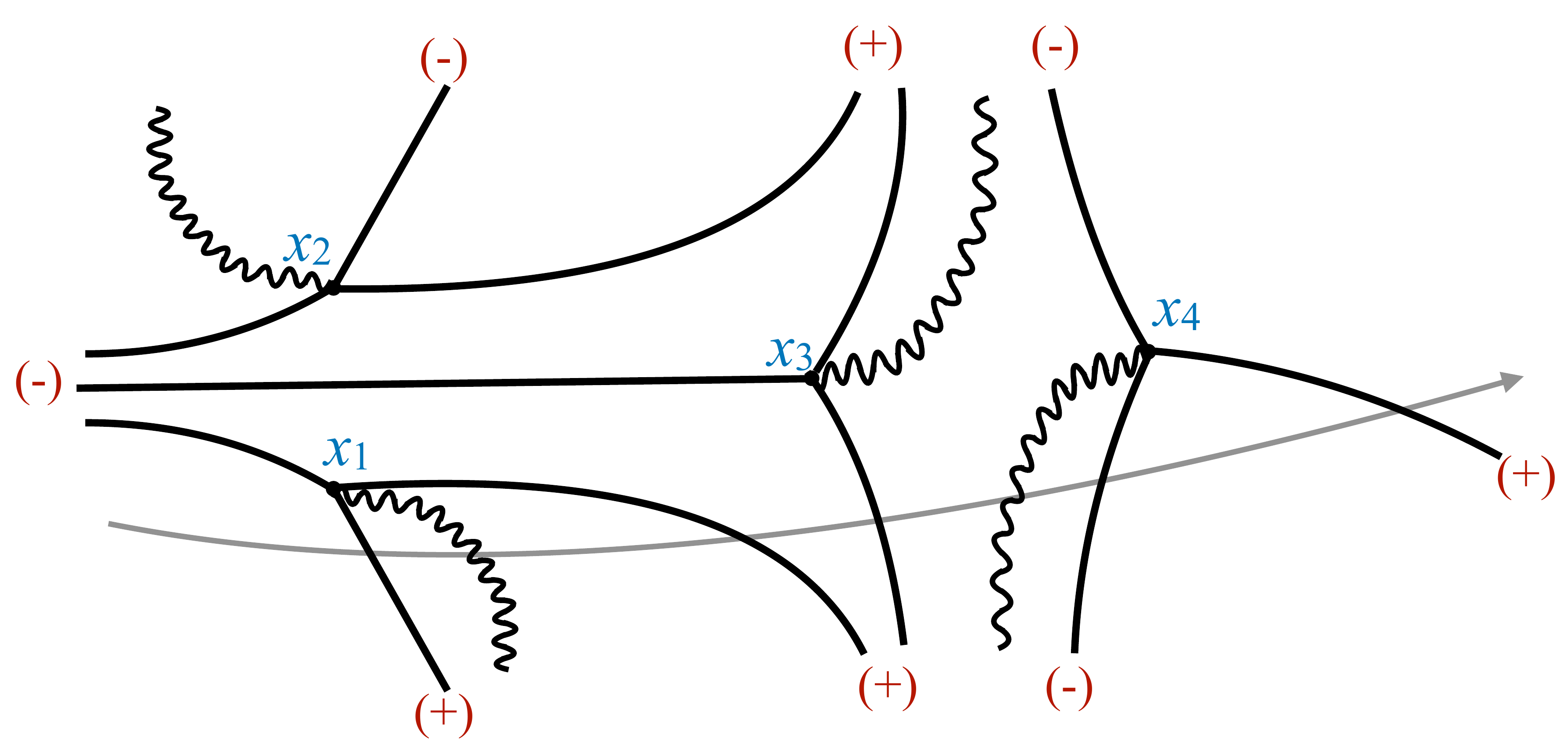}
	\end{subfigure}
	\vspace{6pt}
	\caption{Stokes diagrams for the TDW potential in the three spectral sectors. While $u_\mrmBT< u$ and $u_\mrmFV< u < u_\mrmBT$ are continuously connected to each other, the transition to $u_\mrmTV< u < u_\mrmFV$ is a discontinuous one.} \label{Figure: TDW_Diagrams}
\end{figure}

Let us start with the discussion on the Stokes geometries of the TDW potential in \eqref{TDW_example}. They are illustrated in Fig.~\ref{Figure: TDW_Diagrams} for all three sectors. Note that for convenience, we change the orientation of the branch cuts emerging from $x_1$ and $x_2$ turning points in $\t_\mrmBT = 0^+$ case. This diagram is related to the choice in $\t_\mrmBT=0^-$ by inversion of the $A_1$-cycle and the associated WKB action. (See Appendix A of \cite{Misumi:2024gtf} for details of this connection.) The effect of this inversion is seen in the quantization condition at $\t_\mrmBT=0^+$ and the corresponding analytic continuation $\t_\mrmBT = +\pi^-$, but the physical properties of the quantum system are not affected.

Apart from the branch cut choice, the Stokes geometry of the sector $u_\mrmFV < u <u_\mrmBT$ is a quite well-known one: Topologically it is exactly the same as the symmetric double-well potential \cite{Sueishi:2020rug,Misumi:2024gtf}. The only difference in the present case is that the contributions of $A_1$ and $A_2$ cycles are no longer the same. We elaborate its effects on the spectrum later.

As before, to perform the transition across $u_\mrmBT$, we introduce a phase to it 	 in the Airy-type curve:
\begin{equation}
	P_\mrmTDW = 2 \left(V_\mrmTDW - u_\mrmBT + \left|\l_\mrmBT\right|\,  e^{i\t_\mrmBT} \right)\, . \label{TDW_curve_BT}
\end{equation}
At $\t_\mrmBT = 0^\mp$, the Stokes diagrams for $u_\mrmFV < u < u_\mrmBT$ are illustrated in Fig.~\ref{Figure: TDW_StokesDiagram1_Minus} and Fig.~\ref{Figure: TDW_StokesDiagram1_Plus}. The analytic continuations to $\t_\mrmBT = \mp \pi^\pm$ are the same as the symmetric case, which was discussed in \cite{Misumi:2024gtf} and the resulting diagrams for $u_\mrmBT<u$ sector are illustrated in Fig.~\ref{Figure: TDW_StokesDiagram2_Minus} and Fig.~\ref{Figure: TDW_StokesDiagram2_Plus}. Since the same Stokes lines should be crossed in order to connect $x=\pm \infty$, the quantization conditions in both sectors for the respective analytic continuations become the same:
\begin{align}
	D_{\t_\mrmBT = 0^-} = D_{\t_\mrmBT = -\pi^+} &= \O_\mrmTDW  \Bigg\{\left(1+\Pi_A^{(1)}\right)\left(1+ \Pi_A^{(2)}\right) + \Pi_B \Bigg\}\, , \label{TDW_EQC_Minus1}\\
	D_{\t_\mrmBT = 0^+} = D_{\t_\mrmBT = -\pi^+} &= \O_\mrmATW \Bigg\{\left(1+\left(\Pi_A^{(1)}\right)^{-1}\right)\left(1+ \Pi_A^{(2)}\right) + \Pi_A^{(2)}\Pi_B \Bigg\}\, \label{TDW_EQC_Plus1} , 
\end{align}
where $\O_\mrmATW = \left(\Pi_A^{(1)}\Pi_A^{(2)}\Pi_B\right)^{-1/2}$ and we incorporate the effect of the branch cut choice at $\t_\mrmB = 0^-,-\pi^+$ in \eqref{TDW_EQC_Plus1}.

The transition across $u_\mrmFV$,  on the other hand, is a different one from our previous discussions. Note that in this case, the transition point corresponds to a locally harmonic minimum rather than a maximum. The transition is handled by introducing a complex phase into the Airy-type curve as
\begin{equation}
	P_\mrmTDW = 2\left(V_\mrmTDW - u_\mrmFV - \left|\l_\mrmFV\right|\, e^{i\t_\mrmFV} \right)\, ,
\end{equation}
and performing analytic continuations $\t_\mrmFV:\, 0^\mp \rightarrow \mp \pi^\pm$. For $\t_\mrmFV = 0^\mp$, the Stokes diagrams are the same for $\t_\mrmBT=0^\mp$; then we have
\begin{equation}
	D_{\t_\mrmFV = 0^\mp} = D_{\t_\mrmBT = 0^\mp} \, . \label{TDW_EQC_falseVacuum}
\end{equation}
As we discuss in Section~\ref{Section: EWKB_review}, during this transition, we encounter two Stokes phenomena, which induces the following Stokes automorphisms:
\begin{align}
	\mfrS_B &= \Pi_B \mapsto \Pi_B \left(1+\left(\Pi_A^{(1)}\right)^{\mp 1}\right)^{\pm 1} \, , \label{TDW_StokesAuto_1} \\
	\mfrS_{A_i} &= \Pi_{A}^{(i)} \mapsto \Pi_{A}^{(i)} \left(1+ \Pi_B\right)\, . \label{TDW_StokesAuto_2}
\end{align}
As a result, the exact quantization conditions at $\t_\mrmFV=0$ and $\t_\mrmFV = \mp \pi$  differ from each other.

Let us now compute the quantization conditions for $u_\mrmTV < u <u_\mrmFV$ sector by using \eqref{TDW_StokesAuto_1} and \eqref{TDW_StokesAuto_2}. In light of the transition demonstrated\footnote{Since in TDW, only $A_1$-cycle's orientation changes during the transition, the part of $A_2$-cycle stays the same and the analytic continuation of the Stokes diagrams occur exactly as in Fig.~\ref{Figure: Transition_BelowWell}. If there were a simultaneous transition of another cycle, it should have been taken into account as well.} in Fig.~\ref{Figure: Transition_BelowWell}, we first encounter a degenerate Stokes line connecting $x_1$ and $x_2$ at $\t_\mrmFV = \mp \t_1$, inducing 
\begin{equation}
	D_{\t_\mrmFV = \mp \t_1^{\mp}} = \mfrS_B^{\mp 1}\, D_{\t_\mrmFV = \mp \t_1^\pm}\, , \label{TDW_StokesJump1}
\end{equation} 
where we also use $ D_{\t_\mrmFV = \mp \t_1^\pm} =  D_{\t_\mrmFV = 0^\mp}$. Then, this first Stokes jump leads to the following quantization conditions:
\begin{align}
	D_{\t_\mrmFV = \t_1^+} &= \O_\mrmTDW \sqrt{1+\Pi_A^{(1)}}\left(1+ \Pi_A^{(2)} +\Pi_B \right) \, , \label{TDW_EQC_Minus2} \\
	D_{\t_\mrmFV = \t_1^-} &= \O_\mrmTDW \sqrt{1+\left(\Pi_A^{(1)}\right)^{-1}} \left(1+\Pi_A^{(2)} + \Pi_B\Pi_A^{(2)}\right) \label{TDW_EQC_Plus2} \, .
\end{align}
We observe that in both cases, $\Pi_A^{(1)}$ is disconnected from the other action $\Pi_A^{(2)}$ and $\Pi_B$. This already indicates a significant qualitative difference with respect to $D_{\t_\mrmFV = 0^\mp}$.

After $\t_\mrmFV = \mp \t_1$, the quantization conditions \eqref{TDW_EQC_Minus2} and \eqref{TDW_EQC_Plus2} remains the same until $\t_\mrmFV= \mp \t_2$, where we encounter the second Stokes phenomenon. This time, the Stokes line that connects $x_2$ and $x_3$ becomes degenerate and it induces
\begin{equation}
	D_{\t_\mrmFV = \mp \t_2^\mp} =  \left(\mfrS_{A_1}\right)^{\pm 1} \left(\mfrS_{A_2}\right)^{\pm 1} D_{\t_\mrmFV = \mp \t_2^\pm}\, . \label{TDW_StokesJump2}
\end{equation}
After this point, there is no degeneracy in the Stokes diagrams. Therefore, the quantization conditions at $\t_\mrmFV = \mp \pi$ are given by \eqref{TDW_StokesJump2} as
\begin{align}
	D_{\t_\mrmFV = -\pi} &= \O_\mrmTDW \sqrt{1+\Pi_A^{(1)} + \Pi_B\Pi_A^{(1)}} \left(1 + \Pi_A^{(2)}\right)\, , \label{TDW_EQC_Minus3}\\
	D_{\t_\mrmFV = \pi} &= \O_\mrmTDW \sqrt{1+\left(\Pi_A^{(1)}\right)^{-1} +\Pi_B \left( \Pi_A^{(1)}\right)^{-1}} \left(1+ \Pi_A^{(2)}\right)\, . \label{TDW_EQC_Plus3}
\end{align}
Note that this time $\Pi_A^{(2)}$ is disconnected from other actions. Setting $D_{\t_\mrmFV = \mp\pi} = 0$, it is possible to get two equations to solve. However, only $1+\Pi_A^{(2)} = 0$ has a physical meaning. In principle, the square-root term might be investigated as an exotic system's quantization condition for mathematical curiosity, but we do not carry out such a discussion here.

\subsection{Discontinuity in the Spectrum}
\begin{figure}
\centering
	\includegraphics[width=1\textwidth]{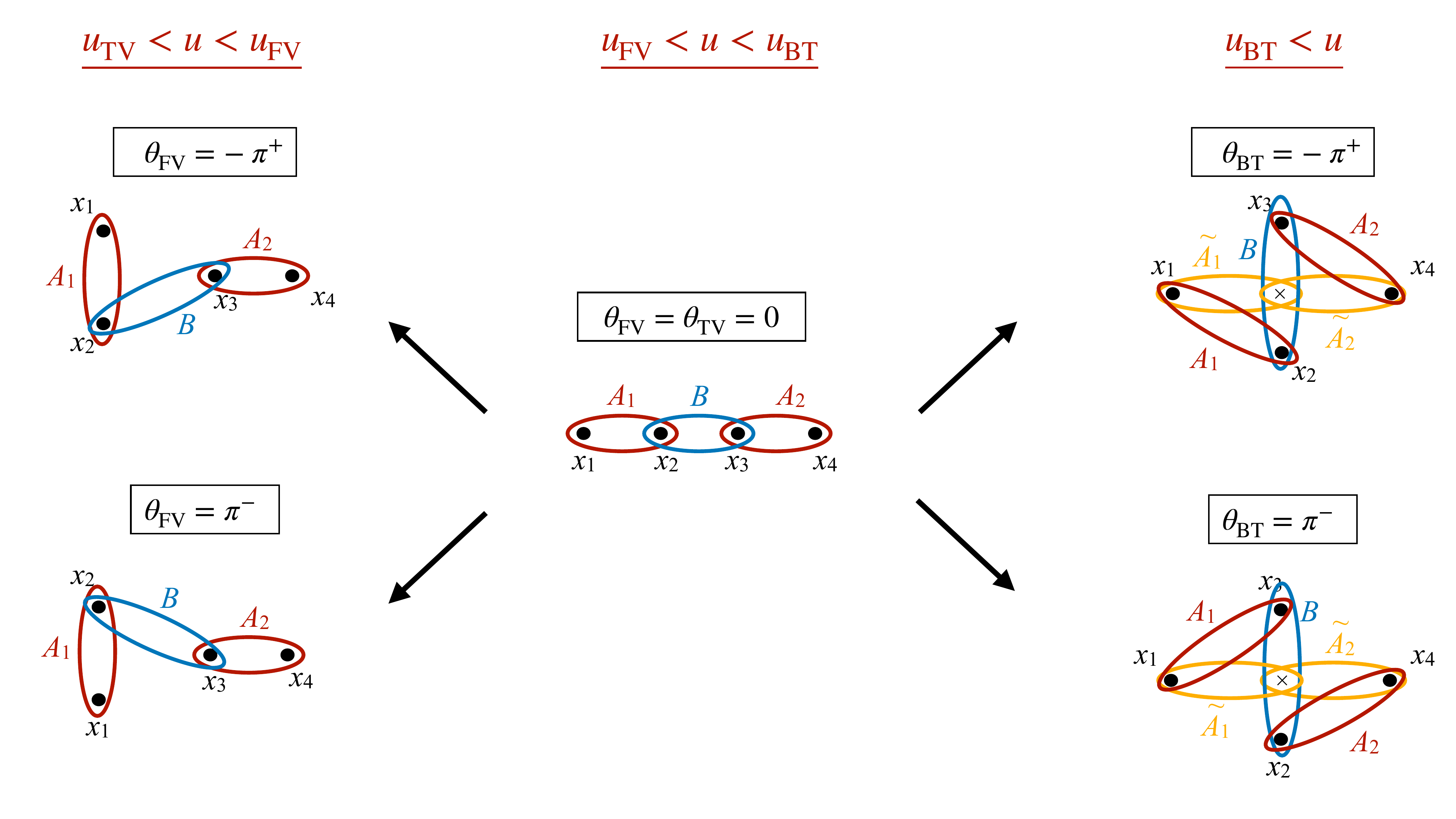}
	\caption{WKB cycles in all sectors of the TDW potential. Same coloured cycle in different sectors are direct analytic continuations of each other. In $u_\mrmBT<u$ sector, the $A_{1,2}$-cycles are redefined as $\tA_{1,2}$-cycles, which are part of the homology basis of the cycles. These re-definitions also keep the median QCs in both sectors intact. The genus-1 geometry is preserved in $u_\mrmTV <u< u_\mrmFV$ sector as well although $A_1$ and $B$ cycles have no contribution to the exact spectrum here.}	\label{Figure: TDW_CyclesAll}
\end{figure}

The transition across different sectors causes changes in the geometry of the WKB cycles as well. We illustrate the geometry in each sector in Fig.~\ref{Figure: TDW_CyclesAll}. Note that although genus-1 geometry is preserved, the $B$-cycle contributes to the physics in the $u_\mrmTV < u <u_\mrmFV$ sector, where the exact quantization condition reads
\begin{equation}
	D_\mrmTV = 1 + \Pi_A^{(2)} = 0 \, . \label{EQC_TrueVacuum}
\end{equation}
This is nothing but the Bohr-Sommerfeld quantization condition
\begin{equation}
	\mcalF_2(\tu,g) = N_\mrmTV + \frac{1}{2} = 0\, ,  \label{TDW_BohrSommerfeld_TV}
\end{equation}
and it shows that the spectrum is perturbatively exact. Note that the associated series still diverges but it is known to be Borel summable \cite{Brezin:1976wa,Cavusoglu:2023bai}. Therefore, there is no need for a non-perturbative cycle to cancel Borel ambiguities. In other words, the WKB cycles other than $A_2$ have zero Stokes constants, meaning that the associated Lefschetz thimbles have no intersection number. Therefore, they play no role in the quantization procedure. 



As a result of these disconnected cycles, \eqref{EQC_TrueVacuum} is also the median quantization condition and it can also be seen from the Stokes diagrams in Fig.~\ref{Figure: TDW_Diagrams} at $\t_\mrmFV = \mp \pi$, which are non-degenerate. In this sense, the perturbative solution of \eqref{EQC_TrueVacuum} corresponds to the entire energy trans-series for $u_\mrmTV < u < u_\mrmFV$.


For $u_\mrmFV < u$, the quantization conditions \eqref{TDW_EQC_Minus1} and \eqref{TDW_EQC_Plus1} can be expressed as
\begin{equation}
	D^\pm_\mrmFV = \left(1+\left(\Pi_A^{(1)}\right)^{\pm 1}\right)\left(1+ \left(\Pi_A^{(2)}\right)^{\pm 1}\right) + \Pi_B =0\, , \label{EQC_FV}
\end{equation}
where contrary to $u<u_\mrmFV$ sector, $B$-cycle contributes to the exact spectrum and as in the symmetric double-well case, it crucially cancels the ambiguities. 

The continuity of the spectrum throughout $u_\mrmFV<u$ can be verified by obtaining the median QC in both sectors. The computations can be carried out in the same way as the symmetric case \cite{Misumi:2024gtf} as well as ATW potential in Section~\ref{Section: ATW_Transition}. Therefore, we do not discuss it here in detail. We express the resulting median QC in $u_\mrmFV<u<u_\mrmBT$ and $u_\mrmBT < u$ sectors as
\begin{equation}
	D_{u_\mrmFV < u}^\med = \sqrt{1+\Pi_B}\left(1 + \Pi_A^{(1)}\Pi_A^{(2)}\right) + \Pi_A^{(1)} \Pi_A^{(2)} = 0 \, , \label{TDW_MedianQC}
\end{equation}
where we incorporate the change in the geometries of $A_{1,2}$-cycles in Fig.~\ref{Figure: TDW_CyclesAll} in view of the discussion in Section~\ref{Section: ATW_Transition} and \cite{Misumi:2024gtf}.

\subsection{Spectrum and Non-perturbative saddles}

The TDW potential \eqref{TDW_example} is a stable system (bounded from below) with a real spectrum. As mentioned above, \eqref{EQC_TrueVacuum} is perturbatively exact, yielding the spectrum to be represented by the exact series
\begin{equation}
	u_\mrmTV = \sum_{g=0}^\infty u_k g^k\, .  \label{TDW_TransSeries_TV}
\end{equation}
Since it is Borel summable, it is straightforward to verify the reality of the corresponding eigenvalues. This exactness comes to a sudden halt at $u=u_\mrmFV$. Above this level, we need to incorporate the non-perturbative shifts into the Weber-type perturbative actions as
\begin{equation}
	\mcalF_1 = N_\mrmFV + \frac{1}{2} + \d_\mrmFV \, , \qquad \mcalF_2 = N_\mrmTV + \frac{1}{2} + \d_\mrmTV \, ,\label{TDW_Ansatz_NP_shift}
\end{equation}
and solve \eqref{EQC_FV} accordingly. As in ATW case, we take one of the perturbative saddles as the basis and solve the quantization conditions using one of the ansatz in \eqref{TDW_Ansatz_NP_shift}. Note that in the cubic deformation of the symmetric double-well, the perturbative levels above $u>u_\mrmF$ are not degenerate, and $\mcalF_1$ and $\mcalF_2$ are related to each other via a perturbative shift as
\begin{equation}
	\mcalF_2 = \mcalF_1 + \D_\mrmTDW \, . \label{TDW_DifferencePerturbative} 
\end{equation}
Then, the solutions \eqref{EQC_FV} for the two cases in \eqref{TDW_Ansatz_NP_shift} differ only by the sign of $\D_\mrmTDW$. Thus, in the following, we only solve \eqref{EQC_FV} to see corrections to the perturbative states around $x=z_\mrmFV$.

A straightforward implementation of the Weber-type expressions \eqref{Dictionary_NP_Well_Bounce} to \eqref{EQC_FV} yields
\begin{equation}
	\frac{1}{\Gamma\left(-N_\mrmFV -\d_\mrmFV\right)} = -\mcalK_\mrmTDW\,  \, e^{\pm 2\pi i \left(N_\mrmFV + \frac{1}{2}\right)}\, e^{\pm i \pi \D_\mrmTDW} \sum_{k=0} \tH_k \d_\mrmFV^k\, ,  \label{TDW_intermediateSoln}
\end{equation}
where the fugacity term is
\begin{equation}
	\mcalK_\mrmTDW = \frac{e^{-\frac{S_\mrmTDW}{g}}}{2\sqrt{2\pi}}\left(\frac{c}{g}\right)^{N_\mrmFV +\frac{1}{2}}\, ,\quad  c= \frac{8}{\g (\g+2)}\, ,
\end{equation}
and to obtain the series in $\d_\mrmFV$, we use \eqref{expansion_bionATW} and define
\begin{equation}
	\sum_{k=0}^\infty \tH_k \d^k_\mrmFV := \frac{e^{\d \left(\log\frac{c}{g} \pm 2 \pi i \right)}}{\cos\left[\pi \left(\frac{1}{2}+N_\mrmFV + \D_\mrmTDW + \d_\mrmFV\right)\right]} \sum_{k=0}^\infty  H_k \d_\mrmFV^k\, . 
\end{equation}
Then, as in ATW case, using a series ansatz
\begin{equation}
	\d^\pm_\mrmFV = \sum_{n=0}^\infty \a_n^\pm \left(\mcalK_\mrmTDW\right)^n \,  , \label{TDW_seriesAnsatz}
\end{equation}
we solve \eqref{TDW_intermediateSoln} and get 
\begin{equation}
	\d^\pm_\mrmFV \sim -\frac{1}{N_\mrmFV !} \mcalK_\mrmTDW\, e^{\pm \pi i N_\mrmFV } e^{\mp \frac{i\pi}{2}} \frac{e^{\mp \frac{i\pi}{2}} e^{\pm i \pi\left(N_\mrmFV +  \D_\mrmTDW\right)} }{\cos\left[\pi \left(\frac{1}{2}+N_\mrmFV + \D_\mrmTDW \right)\right]}\, . \label{TDW_NP_LeadingOrder}
\end{equation}
Then, we obtain the leading order non-perturbative real and imaginary contributions as
\begin{align}
	\Re \d_\mrmFV &\simeq -\frac{1}{N_\mrmFV !}\frac{e^{-\frac{S_\mrmTDW}{g}}}{\sqrt{\pi}}\, \left(\frac{1}{g\g \left(\g+2\right)}\right)^{N_\mrmFV+\frac{1}{2}}\, \cot\left[\pi \left(N_\mrmFV + \D_\mrmTDW\right)\right] \, ,\label{TDW_realshift} \\ \nonumber \\
	\Im \d^\pm_\mrmFV &\simeq \pm \frac{e^{-\frac{S_\mrmTDW}{g}}}{\sqrt{\pi}\, N_\mrmFV ! } \left(\frac{1}{g\, \g\left(\g+2\right)}\right)^{N_\mrmFV + \frac{1}{2}} \, . \label{TDW_imaginaryNP}
\end{align}
Both parts directly lead to the first non-perturbative sector in the trans-series expression using
\begin{equation}
	u_\mrmFV \simeq \ve^\mrmFV(N_\mrmFV) + \d_\mrmFV \frac{\dee \ve^\mrmFV}{\dee \mcalF}\Big|_{\mcalF=N_\mrmFV + \frac{1}{2}} + O(\d_\mrmFV^2) \, , \label{TDW_energy_TransSeries}
\end{equation}
where $\ve^\mrmFV$ is the perturbative series around $x=z_\mrmFV$ and $\frac{\dee \ve^\mrmFV}{\dee \mcalF} \simeq \o_\mrmFV = 1$. Since TDW is a stable quantum system, the imaginary contribution should be canceled by the Borel ambiguity arising from $u_\mathrm{pert}$. The divergent behaviour for the ground state perturbation series is
\begin{align}
	u_k 
	&\sim \frac{1}{\pi \sqrt{\g\left(\g+2\right)}}\, \frac{\Gamma\left(k+\frac{1}{2}\right)}{S_\mrmTDW^{k+\frac{1}{2}}}\, , \label{TDW_largeOrder}
\end{align}
and it perfectly coincides with the divergent series associated with \eqref{TDW_imaginaryNP}, verifying the leading order ambiguity cancellation. 

The cancellation between \eqref{TDW_imaginaryNP} and \eqref{TDW_largeOrder} is an old and well-known fact \cite{Brezin:1976wa}, which was also extended recently up to NNLO by using the fluctuation series around the \eqref{TDW_imaginaryNP} \cite{Cavusoglu:2023bai}. At a first glance, it seems that the cancellation is due to the direct link between the divergent perturbation series and the bounce configurations emerging from $x=z_\mrmFV$. However, via \eqref{TDW_NP_LeadingOrder}, our discussion reveals a more intricate structure for the non-perturbative spectrum around $u=u_\mrmFV$. Let us investigate \eqref{TDW_NP_LeadingOrder} more carefully:


\begin{itemize}[wide]
	\item Our first observation is that $\Im \,\d_\mrmFV$ originates from the exponential $e^{\pm i\pi\D_\mrmTDW}$, where  $\D_\mrmTDW$ is an analogue of HTA, which is known from the SUSY version of TDW potential \cite{Behtash:2015zha, Behtash:2015loa,Behtash:2017rqj}. Consequently, the trans-series \eqref{TDW_energy_TransSeries} is in the same form with the SUSY TDW with $\z$ deformation. This is expected since the Stokes geometries in both cases are the same\footnote{Comparing to TDW, we observe that geometrically the SUSY DW is restricted in the $u>u_\mrmFV$ sector. This is simply because in the latter case, the deformation is a quantum one and as a result the Stokes geometry is not effected by the deformation at all and the transition to $u< u_\mrmFV$ sector we discuss here does not happen.} \cite{Kamata:2021jrs}. In the SUSY case, however, the HTA could have integer values leading to vanishing imaginary non-perturbative contributions. These are either SUSY or quasi-exactly solvable cases specific to the linear deformation of the double-well potential and the cubic deformed case does not enjoy such symmetries\footnote{More precisely, this is true when the deformation parameter is real. If the cubic deformation is allowed to be pure imaginary, the spectrum of the corresponding PT-symmetric Hamiltonian can be quasi-exactly solvable \cite{Bender:1998kf}.}.
	\item For the SUSY potential, such an imaginary contribution is associated with a complex bion, which originally arises from the fermionic degrees of freedom \cite{Behtash:2015loa}. After integrating out the fermion, the TDW type potential arises and can be interpreted as a quartic elliptic curve, which encodes the quantum dynamics in the same way with $V_\mrmTDW$ in \eqref{TDW_example} . In this reduced bosonic setup, the complex bion was shown to be linked to the analytic continuation of the real bounce to the $2^\nd$ Riemann sheet of this curve.
	\item In \cite{Behtash:2015loa}, the authors anticipated a similar structure for the classical cubic deformation of the double-well, but didn't discuss the details. To our knowledge, except this small note in \cite{Behtash:2015loa}, the existence of HTA and its consequences has not been addressed in the literature and the resurgence structure of TDW has been restricted to the link between real bounce solution and the divergence of the pertubation series, as we mention above. The structure of \eqref{TDW_NP_LeadingOrder}, however, suggests otherwise: 
	\begin{itemize}
		\item First, we observe that the bounces are known to have imaginary factors due to the negative modes appearing in the fluctuation determinants. For TDW potential, such contribution should appear as $\sqrt{-\frac{1}{g\, \g(\g+2)}}$ in $\mcalK_\mrmTDW$.
		\item  In \eqref{TDW_NP_LeadingOrder}, the first $e^{\pm \frac{i\pi}{2}}$ might have created this contribution, and it eventually cancels against the other $e^{\pm \frac{i\pi }{2}}$ term. As a result, the imaginary non-perturbative contribution arises from the HTA term via \[\Im\, e^{\pm i\pi\left(N_\mrmFV + \D_\mrmTDW\right)} = \sin\left[\pi \left(N_\mrmFV + \D_\mrmTDW\right)\right]\, ,\] as in the SUSY case. Moreover, the real non-perturbative part in \eqref{TDW_realshift} also cannot stem from the standard real bounce contribution, which is purely imaginary.
	\end{itemize}
	
	\item Guided from the non-perturbative structures in the SUSY case, it is expected that the latter imaginary contribution and the HTA parts arise from an analytic continuation of the real bounce. In fact, in Section~\ref{Section: P_NP_revisit} we discussed such an analytic continuation in the context of P-NP relations. Now examining Fig.~\ref{Figure: Deformation_InstantonCycle}, we observe that the analytic continuation $\g \rightarrow \g e^{2\pi i}$ indeed incorporates $A_2$ cycle into the $B$-cycle in the $2^\nd$ Riemann sheet\footnote{As illustrated in Fig.~\ref{Figure: Deformation_Linear}, in the linearly deformed case, such an analytic continuation does not change the geometry of the cycles. However, this does not contradict with our observation as in the SUSY case, the HTA appear as a real number rather than being related to the perturbative quantization of $A_2$-cycle.}, which induces the HTA contribution through the identification of the action for $A_2$-cycle in \eqref{TDW_DifferencePerturbative}. 
	
	\item The picture we discuss above predicts the existence of a complex saddle contributing to the path integral and playing the role of the complex bion for TDW potential. Note that the analytic continuation that yields the HTA is not the same as the one in the SUSY case in \cite{Behtash:2015zha, Behtash:2015loa}. In the latter, the analytic continuation maps the false-vacuum to the true vacuum, which possesses a complex bion. An equivalent mapping in our case, i.e. $\g \rightarrow -\g$, yields a complex bounce. Its action, however, does not have the same real part with the real bounce and the corresponding imaginary part cannot be related to the HTA defined in \eqref{TDW_DifferencePerturbative}. In fact, the exact quantization condition of the true-vacuum sector in \eqref{EQC_TrueVacuum} indicates that the complex bounce does not contribute to the spectrum. This leaves $\g\rightarrow \g e^{2\pi i}$ as the only option for such an analytic continuation, which fits our observation. To settle the issue, a precise path integral analysis in this line might be very illuminating in identifying all contributing thimbles to the exact quantization around the false-vacuum.
\end{itemize}

\subsection{P-NP relations for all saddle points}\label{Section: TDW_PNP}
Finally, let us complete the discussion on the P-NP relation of TDW potential. In Section~\ref{Section: P_NP_revisit}, we have already shown that the parameters $f_{2,3}(\g)$ are the same for all P-NP relations in TDW potential. The $f_1(\g)$ term, on the other hand, no longer transforms as in \eqref{f1_FreqDependence}. To quickly observe that latter, let us express the perturbative expansions around $x=z_\mrmFV$, $x=z_\mrmTV$ and $x=z_\mrmBT$ up to $O(g^1)$:
\begin{align}
	\tu_\mrmFV & = \o_\mrmFV \mcalF - \frac{1}{16\o_\mrmFV} \Big[4+7 \g (\g+2)+12(4+5 \g (2+\g))\mcalF^2 \Big] g \, , \label{TDW_energyFV} \\
	\tu_\mrmTV & = \o_\mrmTV \mcalF -  \frac{11 + 43 \g (\g+2)+  5(\g+1)\D + 12\Big(9 + 41 \g (\g+2) + 7 (\g+1) \D\Big)\mcalF^2}{8\, \o_\mrmTV^2}\, g \, , \label{TDW_energyTV} \\
	\tu_\mrmBT & = \o_\mrmBT \mcalF + \frac{11 + 43 \g (\g+2) -  5(\g+1)\D + 12\Big(9 + 41 \g (\g+2) - 7 (\g+1) \D\Big)\mcalF^2}{8\, \o_\mrmBT^2}\, g \, , \label{TDW_energyBT}
\end{align}
where $\D = \sqrt{1 + 18 \g + 9\g^2}$ and the frequencies are given as 
\begin{align}
	\o_\mrmFV &=1\, , \qquad \; \o_\mrmTV = \frac{1}{2}\sqrt{1+ 9 \g(\g+2) + 3(\g+1)\D} \, , \nonumber \\
	\o_\mrmBT &= \frac{1}{2}\sqrt{1+ 9 \g(\g+2) - 3(\g+1)\D}\, .
\end{align}
Note that there is an overall frequency dependence in $\mcalF^2 g$ term for all cases. However, the numerators are no longer equal or even proportional to each other; and therefore, the duality transformations of the classical actions, i.e. \eqref{proportional_ClassicalActions} and \eqref{S-duality_ClassicalActions}, do not hold except the leading order, which is the exact classical limit. 

Despite the loss of the duality, however, the non-perturbative fluctuations, $\mcalG^{(i)}_n$ remain in the same form for all saddles. For the potential \eqref{TDW_example}, we write the order by order solution to the deformed P-NP relation as
\begin{align}
	\mcalG^{(i)}_n(\mcalF) &= \frac{2^n}{\Big(\g(\g+1)(\g+2)\Big)^n} \left\{c_n \,+ \, \frac{1}{3}\,\int \mrmd g\, \left(\frac{2}{\g(\g+1)(\g+2)}\right)^n  \frac{\dee \tu^{(i)}_{n+1}}{\dee \mcalF} \right\} \, , \label{TDW_NP_corrections}
\end{align}
where $c_n$ is an integration constant which should be determined at each order. Note that due to $R^{(i)}(\g) = \g(\g+1)(\g+2)$ in the denominator, each term is singular at particular values of $\g$, where $\mcalG_n^{(i)}$ is known to be analytic. Then, $c_n$ can be determined  by canceling these singular contributions. This procedure was explained in  \cite{Cavusoglu:2023bai,Cavusoglu:2024usn} and performed for the P-NP relation for the saddle around $x=z_\mrmFV$. Our computations generalize their discussion to other saddle points, which has not been analyzed before.

Note that for clarity in the expressions, we set $\g=\frac{1}{20}$ without losing any generality in our discussion. Then, the action of perturbative saddles are found as
\begin{align}
	\mcalF_\mrmFV &= \tu+\left(0.294844\, +3.38438 \tu^2\right) g +\left(9.1647 \tu+46.5209
	\tu^3\right) g^2 \dots \, , \\
	\mcalF_\mrmTV &= 0.797445 \tu+\left(0.114314\, +0.899507 \tu^2\right) g +\left(0.993087
	\tu+3.82478 \tu^3\right) g^2\dots \, , \\
	\mcalF_\mrmBT &= 1.27903 i \tu+\left(-0.434727 +3.16183 \tu^2\right)i g +\left(11.9490 \tu-38.1661  \tu^3\right)i g^2 \dots   \, , 
\end{align}
As expected there is no symmetry between these actions. Inverting them provides the series for $\tu^{(i)}$, whose first two terms were given in \eqref{TDW_energyFV}-\eqref{TDW_energyBT}. Finally, inserting each expression in \eqref{TDW_NP_corrections}, we obtain the fluctuations around instantons. Then, combining them with the instanton actions in \eqref{BounceAction_TDW_General} with the turning points \eqref{TurningPoints_FV}-\eqref{TurningPoints_BT}  at $\g=\frac{1}{20}$, we write the instanton functions as 
\begin{align}
	\mcalG_\mrmFV &= -\frac{0.235997}{g}+\left(-1.05863-28.0627 \mcalF^2\right) g
	+\left(-68.4749 \mcalF-106.842 \mcalF^3\right) g^2 \dots \, , \\
	\mcalG_\mrmTV &= -\frac{0.77453\, +0.169057 i}{g}-\left(1.50773-1.16155 \mcalF^2\right) g
	- \left(22.2423 \mcalF-30.4083 \mcalF^3\right) g^2 \dots  \, , \\
	\mcalG^{(\mrmL/\mrmR)}_\mrmBT &= 
	\frac{S_\mrmBT^{(\mrmL/\mrmR)}}{g}+ \left(1.70125 + 6.06196 \mcalF^2\right)ig +\left(35.5138 \mcalF+44.5035 \mcalF^3\right) g^2 \dots  \, . \label{NP_barrierTop}
\end{align}
where in the last equation $(\mrmL/\mrmR)$ refers to Left/Right non-perturbative cycles and the associated actions are found as
\begin{equation}
	S_\mrmBT^{\mrmLeft} = -0.194511 i\; \qquad S_\mrmBT^{\mrmRight} = -0.532625 i\, . 
\end{equation}
We independently verified these results via Weber-type EWKB formula in Section~\ref{Section: EWKB_review}. They confirm the invariance of $f_{2,3}$ terms in the P-NP relation for the TDW potential. Although the details of our computations use the specific form in \eqref{TDW_example}, we expect the invariance property of $f_{2,3}$ to hold for any TDW potential and its dual. A generalization to other deformed genus-1 cases and higher genus potentials, if/when a resurgent P-NP relation is established, would be interesting to study in the future. 

\paragraph{\underline{Remarks}:} Before closing this section, let us comment on the barrier-top (or equivalently the dual) case, which has two bounces configurations. 
\begin{itemize}[wide]
	\item  Since \eqref{TDW_NP_corrections} depends only on the perturbative fluctuations, the expression \eqref{NP_barrierTop} is the only fluctuation term for both bounce configurations. From the perspective of the P-NP relation, this interesting result is simply due to the invariance of $f_{2,3}$ terms. Note that it is also different from the undeformed triple-well in Section~\ref{Section: ATW_SymmetricLimit}, where $f_2(0)$ term changes accordingly to the changes in the frequency of the perturbative saddle or the bion/bounce action. (See \eqref{transformationSTW_f2}). In some sense, turning on the deformation parameter balances the change of the action and keep $\frac{\det M^{(i)}}{\o_i}$, so that $f^{(i)}_2$ and $f^{(i)}_3$ invariant for P-NP relations around all saddle points.
	
	\item We can also examine having only one independent fluctuation for two bounces in EWKB language: Using the Weber-type formulas, the corresponding actions for the bounces on the right and the left are written as 
	\begin{equation}\label{NPactions_BarrierTop}
		\Pi_B^\mrmLeft = \frac{\sqrt{2\pi}\, e^{-\mcalG^\mrmLeft_\mrmBT(\mcalF,g)}}{\Gamma\left(\frac{1}{2}+\mcalF_\mrmBT\right)}\left(\frac{\mcalC_\mrmBT}{g}\right)^{\mcalF_\mrmBT}  , \;\, \Pi_B^\mrmRight = \frac{\sqrt{2\pi}\, e^{-\mcalG^\mrmRight_\mrmBT(\mcalF,g)}}{\Gamma\left(\frac{1}{2}+\mcalF_\mrmBT\right)}\left(\frac{\mcalC_\mrmBT}{g}\right)^{\mcalF_\mrmBT}
	\end{equation}
	where
	\begin{align}
		\mcalG_\mrmBT^\mrmLeft &= \frac{S_\mrmBT^\mrmLeft}{g} + \sum_{n=0}^\infty \mcalG_n\, g^n\, , \label{NP_function_BT_Left} \\
		\mcalG_\mrmBT^\mrmRight &= \frac{S_\mrmBT^\mrmRight}{g} + \sum_{n=0}^\infty \mcalG_n\, g^n \, , \label{NP_function_BT_Right}
	\end{align} 
	and $\mcalC_\mrmBT\simeq 3.65411i$ for both cases. We observe that the only difference for the expressions in \eqref{NPactions_BarrierTop} is due to the bounce actions and it is the reason behind the broken parity symmetry of the corresponding dual quantum action $a^\mrmD$. However, this is limited to the classical term and the remaining parts are the same.
\end{itemize}

\section{Summary and discussion}
The analysis in this paper extends the application of EWKB formalism in all spectral sectors to general locally harmonic one-dimensional quantum systems without any symmetries. This complements our previous study \cite{Misumi:2024gtf}, which only dealt with the symmetric potentials with degenerate saddles. Such an extension reveals significant qualitative differences from the symmetric cases, which we verified quantitatively using the Weber-type EWKB formulas reviewed in Section~\ref{Section: Weber-EWKB}. Let us further elaborate the details of our findings in connection with possible future directions: 

\begin{itemize}[wide]
	\item {\bf Analytic continuation with Stokes phenomenon:} In Section~\ref{Section: Transition}, we discuss the analytic continuation across the minima of a locally harmonic perturbative well and showed the existence of Stokes phenomena during such a transition. Contrary to a transition across a barrier top, this signifies a discontinuous analytic continuation, leading to multiple exact quantization conditions; thereby multiple trans-series representations throughout the spectrum. We illustrated a generic qualitative picture in Fig.~\ref{Figure: GenericPotential_DifferentSectors} for a generic potential.
	\item {\bf Transformation of P-NP relations:} In Section~\ref{Section: P_NP_revisit}, using the general formula for the genus-1 P-NP relation and its order by order series solution, we established an explicit dependence on $\o_i$ (frequency) and $S_\mcalB^{(i)}$ (bion/bounce) action for $f_{2,3}(\g)$ terms. We also showed the $\o_i$ dependence for $f_1$ term in the undeformed limit, i.e.  $\g\rightarrow 0$. In this way, we provided explicit links between the P-NP relations of different pairs of WKB cycles of a given (deformed or undeformed) genus-1 potential via only classical parameters $\o_i$ and $S_\mcalB^{(i)}$. Later, in Section~\ref{Section: ATW_SymmetricLimit} and Section~\ref{Section: TDW_PNP}, we verify these rules, by providing numerical data for symmetric triple-well and tilted double-well potentials, respectively. Note that in the original formulation, a P-NP equation shows a link between the perturbative and non-perturbative data of neighbouring cycles. Then, in some sense, these transformation rules connect all the perturbative and non-perturbative data even at distinct local patches of the classical algebraic curve as well as the dual curves. Finally, the generalization of the P-NP relations to higher genus curves is still an open problem. We anticipate that similar transformation properties would hold in such generalizations and possibly, they can be used as a guide together with the deformation of symmetric potentials in line of \cite{Cavusoglu:2024usn}.
	 
	\item {\bf Asymmetric triple-well and PT-symmetric dual:} In Section~\ref{Section: ATW}, we focused on ATW, which has non-degenerate saddles and multiple independent non-perturbative bions governing its spectrum. Performing an exact quantization and using their medianization, we found an explicit condition that relates the actions $A$-cycles as $\Pi_A^{(1)} \Pi_A^{(3)} = \Pi_A^{(2)}$, which must be satisfied for the reality of the spectrum in all sectors. Later, we numerically verified the spectral reality via the large-order low-order resurgence relations, which also verifies this condition. This provided a first-time demonstration of the reality of the spectrum via resurgence cancellation in an asymmetric system with independent non-perturbative contributions.  Finally, we presented the exact quantization of the dual PT-symmetric system and verified the reality of its spectrum via perturbative non-perturbative ambiguity cancellations. Note that for both ATW and its dual potential, the associated trans-series are organized accordingly to the cluster expansion of the weakly interacting instanton gas picture. Generalizing this property to asymmetric cases indicates that it is indeed a significant organizing principle for the semi-classical quantization as it was argued in symmetric systems before \cite{Behtash:2018voa}.

	\item {\bf Tilted double-well and complex saddles:} In Section~\ref{Section: TDW}, using the analytic continuation techniques developed in Section~\ref{Section: Transition}, we showed the existence of distinct trans-series solutions representing the exact spectrum of TDW potential in false vacuum and true vacuum sectors. We linked the sudden change in the trans-series structure to the Stokes phenomena during the analytic continuations, which leads to vanishing $B$-cycle contributions around the true vacuum. We discussed the trans-series solution around the false vacuum and presented a non-perturbative imaginary ambiguity associated with an analogue of hidden topological angle in SUSY theories. While the cancellation of the ambiguous contribution against the perturbative one is a well-known fact, we argued that it should be linked to a complex non-perturbative saddle rather than the real bounce. We demonstrated this link via analytic continuation of the $B$-cycle to the $2^\nd$ Riemann sheet in light of the complex bion discussion in \cite{Behtash:2015loa}. A follow-up based on the path integral formalism via Picard-Lefschetz theory would be interesting and necessary to solidify this relationship. Such an analysis would also strengthen and expand the link between EWKB formalism and semi-classical path integrals, which was recently discussed in relation to the Gutzwiller type quantization \cite{Sueishi:2020rug,Sueishi:2021xti,Ture:2024nbi}.
\end{itemize}

\section*{Acknowledgements}
We thank Syo Kamata for correspondence and Alireza Behtash for a very fruitful discussion on this work.
This work of T. M. is supported by the Japan Society for the Promotion of Science (JSPS) Grant-in-Aid for Scientific Research (KAKENHI) Grant Numbers 23K03425 and 22H05118. C.P. thanks Kindai University for the hospitality of the university where the initial steps of the project took place.
The beginning of this work is also owed to the discussion in the workshop ‘Invitation to Recursion, Resurgence and Combinatorics’ at Okinawa Institute of Science and Technology Graduate University (OIST).

\bibliographystyle{JHEP}
\bibliography{EWKB.bib}

\end{document}